\documentclass[pdflatex,iicol]{sn-jnl}%
\usepackage{url,hyperref}
\usepackage[sort, numbers]{natbib}
\jyear{2023}%

\begin{document}

\let\WriteBookmarks\relax
\def\floatpagepagefraction{1}
\def\textpagefraction{.001}

\title [mode = title]{Simulation of Lateral Impulse Induced Inertial Dilation at the Surface of a Vacuum-Exposed Granular Assembly\footnotemark[2]}

\author*[1]{\fnm{Eric S.} \sur{Frizzell}}\email{efrizz@umd.edu}

\author[1]{\fnm{Christine M.} \sur{Hartzell}}

\affil[1]{\orgdiv{University of Maryland}, \orgname{Aerospace Engineering}, \orgaddress{\street{3178 Glenn L. Martin Hall}, \city{College Park}, \state{MD} \postcode{20740}, \country{United States}}}

\abstract{We demonstrate for the first time that a lateral impulse experienced by a granular channel can induce an inertial bulk dilation over long distances across a granular medium with a mechanically free surface. The surface dilation requires zero overburden pressure (exposure to vacuum) and is precipitated by the passing of waves traveling barely above the sound speed (> Mach 1.05). We simulate this phenomenon using open source Soft Sphere Discrete Element Method (SSDEM) software. We prepare channels of monodisperse, cohesive spherical particles exposed to vacuum and modeled as Hertzian springs. We validate our model by recreating acoustic wave, strong shock, and shear dilation behavior. We then create shocks within the channel to determine the sensitivity of surface dilation to wave speed, wave type, initial packing fraction, and boundary effects. The shocks we create undergo a rapid decay in strength and appear to propagate as solitary waves that can be sustained across the channel. We find that an inertial surface dilation is induced by compressive solitary waves, is insensitive to channel length, increases with bed height, and increases substantially with initial packing fraction. A hard subsurface floor is required to maintain this wave over the entire channel. Free surface dilation induced by laterally propagating impulse loading could be implicated in the formation of Lunar Cold Spots, distal regions of low thermal inertia surrounding young craters on the Moon.}

\keywords{Inertial dilation, microgravity, SSDEM, granular shock, solitary wave, Lunar Cold Spots}

\maketitle

\begingroup
\renewcommand\thefootnote{\fnsymbol{footnote}} 
\footnotetext[2]{This version of the article has been accepted for publication, after peer review but is not the Version of Record and does not reflect post-acceptance improvements, or any corrections. Version of Record: https://doi.org/10.1007/s10035-023-01363-6. Use of this version subject to the publisher’s terms: https://www.springernature.com/gp/open-research/policies/accepted-manuscript-terms.}
\endgroup

\section{Introduction} \label{sec:DilationInGrains}

Granular dilation is the property of a compacted granular material (packing fraction $\phi >$ 0.58, \cite{royer2011role}) to undergo bulk volume increase as a result of shear-induced grain deformation, first documented by \cite{reynolds1885lvii}. This dilation is commonly studied in applications ranging from natural (tectonic-induced dilation, \cite{tillemans1995simulating}) to industrial (dilation waves in grain silos, \cite{wensrich2002experimental}). Bulk density increase via granular dilation can occur gradually as a result of inter-grain contacts (through vibration (\cite{philippe2002compaction}) or at a shear surface interface (\cite{sture1998mechanics})) or as the result of a more kinematic rearrangement of particles due to large strain rates (\cite{tournat2005experimental}, \cite{van2012auto}). The latter (`inertial dilation') occurs during rapid granular flow, where grain movement can be modeled by the kinetic theory of gasses (\cite{campbell2005stress}) and particle motion is dominated by collisions. However, the movement of grains in this regime is not well characterized and few studies exist that examine this behavior in microgravity (\cite{daniels2013rubble}). Recently we have seen that, often, granular materials in microgravity and low confinement pressure do not behave as expected. Some examples include the unexpected properties of Bennu's surface as revealed by  OSIRIS-REx  (\cite{walsh2022near}) and parabolic-flight experiments that reveal microgravity suppression of secondary granular flow (\cite{murdoch2013convection}). Examples specific to high speed impacts include the ability of a single pulse from a granular wave to produce previously unknown effects. \cite{shinbrot2017size} showed that, while long thought to only occur as the result of shaking, the Brazil Nut Effect can occur as the result of a single granular pulse. \cite{wright2020boulder} also saw that boulder stranding will occur far from the impact source as the result of only a single impulsively generated wave.  Given these examples, inertial dilation under microgravity and unconfined upper boundary deserves further exploration.

In this paper we are particularly interested in the behavior of surface grains far away from the source of impact-triggered ground shocks (\cite{farr1986loading}) on the Moon (\cite{holsapple1993scaling}). Do surface grains on the Moon exhibit a different response to the regimes of wave strength experienced as an impulse travels and evolves than they do on the Earth? We have found no prior work that quantifies vertical bulk volume changes induced by laterally propagating granular waves at the surface of a mechanically free interface under reduced gravity. The literature largely focuses on waves traveling vertically toward the free surface (\cite{aoki1995simulation}, \cite{wensrich2002experimental},  \cite{gusev2008reflection}, \cite{sanchez2022transmission}, \cite{tancredi2022lofting}) or shock studies that quantify the lateral extent of dilation behind the shock front (\cite{goldshtein1996mechanics}) at atmospheric or greater confinement pressures (\cite{omidvar2012stress}, \cite{lu2018state}). Furthermore, experiments examining lateral shock propagation can be limited given difficulty in preparing long samples and uncertainty of results caused by invasive measurement techniques (\cite{omidvar2012stress}). Given the challenges of experiments studying long range wave propagation and recreating the Lunar environment, we employ particle based modeling to study laterally propagating surface waves. Numerical simulations of terrestrial impact cratering response by \cite{collins2014numerical} showed that a narrow dilation band exists in the near surface as a result of an impact, but their modeling method is unable to account for individual grain motion. Dilation is restricted to near surface depths since the confining pressure of the overburden (grains supported by a particle at depth) quickly turns shear forces towards compaction (\cite{gowd1980effect}, \cite{brown2012role}). Experiments indicate that both the magnitude (\cite{sture1998mechanics}) and vertical extent (\cite{collins2014numerical}) of the dilation band will increase in the Moon's microgravity environment. Experiments by \cite{royer2011role} confirmed that exposure to vacuum maximizes post impact dilation. We investigate how the surface grains in this dilation band region respond to laterally propagating waves. Our motivation for exploring this topic is rooted in the common appearance of the mysterious Lunar Cold Spots (LCS), distal regions of reduced bulk density surrounding craters on the Moon (\cite{bandfield2014lunarCS}).

While LCS are a common feature of young craters (\cite{williams2018lunarCS}) and occur across all terrains of the Moon, their formation mechanism remains unknown. The LCS is apparent as a relatively cold halo (as compared to background regolith) and is the result of reduced surface bulk density ($\rho_{bulk}$) in the near surface (to 40 $cm$ depth). The cold spot begins about 10 crater radii away from the source crater and can extend radially to more than 100 crater radii, abruptly terminating in some regions while gradually fading in others. The regions nearer to the crater (< 10 crater radii) show signs of lateral transport while there is no observable surface modification in the distal reduced thermal inertia region. The expected dilation varies between 1\% and 10\% over the first 40 $cm$ of depth. Taking the percent difference of the average LCS and background bulk densities at the surface results in just over 4\% dilation which is about the same as that caused by silo rarefaction waves on Earth (up to 5\%, \cite{wensrich2002experimental}). Hemispherical shock waves propagating outward from an impact \cite{holsapple1993scaling} are essentially lateral waves at the surface and represent a potentially significant mechanical disturbance in the same region where LCS form (i.e., very low confinement pressure). Nearest to the impact site, regolith will undergo fragmentation, melting, or even vaporization depending on the strength of the initiated shock, but the waves decay as they propagate outward. The distal regions surrounding a LCS would be subjected to waves of varying strengths following an impact. Given that the LCS region of expression is limited to a narrow surface band of regolith exposed to vacuum we hypothesize that inertial dilation induced by lateral surface waves is at least involved in the LCS formation process. However, a surface wave would need to travel over large distances ($\sim$100 km) to be relevant in LCS formation. In this paper, we study the dynamics of surface grains under microgravity following the passage of a single, impulsively generated granular wave. Our objective is to simulate laterally propagating waves and determine if they can reduce surface bulk density  when waves are active at the grain-vacuum interface.

We present a simulation of a laterally propagating granular wave in a long, open-to-vacuum channel generated by piston impact.  We consider an idealized assembly of randomly packed monodisperse spheres as an initial proof of concept. Since we are interested in the ability of surface grains far from the source crater to undergo dilation, we generate a single wave via an impulse and allow it to evolve laterally across the length of the channel. The channel is filled with deformable particles and we measure the assembly's dilation response to both acoustic and shock waves by measuring the at rest bed height following the passage of a wave. We provide a brief overview of waves in granular media (sec. \ref{sec:ShocksInGrains}) in the literature before describing our simulation methodology (sec. \ref{sec:Methodology}). We then provide results from simulations that explore the influence of wave speed, wave type, boundary effects, and initial conditions on volumetric changes in the bed and characterize the observed dilation (sec. \ref{sec:Results}).

\subsection{Waves in granular media}  \label{sec:ShocksInGrains}

Waves propagating in a granular medium as the result of an impact fall into two distinct regimes: weak and strong. Weak impacts produce a pressure wave with propagation speed dependent on confinement pressure, whereas strong impacts produce a pressure wave with a non-linear power law dependence on impact force due to geometrical effects of the particle contacts (\cite{duffy1957stress}). Furthermore, depending on both the duration of an initial impulse as well as the pre-compression in the grains, the generated waves can exhibit stationary, oscillatory, dispersive, and even non-decaying responses. While a comprehensive overview of the different responses is provided in \cite{nesterenko2013dynamics}, we briefly introduce three important categories here. Acoustic waves are generated via weak impacts and result in waves traveling at the sound speed, $c_0$, but with high dispersion. Strong impacts produce waves with pressure-independent speed and when there is high dissipation, these waves are shocks. Between these two extremes are solitary waves, which travel while maintaining strength and shape (the result of a balance between dispersion and nonlinear effects). These waves can be generated on impact or as the decay product of a shock. Solitary waves are the most compelling for our purposes here given the large extent of LCS. \cite{nesterenko1984propagation} first characterized propagating solitary waves in a 1D granular chain subject to an impulse and confirmed the results experimentally. When the pre-compression in the grains is zero (or very low), the assembly is considered to be weakly compressed. In other words, when the compression induced by an impulse is much greater than the initial pre-compression, the assembly is weakly compressed. This leads to a scenario where waves propagating in the chain are dominantly influenced by nonlinear effects, a regime of propagation dubbed the `sonic vacuum' by \cite{nesterenko1984propagation}. Following their introduction, solitary waves in 1D chains have been investigated by many, both experimentally (\cite{coste1997solitary}) and numerically (\cite{sen1998solitonlike}, \cite{hong2000characterization}, \cite{shoaib2011discrete}, \cite{chakravarty2018possibility}). Solitary waves have been similarly confirmed and explored in 2D (\cite{shukla1987experimental}, \cite{sen1996sound}, \cite{owens2011sound}, \cite{nishida2009simulation}, \cite{awasthi2012propagation}, \cite{leonard2013directional}, \cite{pal2014characterization}, \cite{waymel2018propagation}). The study of solitary waves is still ongoing. One branch of focus lies in the tunability of a solitary wave's existence based on the randomness of the packing (\cite{leonard2014traveling}, \cite{zhang2020pulse}) or the stiffness ratios between particles (\cite{awasthi2012propagation}). Another focus is on granular wave response in the very low confinement pressure regime which is not fully understood (\cite{tournat2010acoustics}, \cite{tell2020acoustic}). \cite{gomez2012shocks} found wave speed relationships for sound waves and shocks in a randomly packed 2D bed of particles near the jamming transition (very low pressure). \cite{sanchez2022transmission} continued in this vein, characterizing seismic wave speeds at confinement pressures as low as 0.1 $Pa$ in a 3D randomly packed bed, but they studied vertically propagating waves and were not concerned with the grain-vacuum interface. In the context of the vacuum and microgravity of the Lunar surface, the first layer of surface grains is close to the zero pre-compression regime as the initial overlap ($\delta_0$) approaches zero at small depths (eq. \ref{eq:coste_delta_v_depth}). However, on the depth scale of the LCS ($40$ $cm$), grains will experience gravitational loading (Fig. \ref{fig:initialdelta}) and are also confined laterally by neighboring particles. The wave propagation regime may therefore vary based on the strength of an applied disturbance.

Even in 3D granular assemblies on Earth, solitary waves can be sustained and propagate in a weakly dispersive regime. \cite{rogers1994location} first used a large area impulse to generate a solitary wave which could be used to detect buried objects. \cite{sen2001solitary} mentioned \cite{rogers1994location} as an example when commenting on the robustness of solitary waves in 3D, though this is usually in the context of understanding how granular materials can be used to selectively disperse the waves \cite{sen2005using}, \cite{sen2017impact}. \cite{hostler2005pressure} may have created solitary waves in their 3D granular bed but had a channel of insufficient length to make a distinction between a solitary and semi-permanent wave. While experiments by \cite{quillen2022propagation} found that a single granular pulse is generated by low speed normal impacts, the resultant lateral signal attenuated as the pulse expanded. However, the tests of \cite{quillen2022propagation} were conducted at atmospheric confinement pressures which leads to increased pre-compression on particles at the surface as compared to the Lunar (vacuum exposed) case.  The most recent work on this topic (\cite{jiao2023revisiting}) studied the effects of increasing pre-compression on the behavior of solitary waves. A solitary wave that was highly stable at some small initial loading transitioned to acoustic-like propagation as the pre-compression was increased. While the investigation in our paper builds on these prior works, we are most interested in observing grain-scale dynamics and the resultant dilatory response in the bulk due to lateral waves propagating in different regimes (acoustic, solitary, shock). 

The sound speed of compressive waves at (45 $m/s$, Surveyor landing data - \cite{sutton1970elastic}) and near (100 $m/s$, 4 $m$ depth, Apollo seismic experiments - \cite{cooper1974lunar}) the Lunar surface are known, but these in-situ experiments did not examine the response of the near surface grains distal to impacts. To investigate grain motion far from the impact, we must understand how a randomly packed 3D granular medium influences the expression of granular wave fronts. For waves traveling laterally, there is an increase in the wave speed with depth (\cite{hostler2005pressure}), increased friction (\cite{mouraille2009sound}), increased cohesion (\cite{botello2016ultrasonic}), denser initial packing (\cite{agui2018high}), and increased gravity (\cite{zeng2007wave}).  \cite{mouraille2009sound} also found that the assembly's preparation history has considerable effect on wave propagation in the medium. Both sound speed (\cite{el2008acoustic}, \cite{burgoyne2015guided}) and dispersion (\cite{hostler2005pressure}, \cite{sen2017impact}) increase in a 3D assembly as compared to the one and two dimensional cases. However, dispersion also decreases with decreasing confinement pressure (\cite{fonseka2022shockwaves}, \cite{fonseka2022effect}) and the Lunar surface represents an environment with a unique balancing of these competing factors. 

Our focus differs from other recent works exploring waves in granular media: we characterize bulk volume change induced by laterally propagating surface waves far away from the impact site. To the best of our knowledge, no work has yet investigated the behavior of randomly packed, cohesive grains at and near the surface of a low gravity and mechanically free surface after experiencing a lateral impulse. We use these recent studies to guide our wave generation process and to serve as a validation case for our simulations before measuring bulk volume changes in the assembly as the result of a passing wave. 

\section{Methodology}  \label{sec:Methodology}
Here we describe our modeling approach (a simulation domain representative of an idealized granular assembly exposed to vacuum) and our particle seeding and settling routine used to generate the assembly. We also describe our wave generation procedure and our data analysis methods. We demonstrate that the grain rearrangement we observe as well as the waves created within our assembly are physical.

\subsection{Soft Sphere Discrete Element Method and contact model} \label{sec:SSDEM}

To investigate surface dilation as a result of passing granular waves, we model a granular channel using LIGGGHTS \footnote{https://www.cfdem.com/media/DEM/docu/Manual.html} (\cite{kloss2012models}), an extensively validated open-source Soft Sphere Discrete Element Method (SSDEM) solver. While wave propagation in a granular medium can be modeled using Effective Medium Theory (EMT) (\cite{jia1999ultrasound}), EMT cannot account for grain rearrangement leading to dilation (\cite{goddard1998computations}, \cite{makse1999effective}), variation in packing parameters (\cite{botello2016ultrasonic}) or relaxation mechanisms (\cite{makse2004granular}).  Discrete Element Method (DEM) (\cite{cundall1979discrete}) integrates the equations of motion for every particle in the system to resolve contact forces. \cite{thornton2000numerical} showed that 3D DEM simulations produce realistic stress-strain dilation behavior. DEM shows good agreement with experiments when studying wave propagation through a granular material (\cite{aoki1995simulation}, \cite{tanaka2002discrete}, \cite{mouraille2009sound}, \cite{awasthi2012propagation}, \cite{ning2015shear}). SSDEM integrates equations of motion at a very small time steps to allow for deformation by way of particle overlap. The time step must be much less than the Hertzian collision time ($\tau_c$) to accurately resolve elastic forces (\cite{schwartz2012implementation}). SSDEM has been used by many to study grain dynamics in low gravity environments (\cite{sanchez2011simulating}, \cite{schwartz2012implementation}, \cite{tancredi2012granular}, \cite{sanchez2016disruption}, \cite{demartini2019using}, \cite{zhang2021creep},  \cite{sanchez2022transmission}). We select LIGGGHTS due to its open source nature, ease of contact force addition and modification, ease of scalability for parallel processing, and ability to simulate non-spherical particles. LIGGGHTS' predecessor LAMMPS has been used to study waves in a granular media (\cite{awasthi2012propagation}, \cite{o2016micromechanics}, \cite{fonseka2022shockwaves}). LIGGGHTS has been used to study regolith and regolith simulant behavior (\cite{berger2017predicting}, \cite{otto2018cfd}) and to model inertial dilation during shear flow (\cite{hurley2015friction}).

Our granular assembly is a collection of spheres subject to Hertz contact law (\cite{mindlin1953elastic}) which accurately captures wave propagation in granular media (\cite{coste1999validity}). We also include static and rolling friction (\cite{ai2011assessment}) and cohesion (\cite{johnson1971surface}). Our model is quite similar to \cite{sanchez2016disruption} with the addition of cohesion. We summarize our force model with the equations below, a full description is provided in appendix \ref{sec:appendix_contact}. This example considers a single contacting pair of monodisperse particles (radius $R$), for multiple contacts the force will be summed for all neighbor particles (any particle within cutoff radius $R/2$). When two particles come into contact they begin to experience a nonlinear restorative elastic force. The force experienced by the particle depends on its stiffness, $k$, and the amount of overlap $\delta$. The overlap is the sum of the particle radii less the distance between their centers,  $\delta = R_1+R_2 - \lVert \textbf{x}_1 - \textbf{x}_2 \rVert$, with $\textbf{x}_i$ the particle state. Subscripts $n$ and $t$ refer to the normal (eq. \ref{eq:normal_force_total_main}) and tangential (eq. \ref{eq:tangential_force_total_main}) components in the collision frame, respectively. There is also a damping contribution which depends on damping coefficient, $\gamma$, and relative velocity, $v$. The normal direction includes a cohesive force dependent on the cohesion energy density, $k_c$, and contact area, $A_c$. The tangential component is limited to a portion of the normal force as determined by the friction coefficient $\mu_s$.

\begin{equation}\label{eq:normal_force_total_main}
    \textbf{F}_n = - k_n \delta_n^{3/2} - \gamma_nv_n + k_cA_c  \\
\end{equation}

\begin{align}
     \textbf{F}_t = - k_t \delta_t^{3/2} - \gamma_tv_t  \nonumber  \\ \mid F_t \mid \leq \mu_s \mid F_n  \mid \label{eq:tangential_force_total_main}
\end{align}

Our rolling friction model (provides packing support and mimics aspherical particles) is an elastic-plastic-spring-dashpot (EPSD) model which includes a spring torque ($\textbf{M}_{r,k}$) and a viscous damping torque ($\textbf{M}_{r,d}$). We show the full rolling torque in eq. \ref{eq:rolling_friction_main}, though the viscous damping component is neglected in this work.

\begin{equation}\label{eq:rolling_friction_main}
    \textbf{M}_r = \textbf{M}_{r,k} + \textbf{M}_{r,d}
\end{equation}

In equation \ref{eq:governing_eq_main} we give the governing equations which are integrated at each step. Torque, normal and tangential contact forces, and gravity are included. Gravitational force is $\textbf{F}_g = m\boldsymbol{g}$, with $g$ the gravitational acceleration vector. $\textbf{r}_{c,i}$ is the vector connecting the centers of the colliding particles, $I_i$ the equivalent moment of inertia of the contact, and $\boldsymbol\omega_i$ the angular velocity:

\begin{align}
    m_i \ddot{\textbf{x}}_i = \textbf{F}_{n,i} + \textbf{F}_{t,i} + \textbf{F}_{g,i} \nonumber \\ I_i \frac{d \boldsymbol\omega_i}{dt} = \textbf{r}_{c,i} \times \textbf{F}_{t,i} + \textbf{M}_{r,i} \label{eq:governing_eq_main}
\end{align}

Hertz theory is valid when the contact deformation is elastic. As an example, impact velocity for steel grains is limited to 1 $m/s$ (see estimates in \cite{nesterenko2013dynamics} , page 80). Above this speed, there are effects from possible plastic deformation at the contact site. The study by  \cite{gomez2012uniform} used Hertz theory with deformable particles to simulate granular shocks at low pressures. \cite{gomez2012uniform}  point out that the shock speed in the bulk is limited by the material sound speed of the particles themselves ($\sim$6000 $m/s$ for steel). We give the equation for solid sound speed ($c_m$, \cite{mase2009continuum}) in eq. \ref{eq:materialsoundspeed}, which depends on the density ($\rho$), Young's modulus ($E$), and Poisson ratio ($\nu$). 

\begin{equation}\label{eq:materialsoundspeed}
    c_m = \sqrt{\frac{(1-\nu)E}{(1+\nu)(1-2\nu)\rho}} \\
\end{equation}

\subsection{Model validation} \label{sec:model_validation}
We inspected a few simple test cases to ensure that our model accurately captures the underlying physics. First, we repeated the unit test cases of \cite{sunday2020validating} which confirm single particle behavior (sphere-sphere, sphere-wall). While the ability of LIGGGHTS to reproduce experiments has been extensively demonstrated, issues with the force model can be masked by bulk behavior. All tests were ultimately successful, though we did uncover and resolve one such issue. Initially a test evaluating the stopping distance of a particle sliding on a wall saw the particle stopping sooner than expected. As of the most recent version of LIGGGHTS (release 3.8.0) the Hertz tangential force model implements the threshold and rescaling for Coulomb friction based on only the tangential spring force (eq. \ref{eq:tangential_elastic}) as opposed to the full tangential force (eq. \ref{eq:tangential_force_total}) which includes damping (eq. \ref{eq:tangential_damp}). We provide the corrected model in section \ref{sec:computationdetails}.

We then validated our ability to produce expected shock wave speeds in a 1D particle chain subject to normal impact at speed $v_{im}$. We used the same simulation setup as \cite{abd2010force} (validated experimentally in \cite{abd2009contact}), creating a 5x5 square packed grid of particles into which a normal incident particle impacts a single particle in the packing. This creates a pressure wave within the chain which we track by finding the peak force within each particle and then fit the resultant particle position vs time data to find the wave speed in the chain. We performed this test for the same steel particle properties used in \cite{abd2010force} as well as for softer particles which have properties closer to those used in our 3D simulation. The parameters are summarized in Table \ref{tab:1Dproperties}. 

\begin{table}[h]
\begin{center}
\begin{minipage}{174pt}
\caption{1D Particle Properties. Steel values are those used in \cite{abd2010force}. The soft particles are the same size as the steel particles, but with the material properties used in our 3D simulations (see section \ref{sec:bed_prep}).}\label{tab:1Dproperties}
\begin{tabular}{@{}ccc@{}}
\toprule
\begin{tabular}{l}Particle Type \end{tabular}
 & \begin{tabular}{l}Steel \end{tabular}
 & \begin{tabular}{l}Soft \end{tabular} \\
\midrule
$E$  $[Pa]$ & $2.1x10^{11}$ & $5x10^6$ \\
$\Delta t$ $[s]$ & $10^{-10}$  & $10^{-6}$  \\
$\nu$  & 0.28 & 0.2 \\
$\mu_s$  & 0.5 & 0.7 \\
$R$  $[mm]$ & 10.32 & 10.32  \\
$\rho$  $[kg/m^3]$ & 7870 & 2500 \\
$c_m$ $[m/s]$ & $5.84x10^3$ & 47.14 \\
$c_0$ $[m/s]$ & $361.87 \pm 9.09$ & $8.57 \pm 0.03$ \\
\bottomrule
\end{tabular}
\end{minipage}
\end{center}
\end{table}

\begin{figure}[h]
\begin{center}
\includegraphics[scale = .41]{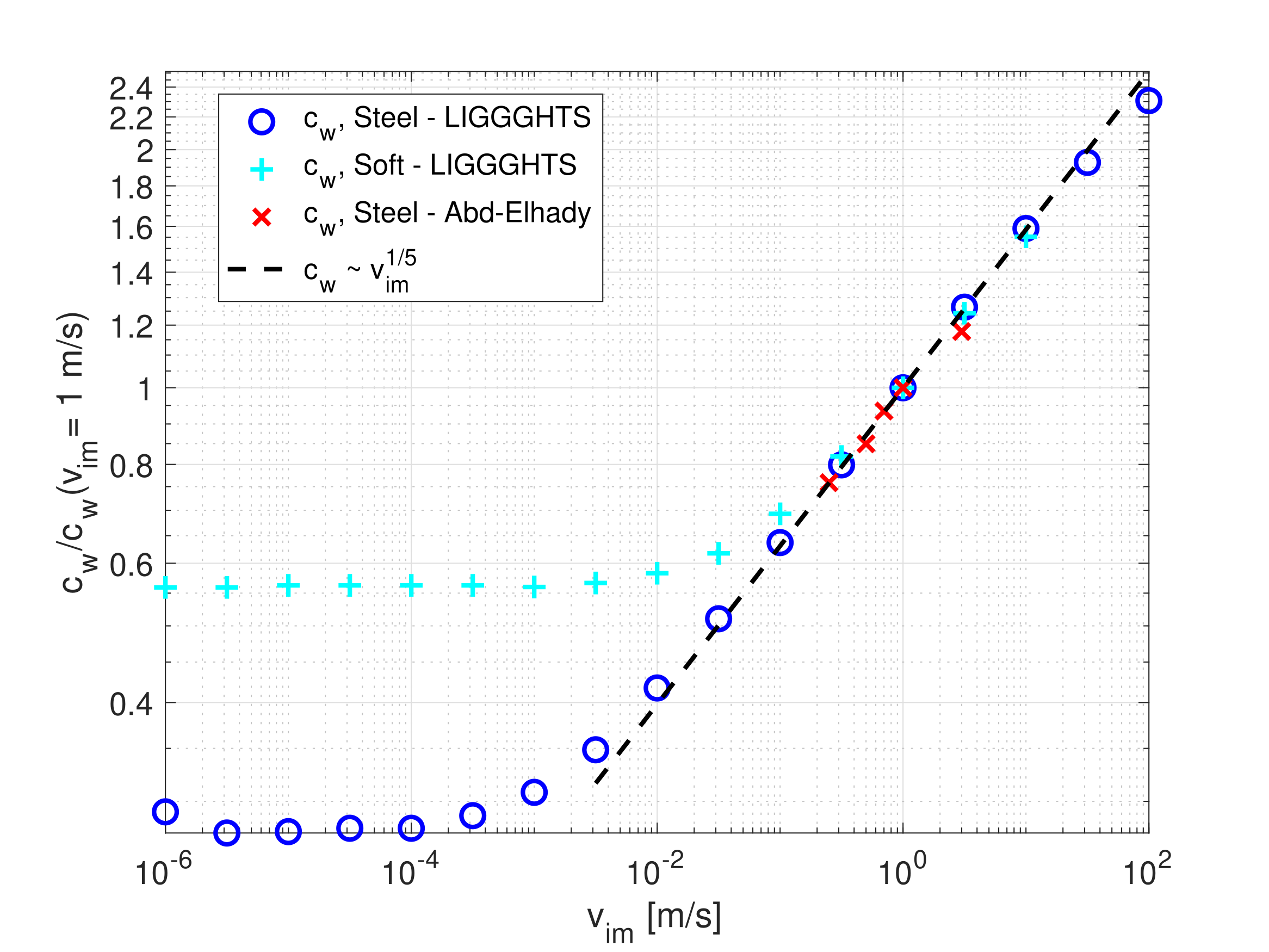}
\end{center}
\caption{\textbf{Normalized wave speed vs impact velocity.} $c_w$ reported for varied $v_{im}$ are normalized by $c_w(v_{im} = 1$ $m/s)$. We normalize $c_w$ at $v_{im} = 1$ $m/s$ instead of at vanishing $v_{im}$ (as in \cite{gomez2012shocks}) in order to include the data points from \cite{abd2010force}. Our tests are in agreement with theirs and recover the expected power law dependence.}
\label{fig:1Dsonicv}
\end{figure}

We conduct the normal impact tests for the same speeds as evaluated in \cite{abd2010force} and extend the range to induce both sound waves and shocks. We should find that weak impacts produce constant speed acoustic waves within the assembly and strong impacts produce shocks that follow a nonlinear power law dependence on the impact velocity. The expected power law dependence for shocks is $c_w \propto v_{im}^{1/5}$ (\cite{gomez2012shocks}), where $c_w$ is the initial wave speed as a result of normal impact. Figure \ref{fig:1Dsonicv} shows the results of these tests, reporting the normalized sound speed against $v_{im}$. We have good agreement with \cite{abd2010force} and we clearly see the expected weak impact-induced acoustic waves (constant $c_w$) and strong impact-induced shocks ($c_w \propto v_{im}^{1/5}$).  Taking the average $c_w$ for $v_{im} \leq 10^{-4}$ $m/s$ in Figure \ref{fig:1Dsonicv}, we find the sound speed, $c_0$ (reported in Table \ref{tab:1Dproperties}). The normalized $c_w$ for the soft particle acoustic waves is larger than that of steel since the normalization at $v_{im} = 1$ $m/s$ is closer to the acoustic wave region for the soft particles than for steel. We reproduce the expected power law dependence for shocks, regardless of material properties. Given the agreement with expectations, we can be confident that our choice of model and its implementation in LIGGGHTS accurately reproduces this 1D physical phenomenon. 

It is important to note that while the shocks resulting from the greatest $v_{im}$ ($> 10$ $ms$) follow the expected power law, in reality the particles experiencing this impulse would be subject to plastic effects and possibly fragmentation too. The Hertz model elastic limit ($1$ $m/s$) for steel particles is about $\sim0.02\%$ of $c_m$. Using that same criteria on the soft particles gives an elastic limit speed of $\sim v_{im} = 0.1$ $m/s$ which is where the soft particles have just turned toward the power law in Figure \ref{fig:1Dsonicv}. The elastic limit for the soft particles is much closer to the sound speed than is the case with the steel particles. Since the experiments in \cite{abd2010force} probe $v_{im}$ that produce shocks traveling in the assembly at speeds $> 2c_0$ we want to explore a similar range. To do so, we use an upper limit to particle impact velocities to 20\% of $c_m$, corresponding to $v_{im}$ of about 10 $m/s$ which produce shocks traveling at speeds $< 3c_0$. However, we must remember the limitations of the model (plasticity, fragmentation) when considering results for $v_{im}$ that are orders of magnitude above the elastic limit, even if they do reproduce the expected shock power law. As we'll see later, in our 3D assembly we are more interested in the decay products of the strongest shocks than the shocks themselves. These decay products are waves that travel at lower speeds than the initial shock and experience particle-particle impact velocities that are mostly less than the elastic limit (sec. \ref{tab:speed_waveDetails}).
 
\subsection{Simulating 3D wave propagation}
\begin{figure*}[h]
\begin{center}
\includegraphics[scale = .77]{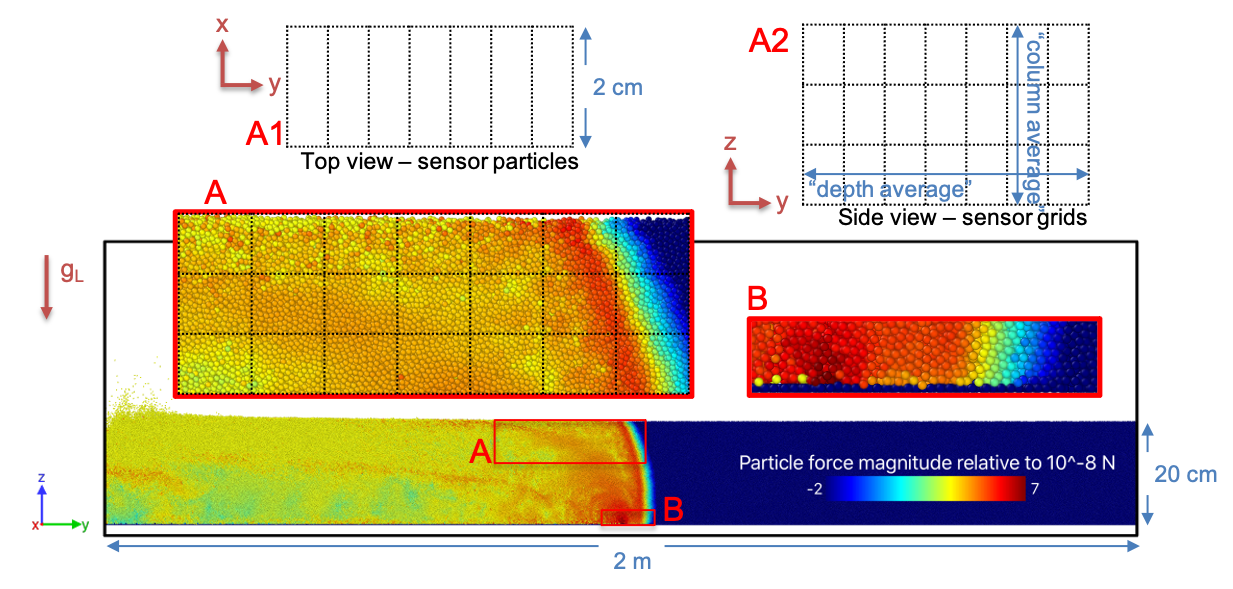}
\end{center}
\caption{\textbf{Channel Schematic.} A channel 2 $m$ in length is filled 20 $cm$ deep with particles (total of ~$5x10^5$ particles) and a shock was initiated at the left wall ($v_p = 10$ $m/s$) in the +y direction at t = 0 $s$. Here we show the channel's response several time steps after shock initiation (t = 0.1 $s$). The bold black line represents the simulation domain with the channel of particles occupying the bottom $\sim$half. Particles are colored with their order of magnitude force relative to $10^{-8}$ $N$ which reveals a clearly evident solitary wave around y = 1 $m$. There are two boxed regions highlighted which are shown in expanded views (not to scale). \textbf{Box A} shows a zoomed in a view of the wave front near the surface, overlaid with a sensor grid (sec. \ref{sec:data_analysis}) in dashed black lines. The shock front is (qualitatively) $\sim$15 particle diameters in size. A single sensor (one grid element in A) would find its properties by averaging the properties of all particles in the sensor at each time step. We provide top view (A1) and side view (A2) graphics of the sensor grid in box A to demonstrate how they are used in data analysis. Sensor particles are collected from the particles that rest on top of the 1D y-grids (A1). Side view elements are gridded in two dimensions (A2). To find average properties at a given radial position we take the average over sensor elements in the same column (``column average"). The ``depth average" is taken over all grid elements at a specified depth. \textbf{Box B} is zoomed in on the floor-wave interface. The sheet of particles frozen to the floor is visible along the bottom. The highest force in the solitary wave is the dark spot slightly behind the front. Data visualization as seen here and in subsequent figures is performed using open source visualization software Ovito (\cite{stukowski2009visualization}).}
\label{fig:channelcartoon}
\end{figure*}

The remainder of this paper focuses on our 3D simulation of waves in granular media and the resultant volume change. We model a radial strip in the path of a pressure wave extending outwards from the source using a randomly packed channel of particles. A schematic of the channel is shown in Figure \ref{fig:channelcartoon}. A fixed floor and end walls keep the particles within the channel which is longer in the direction of wave propagation (+y) than it is in width (x). The fill height of particles (z) and the length of the channel vary while the width of the channel is fixed at 2 $cm$ which is larger than the $\approx5$ particle diameter wall sheath size (\cite{gupta2016cfd}). Periodic boundary conditions in x allow for the simulated strip to be representative of a length of grains that experience a radially expanding pressure wave (\cite{potapov1996propagation}). There is a shrink wrap boundary condition on the ceiling of the simulation to allow for the trajectory of any particles ejected vertically from the bed and we enforce a uniform gravitational field of $g_L = 1.625$  $m/s^2$ in the -z direction.  We generate both compressive (P) and shear (S) waves (sec. \ref{sec:wavegen}). After a single wave pulse is generated, the simulation continues until the wave has terminated and any volumetric changes have reached an at-rest state (sec. \ref{sec:dilationmeasurements}). We evaluate wave speed and volume change using sensor particles as well as a virtual sensor network (seen in Figure \ref{fig:channelcartoon} and described sec. \ref{sec:data_analysis}). First, we describe the preparation of our granular assembly.

\subsection{Granular bed preparation} \label{sec:bed_prep}
This work seeks only to demonstrate that granular impulse-induced dilation can occur at the free surface of a granular medium, it does not seek to recreate the exact conditions and behavior of the Lunar regolith. We use the properties of regolith as a rough guideline for material parameter selection. Our assembly is filled with monodisperse particles of 2.5 $mm$ in size which is within the size range distribution for Lunar regolith. While the actual distribution has a mean particle size around 100 $\mu m$ (\cite{mckay1991lunar}), we use one order of magnitude larger particles to keep computation time tractable. Similarly, we set particle elastic modulus ($E$) low so that our time step is not prohibitively small. Friction, cohesion, and particle density are within the range given in \cite{carrier1991physical}, Tables 9.4, 9.5, and 9.12. There have been a range of values estimated for $\nu$ of Lunar regolith, ours is closest to the estimate in \cite{kovach1973velocity}. Our coefficient of restitution is in the appropriate range determined in experiment (\cite{chau2002coefficient}) and similar to that used in other SSDEM simulations of regolith behavior (\cite{zhang2018rotational},  \cite{wang2020behaviors}). The coefficient of rolling friction ($\mu_r$) is the same as used in \cite{sanchez2016disruption}. We set rolling viscous damping by determining a value that allowed a single spinning particle to come to rest with minimal overshoot and no lingering oscillation. We found this to be $\gamma_{d,r} = 2$ which is close to the value in \cite{holmes2016bending} where they conducted SSDEM simulations of tumbler flow using the same rolling friction model. Mechanical properties are uniform across particles and are set as in Table \ref{tab:BaseSimParam} during both the pouring phase and shocking phase, with exceptions for the coefficient of restitution and rolling viscous damping which are described in the following paragraph. After we set the material parameters and bed geometry, particles are poured into the channel and then settle under Lunar gravity. We chose to pour particles as our primary preparation method over other methods (\cite{gouache2010regolith}) as it allowed for greater control and uniformity of the achieved packing fraction, $\phi$. $\phi$ should have a large influence on granular bed reaction to perturbations (\cite{zhang2005jamming}). In this paper we use $\phi_0 = V_{particles}/V_{bulk}$, where $\phi_0$ is the initial packing fraction and $V_{particles}$ is the volume of particles within the larger bulk volume, $V_{bulk}$. We'll report $\phi_0$ as a percent in this paper.

\begin{table}[h]
\begin{center}
\begin{minipage}{344pt}
\caption{SSDEM Parameters for 3D validation.}\label{tab:BaseSimParam}
\begin{tabular}{@{}cccc@{}}
\toprule
Quantity & Symbol & Value\\
\midrule
Friction Coefficient & $\mu_s$ & 1.0  \\
Rolling Friction & $\mu_r$ & 0.8  \\
Rolling Visc. Damping & $\gamma_{d,r}$ & 2 \\
Coeff. of Restitution & $e$ & 0.5  \\
Cohesion energy density & $k_c$ & 1000 [$J/m^3$] \\
Young's Modulus & E & $5x10^6$ [$Pa$] \\
Poisson Ratio & $\nu$  & 0.2 \\
Particle Density & $\rho$ & 2500 [$kg/m^3$] \\
Particle Radius & $R$ & 1.25 [$mm$] \\
Time Step & $\Delta t$ &  $10^{-6}$ [$s$] \\
\bottomrule
\end{tabular}
\end{minipage}
\end{center}
\end{table}

Particles are poured using the LIGGGHTS ins/stream function. We designate a total number of particles to be inserted, a rate of insertion, and an area over which to insert the particles. The area of insertion matches the channel area (width*length) and is defined as a plane 10 $cm$ above the desired fill height. Through a 20 $cm$ extrusion of this plane, we create an insertion volume in which the particles are created. Since particles are not allowed to overlap, some are inserted and then allowed to exit the insertion region before more particles are created. To study initial packing and boundary effects, we need to fill a range of channel lengths to different fill heights and with varied $\phi_0$. More compact beds are created when the insertion rate approaches one particle per insertion and looser beds are created when the particle insertion rate approaches the total number of particles to be inserted. The insertion routine creates one tenth of the requested particles every 0.1 $s$, so an insertion rate of 100,000 $particles/s$ would see 10,000 particles created in ten 0.1 $s$ intervals over one second. The values we used to produce varied $\phi_0$ are given in Table \ref{tab:fill_param}.  We'll use three levels of packing (loose, medium, compact) to describe our assemblies which correspond to $\phi_0 \sim$ $55$, $59$, and $62$ (respectively). These $\phi_0$  correspond to bulk densities in the range of 1375 - 1550 $g/cm^3$.  Particle insertion takes one second for the loose packing configuration, three seconds  for medium packing, and about 30 seconds of simulation time for the compact packing. To reduce preparation time, most of our simulations occur in loose assemblies.  We note that the coefficient of restitution also influences $\phi_0$, with low $e$ producing looser beds and high $e$ producing more compact beds (\cite{wang2021effects}). Since a lower coefficient of restitution reduces settling time (\cite{sanchez2022transmission}), we set $e = 0$ while particles are poured. We found that a rolling viscous damping set as in Table \ref{tab:BaseSimParam} reduced the range of achievable $\phi_0$ so we also set $\gamma_{d,r} = 0$ during the pouring phase.  

\begin{table*}[h]
\begin{minipage}{344pt}
\caption{Insertion parameters to achieve various $\phi_0$ for varied channel geometries. }\label{tab:fill_param}
\begin{tabular}{@{}ccccccc@{}}
\toprule
\begin{tabular}{c}Channel \\Length \\ ($m$) \end{tabular} & \begin{tabular}{c} Channel \\ Height \\ ($cm$) \end{tabular} & \begin{tabular}{c} Total \\ Particles \\ (\#) \end{tabular} & \begin{tabular}{c} Insertion \\ Rate \\ ($particles/s$) \end{tabular}  & $\langle\phi_0\rangle$ $\pm$ $\sigma\phi_0$ & Type\\ 
\midrule
2 & 20 & 536,180 & 600,000 & $54.72$ $\pm $ $1.67$ & Loose   \\
2 & 20  & 575,410 & 200,000 & $58.65$ $\pm $ $1.31$ & Medium \\
2 & 20  & 607,953 & 18,900 & $61.96$ $\pm $ $1.03$ & Compact  \\
2 & 10  & 270,641 & 600,000 & $54.20$ $\pm $ $1.70$ & Loose   \\
2 & 30  & 814,649 & 600,000 & $54.60$ $\pm $ $1.70$ & Loose  \\
1 & 20  & 267,545 & 300,000 & $54.66$ $\pm $ $1.72$ & Loose  \\
3 & 20  & 804,763 & 900,00 & $54.72$ $\pm $ $1.67$ & Loose  \\
4 & 20  & 1,072,258 & 1,200,000 & $54.72$ $\pm $ $1.65$ & Loose  \\
\bottomrule
\end{tabular}
\end{minipage}
\end{table*}

After particle insertion has ended, $e$ and $\gamma_{d,r}$ are set to the values in Table \ref{tab:BaseSimParam}. The top layer of particles (particles with height greater than the specified fill height) are deleted to ensure a smooth surface. Finally, we allow the assembly to settle under Lunar gravity. At this point the bed is at rest and particles are not moving, though there is substantial energy from the filling procedure resonating within the bed due to the spring-like nature of our particles. As the bed settles energy is slowly dissipated and a lower energy state facilitates wave tracking. To speed this process, we remove kinetic energy from the bed by setting all particle velocities to zero every 25 $ms$. We found that allowing the particles to reach an average force of $10^{-11}$ $N$ was sufficiently settled to be considered steady state, where we could readily identify both acoustic waves and shocks (Section \ref{sec:3dvalidation}). Once we reach a steady state, we save a copy of the particle states to bypass preparing the bed for each simulation run. 

\begin{figure}[h]
\begin{center}
\includegraphics[scale = 0.4]{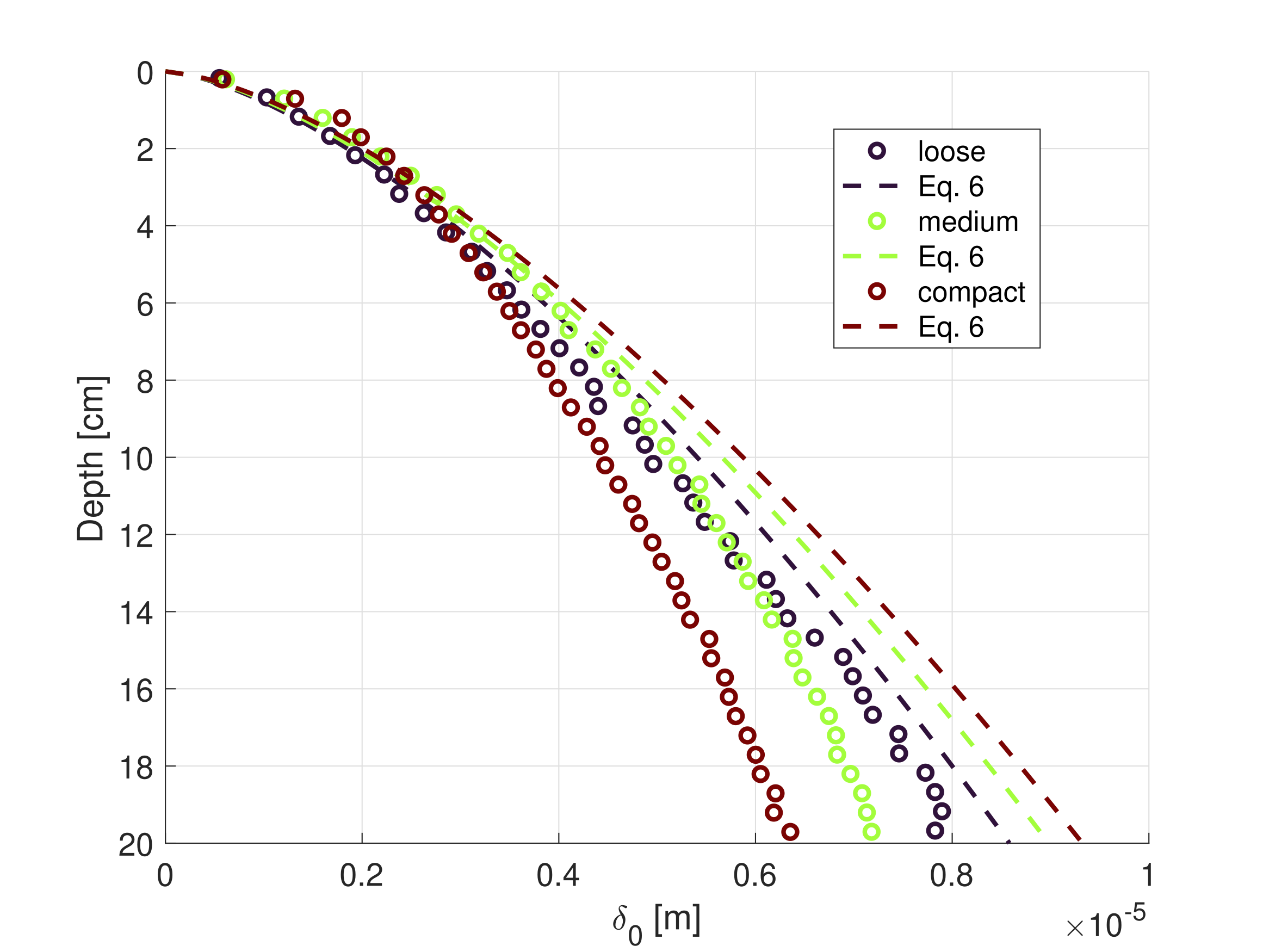}
\end{center}
\caption{\textbf{Depth vs $\delta_0$.} Depth dependence of $\delta_0$ for the loose, medium, and compact 2 $m$ long 20 $cm$ deep channels from Table \ref{tab:fill_param}. Simulated results are represented by circles and the dashed lines show the result of eq. \ref{eq:coste_delta_v_depth} with the $\phi_0$ from the first three entries of Table \ref{tab:fill_param}. These three packings correspond to average $\delta_0$ of 4.82, 4.81 and 4.23 $\mu$m (loose, medium, compact respectively). The compact $\delta_0$ is smaller than the loose since $\delta_0$ is the average overlap here, not the total (as discussed in the text).}
\label{fig:initialdelta}
\end{figure}

We compare the simulated and expected $\delta_0$ (average particle overlap at $t = 0$ $s$) dependence on depth in Figure \ref{fig:initialdelta}. As an approximation we use \cite{coste1997solitary} to find $\delta_0$ as the result of a static force applied to a 1D chain of beads (eq. \ref{eq:coste_delta_v_depth}). If a 1D chain of beads is vertically oriented, then the force on a single bead is equivalent to the weight of the particles above it in the chain. For our 3D bed we define the load as  $F_0 = \rho_{p} \phi * z\pi R^2 * g$, the overburden force which can be thought of as the weight of the column of material supported by a single particle at depth $z$. Figure \ref{fig:initialdelta} shows that this approximation works well near the surface but is an overestimate as depth increases. This is because eq. \ref{eq:coste_delta_v_depth} gives the total overlap on the particle in a 1D column, while our simulated $\delta_0$ is an average over the number of contacts (around $\sim$6 neighbors). A single particle at a depth $z$ in a 3D assembly is supported by several neighbors which distributes the load, resulting in a smaller $\delta_0$ than in the 1D case.  The loosest bed is closest to the estimate since there are less neighbors to average the total $\delta_0$ over than in a more compact bed. The compact bed $\delta_0$ is therefore smaller than the loose bed in Figure \ref{fig:initialdelta}. We confirmed that the total overlap is greatest in the compact bed and smallest in the loose bed as depth increases (not shown), but we found the average $\delta_0$ to produce sound speeds closer to those measured (sec. \ref{sec:waveterminationandspeed}).

\begin{equation}\label{eq:coste_delta_v_depth}
    \delta_0 = \frac{2\left(\frac{3(1-\nu^2)}{4E}F_0\right)^{2/3}}{R^{1/3}}
\end{equation}

\subsection{Bed tapping} \label{sec:Tapping}
For some cases we tap the bed before settling occurs to achieve different packing fractions. We base our tapping procedure on that of \cite{knight1995density} and \cite{philippe2002compaction}, but we use the particles on the floor of the assembly as a virtual shaker instead of inserting a physical base plate into the simulation (inspired by the virtual piston in the following section). We use the same sheet of particles that will be used to make up the rough particle floor (described in the prior section) as our tapping mechanism. Tapping commences when we direct the sheet to oscillate, with each particle in the sheet directed to follow a sine wave: $z_t = z_0 + Asin(2\pi f t)$. We use the same frequency tapping as in \cite{philippe2002compaction}  ($f$ = 30 $Hz$) but a larger tapping displacement amplitude  ($A = 0.5$ $mm$) which allows for a shorter required tapping duration when altering $\phi_0$. The bed is tapped for a specified time, after which the top most particles are again removed. We removed the same amount of particles from all tapped cases so that they were a consistent height which resulted in beds filled 15 $cm$ deep with particles (starting from the 20 $cm$ deep beds). Table \ref{tab:tapping_durations} shows the tapping duration used to achieve various $\phi_0$ starting from the loose, medium, and compact configurations. We tapped the loose bed to several different states and ran a single comparison case in the medium and compact beds. The last row in the table shows that beds with varied $\phi_0$ converge to the same final state, confirming our model accurately captures bulk volume change due to shearing.

\begin{table}[h]
\begin{center}
\begin{minipage}{174pt}
\caption{Tapping Duration. New $\phi_0$ are reported for various tapping duration of the loose, medium, and compact beds at the same amplitude and frequency. The $\phi_0$ designations correspond to those from Table \ref{tab:fill_param} for Loose (54.7), Medium (58.7), and Compact (62.0).  }\label{tab:tapping_durations}
\begin{tabular}{@{}cccc@{}}
\toprule
\begin{tabular}{c} Tapping \\ duration ($s$) \end{tabular} & \begin{tabular}{c}  $\phi_0$ after \\ tapping \end{tabular} & \begin{tabular}{c} $\phi_0$ before \\ tapping  \end{tabular}  \\
\midrule
0.25 & $56.14$ $\pm $ $1.51$ & Loose   \\
0.5 & $56.90$ $\pm $ $1.43$  & Loose  \\
1 & $58.21$ $\pm $ $1.31$ &  Loose \\
2.5 & $59.84$ $\pm $ $1.22$ &  Loose  \\
5 & $61.04$ $\pm $ $1.17$ &  Loose  \\
7.5 & $61.54$ $\pm $ $1.14$   & Loose  \\
10 & $61.83$ $\pm $ $1.14$ & Loose   \\
\hline
10 & $61.86$ $\pm $ $1.16$ & Medium  \\
\hline
10 & $62.17$ $\pm $ $1.17$ & Compact  \\
\bottomrule
\end{tabular}
\end{minipage}
\end{center}
\end{table}

\subsection{Wave generation} \label{sec:wavegen}
We generate horizontal P and S waves within the assembly to capture the bounding directions of shear stresses that may be present as the result of an impact. We use a virtual approach to generate a single wave as opposed to a physical piston (\cite{gomez2012shocks}, \cite{mykulyak2014features}, \cite{sanchez2022transmission}) or bender element (\cite{lee2005bender}). A sheet of particles in the channel within one particle diameter of the left wall (y = [0,2R]) are designated as a virtual piston (\cite{mouraille2006sound}, \cite{li2020grain}) or virtual bender element (\cite{ning2013discrete}). Looking at Figure \ref{fig:channelcartoon}, the sheet would be made up of all particles that reside to the left of an x plane located at y = 2R (all particles directly adjacent to the left wall). To generate a single P wave with the virtual piston, all particles within the sheet are assigned a designated velocity ($v_p$) in the +y direction at the beginning of the simulation. We'll consider the +y direction to be the radial direction pointing away from the crater. As opposed to compression with the piston, a bender element induces strain in an assembly through the small displacement (in the z direction here) of a thin plate.  We generate a single period S wave  (\cite{ning2013discrete}, \cite{o2016micromechanics}) with the virtual bender by directing the sheet of particles to follow oscillatory motion following the same formula from the tapping procedure, $z_t = z_0 + A sin(2\pi f t)$, for one period. The frequency is set high so that we do not have to determine the resonance frequency of the bed, the assembly will damp our signal out to the resonance frequency (\cite{ning2013discrete}). Generating the shear wave with the entire sheet of particles reduces (but does not negate) P wave interference compared to physical bender elements (\cite{ning2015shear}). The velocity that the particles within the sheet obtain is $v_z =  2\pi A f cos(2\pi f t)$, with maximum velocity $v_{z,max} = 2\pi A f$. We will consider this max velocity to be equivalent to $v_p$ for the P waves and can vary $A$ to generate waves of difference speeds ($A = v_p / (2\pi f$)). With $f = 50$ $kHz$, this corresponds to a period of $2x10^{-5}$ s which occurs over 20 simulation time steps.

We then subject the bed to these varied $v_p$ to generate acoustic waves and strong shocks (Section \ref{sec:3dvalidation}). The strong shocks rapidly decay to solitary waves that travel at reduced, though still supersonic, speeds and with very little dissipation. We can call these waves solitary since they propagate over distances (on the scale of meters) much longer than their spatial extent. The size of the shock front is around $\sim$15 particle diameters, (fig. \ref{fig:channelcartoon}), which is larger than the 5 particle diameter size in a 1D chain and less than the 100 particle diameter width in the 3D wave of \cite{hostler2005pressure}. Only a single pulse is generated down the channel which is characteristic of blast loading (\cite{nesterenko2013dynamics})  - the duration of the imparted piston impulse is rapid. This makes sense given that our loading time is essentially the Hertz collision time. On the time step before shock initiation, we fix a sheet of particles ($0\leq z < 2R$, Fig. \ref{fig:channelcartoon}B) to the floor so that the shock travels over a rough surface and bypasses the non-physical particle-wall interaction as has been done in both experiment and recreated in DEM (\cite{shojaaee2012shear}, \cite{sunday2020validating}). The simulation then runs until the generated wave reaches the end wall, reflects, and terminates which then sees the assembly enter an at-rest state, where the only motion present is a result of the slow dissipation of the forces leftover from the passing of the shock within our spring-like grains and the particles can be considered at-rest. The total simulation time is set to allow the lowest induced wave speed to travel across the longest channel. We found a simulated time of 1.5 $s$ to be sufficient for the cases we analyzed. We take measurements every 1 $ms$ for the first 0.25 $s$ and then every 10 $ms$ thereafter. The higher resolution data covers the period when volume changes occur and allows for enough resolution to compute the induced $c_w$ (Section \ref{sec:waveterminationandspeed}). After the wavefront passes and any lofted particles have returned to the surface, a slow gravitational settling begins. The settling is accompanied by small volumetric changes, though we show in Fig. \ref{fig:dilation_vs_time} that this oscillation is about a steady, at-rest value. Further details on the dilation measurement and behavior are discussed in section \ref{sec:dilationmeasurements}.  We reduce the resolution of output data in the settling region to aide with data storage logistics given that the output from a single simulation run can exceed 50 GB.

\subsection{Computation details} \label{sec:computationdetails}
Bed preparation occurs on our standalone 44-core workstation (dual Intel E5-2699V4 processors and 128 $GB$ of 2400 $MHz$ DDR4 RAM) and shock simulations are run using the University of Maryland High Performance computing cluster, Zaratan (128 cores, dual AMD 7763 64-core processors, and 512 $GB$ of AMD EPYC 7763, 2.45 $GHz$ base RAM per node). We verified that there was no difference in results after shocking the prepared bed between our server and the cluster. A summary of computation time is provided in Table \ref{tab:computation_details}. Details from shock runs are given for $v_p = 10$ $m/s$ which is the upper bound for computation time. Run times for lower velocities are slightly reduced from those reported here as there is less energy input into the system, reducing overall particle movement. We provide the LIGGGHTS input files used for bed preparation, tapping, settling, and shocking at our public repository (\cite{frizzell2023code}). The corrected tangential force model (sec. \ref{sec:model_validation}) is included there as well. The restart files that are the prepared beds are hosted at \cite{frizzell2023data} along with the full output from a single shock simulation run. We can provide additional simulation data upon request, however with the input scripts and restart files we have provided any reader could download LIGGGHTS and reproduce our shock simulations.

\begin{table*}[h]
\begin{minipage}{400pt}
\caption{Computation details. Processing time and storage requirements for shocking the different beds. Simulated time for all tests below is 1.5 $s$. Each test used all 128 cores of a single node.} \label{tab:computation_details}
\begin{tabular}{@{}ccccccc@{}}
\toprule
Length ($m$) & Height ($cm$) & Particles (\#) & $\langle\phi_0\rangle$ & Clock time ($hr$) & Storage ($GB$)  \\
\midrule
 2 & 20 & 536,180 & $54.72$ &  1.08 & 49.06    \\
 2 & 20  & 575,410 & $58.65$ &  1.12 & 54.06 \\
 2 & 20  & 607,953 & $61.96$ &  1.29 & 57.85   \\
 2 & 30  & 814,649 & $54.60$ &  1.51 & 72.30   \\
 2 & 10 & 270,641 & $54.20$ &  0.91 & 23.70    \\
 1 & 20 & 267,545 & $54.66$ &   0.83 & 23.83    \\
 3 & 20  & 804,763 & $54.72$ &   1.48 & 108.26  \\
 4 & 20  & 1,072,258 & $54.72$ &   2.36 & 150.87   \\
\bottomrule
\end{tabular}
\end{minipage}
\end{table*}

\subsection{Data analysis} \label{sec:data_analysis}
\begin{figure}[h]
\begin{center}
\includegraphics[scale = .4]{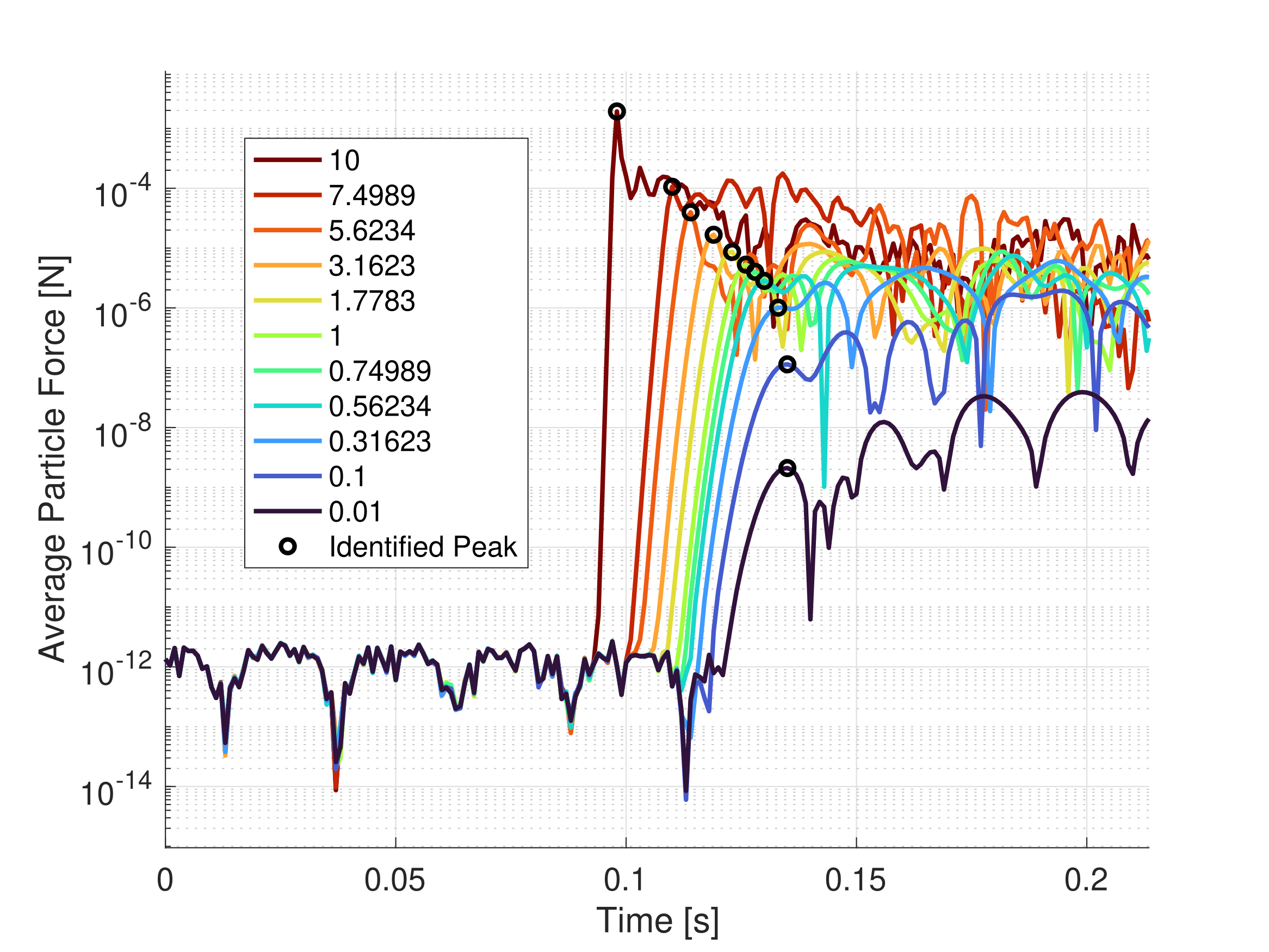}
\end{center}
\caption{\textbf{Average Particle Force (Y) vs time.} Force evolution vs time for P waves created at varied $v_p$ from a sensor at 7.5 $cm$ depth and radial position 1 $m$ in a 2 $m$ channel filled 20 $cm$ deep with particles. The color legend represents different $v_p$ with units in $m/s$. Lines have been added for clarity. Identified peaks are plotted as a black circle and this is taken to be the time at which the wave front passed through the sensor.}
\label{fig:force_vs_time}
\end{figure}

We perform post processing of LIGGGHTS output files in MATLAB, using both sensor particles and a  virtual sensor grid (see section \ref{sec:dilationmeasurements} and Figure \ref{fig:channelcartoon}) to measure volume changes and track pressure waves. We lay a 0.5 $cm$ x 0.5 $cm$ grid over the yz dimension and find average grain properties (e.g., $\phi$, $\delta$, average pressure, etc.) over the grains in that sensor grid element. With the exception of $\phi_0$, the average properties for the bin are the mean value for all particles that are wholly within the grid bounds. We give an example of the time evolution of average particle force in Figure \ref{fig:force_vs_time}, which represents a single grid element from Figure \ref{fig:channelcartoon}, box A/A2. We can then use the sensor grid to find average properties for the entire granular assembly. For example, the first peak force from Figure \ref{fig:force_vs_time} would find the peak force for each sensor grid element and then average across the sensor grid elements at all radial positions at a given depth to find the average value at that depth. Similarly, averaging over the sensor grid elements at each depth for a given radial position gives an average value at that radial position. 
 
 \subsubsection{Packing fraction}
In this work we use a simple calculation to find the packing fractions we have reported so far. Since our particle data is already binned into the 2D sensor grid, we add the volume of all the particles within the cell and then divide by the volume of the cell to find $\phi_0$. Any particles residing within one particle radius of the grid element boundary have their volume split into the correct grid element. If the particle resides along the boundary between two grid elements, we find the volume of the hemispherical cap that resides in the adjacent grid element. For particles near the intersection of four grid elements we compute the volume of the slice residing in the corner grid element in addition to the hemispherical caps in the adjacent bins. Averaging over all of the sensor grid elements at a given depth leads to an average $\phi_0$ at each depth. Then, averaging over all depths (except the very surface grid element) leads to the packing fractions as presented in Table \ref{tab:tapping_durations}. We neglect the particle overlaps in this volume calculation since the impact on our results will be small given the averaging we perform (\cite{zhang2017creep}).
 
\subsubsection{Dilation} \label{sec:dilationmeasurements}
\begin{figure}[h]
\begin{center}
\includegraphics[scale = .7]{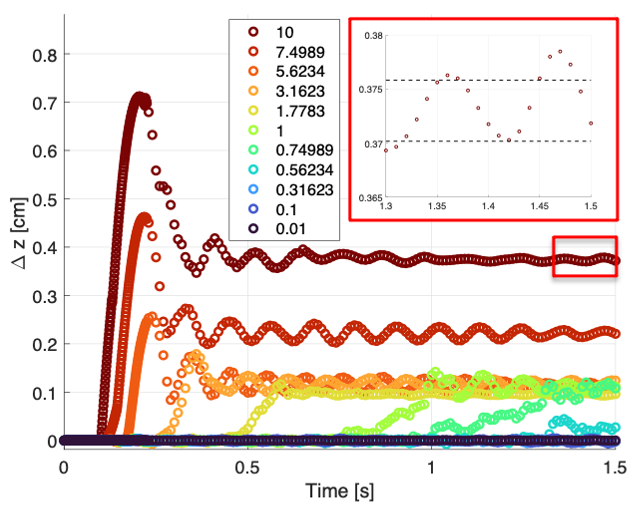}
\end{center}
\caption{\textbf{Height vs Time.} Average sensor particle height over time for a sensor at y = 1 $m$ in a 2 $m$ channel filled 20 $cm$ deep with particles (compact filling). Height increase is initiated by arrival of the wave, falling back down from a peak to a position greater than the initial height when the wave is supersonic. Color legend corresponds to different $v_p$ ($m/s$).  The dashed lines in the zoomed in portion (boxed in red) show the variance in bed height change for the $v_p = $ 10 $m/s$ case taken from a 15 second simulation (not pictured). The  oscillations we see between 1.3 and 1.5 $s$ (our averaging window) occur about the at rest bed height from the longer duration simulation.  }
\label{fig:dilation_vs_time}
\end{figure}

We evaluate the height of the grains in the assembly at each radial position to measure the change in packing fraction as was done in \cite{knight1995density}. Percent change in bed height is equivalent to percent change in packing fraction when the change in bed height is much less than the height of the bed ($\Delta z$ $\ll$ $z$). Instead of the sensor grid used to measure wave speed, we now consider sensor columns. The topmost particles in each column are our sensor particles (Fig. \ref{fig:channelcartoon}, A1). Each column is 0.5 $cm$ in width as before and measuring the height of particles in this column over time leads to Figure \ref{fig:dilation_vs_time}, which shows the height of the top most particles above the initial fill height. Particle height is averaged over the top $N_h$ sensor particles in the column. $N_h \approx \frac{x*0.5}{2\pi R} $ which is the approximate number particles that could fit side by side determined by the area of the column in the xy plane divided by the particle area. We then take a percent difference as compared to the initial bed height to find the volume change that occurred as a result of the wave's passing. We average over the last 0.2 $s$ of values to get the final bed height as there is some small oscillation that occurs. The oscillations are a result of the settling in the bed as the spring-system dissipates the energy of the shock which we discussed in section \ref{sec:wavegen}. We measured the deviation due to the oscillations in our final at rest bed height values to be less than the oscillation observed during sustained tapping tests which found a steady state value that $\phi_0$ oscillated about (\cite{knight1995density}). We report one sample case to confirm the bed is in an at rest height, despite observed oscillations. Using the 2 $m$ long bed filled 20 $cm$ deep with the compact filling the bed is shocked at the same $v_p = 10$ $m/s$ and is allowed to run for ten times the duration of our normal tests (15 seconds). Over the last 0.2 seconds, the mean and deviation change in bed height ($\mu \Delta z, \sigma \Delta z$) in the 1.5 $s$ test are ($0.373$ $cm, 28.32$ $\mu m$) compared to ($0.369$  $cm, 0.121$  $\mu m$) in the 15 $s$ case.  The mean $\pm$ variance from the 15 $s$ case are plotted as dashed lines on the zoomed in portion of Figure \ref{fig:dilation_vs_time}. The bed height change between $t = [1.3, 1.5]$ $s$ is already oscillating about the at rest $\Delta z$ from the 15 $s$ test.

\subsubsection{Wave speed} \label{sec:waveterminationandspeed}
\begin{figure}[h]
\begin{center}
\includegraphics[scale = .4]{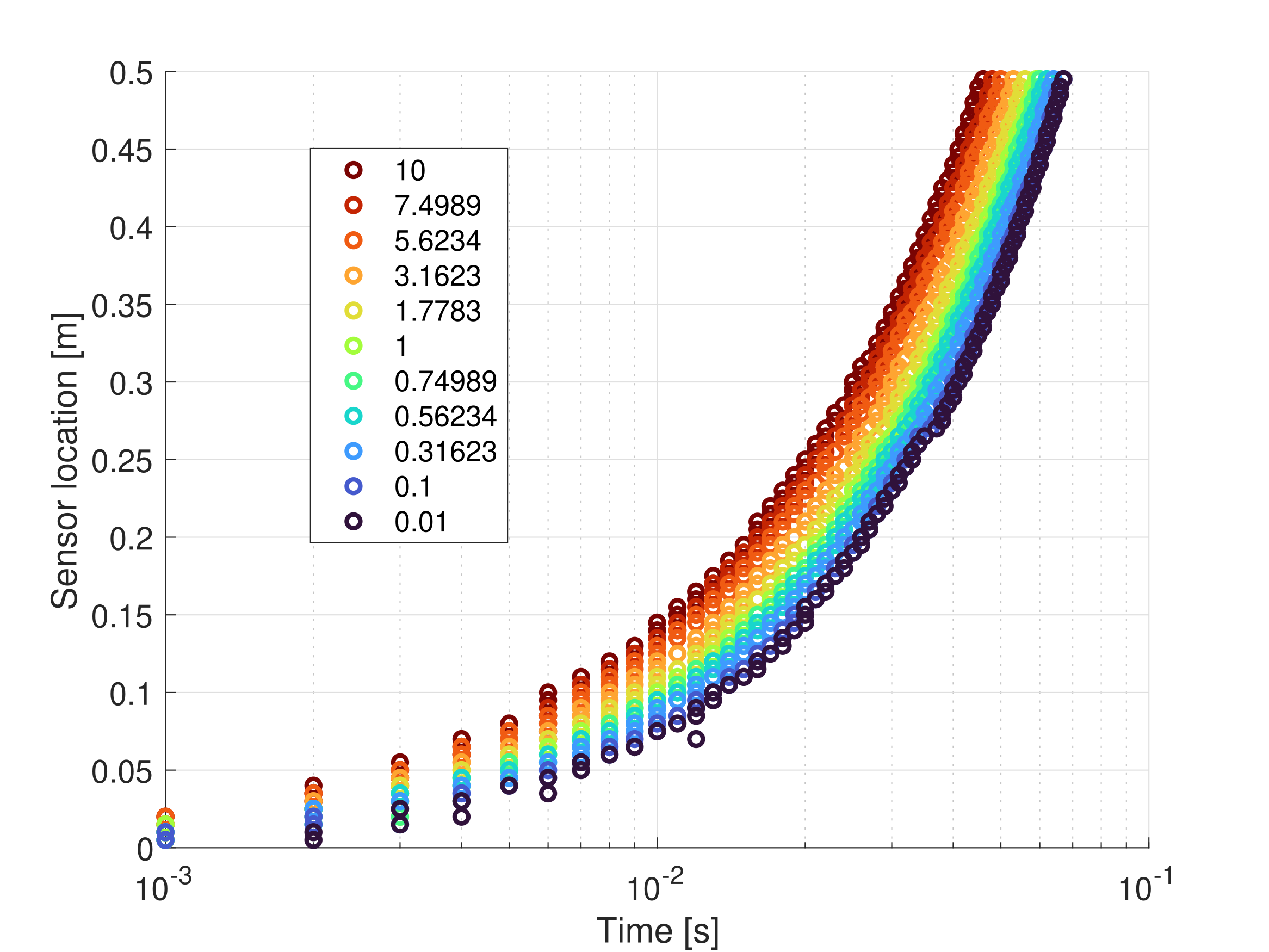}
\end{center}
\caption{\textbf{Sensor location vs logged peak force time.} For varied $v_p$ we show the measurements used to find $c_w$ at a depth of 4 $cm$. A fit of the measurements in the first 15 $cm$ gives $c_w$ (fit lines not shown). The initial linear fit is used to remove outliers more than $2\sigma$ from the mean before finding the final linear fit. Color corresponds to wave $v_p$ ($m/s$). The slope of the linear fits for each trend correspond to reported $c_w$ in Figure \ref{fig:wave_speed_depth_slice}. The propagating solitary wave speed $c_p$ is a linear fit to the data after 30 $cm$. }
\label{fig:measurementfits}
\end{figure}

The generated wave speed ($c_w$) is determined by tracking the first peak force within each bin. Figure \ref{fig:force_vs_time} shows that shocks ($v_p >1$ $m/s$) have an easily identifiable peak, while the acoustic waves ($v_p < 1$ $m/s$) see a smaller and later increase in average force. The first peak is determined as the first maximum found that is one order of magnitude greater than the initial average particle force. This threshold is required to identify sound wave peaks, which can have average particle forces on the same order of magnitude of the initial force. The first peak force gives us the time of the wave's passing through each sensor and, along with the sensor's position, can be used to find wave speeds Figure \ref{fig:measurementfits} shows these results for sensors from y = [0, 50 $cm$] at a single depth within the channel. The plot is semi-log in time (x-axis) to illustrate that there are two different regimes of propagation. The shock wave regime is accompanied by substantial dissipation and this results in an initial brief region of deceleration for the $c_w$ of the strongest impacts. We find $c_w$ by constructing a linear fit to the points in the first 15 $cm$ of the channel (y = [0, 15] $cm$). This will be an underestimate since the wave speed is decelerating. After decelerating ($ t > 0.01$ $s$), the wave speed is still supersonic and appears to travel as a solitary wave with constant speed. We'll call the solitary wave speed the propagating wave speed, $c_p$, which is found with a linear fit to the data from y = [30, 50 $cm$]. 

\begin{figure}[h]
\begin{center}
\includegraphics[scale = .4]{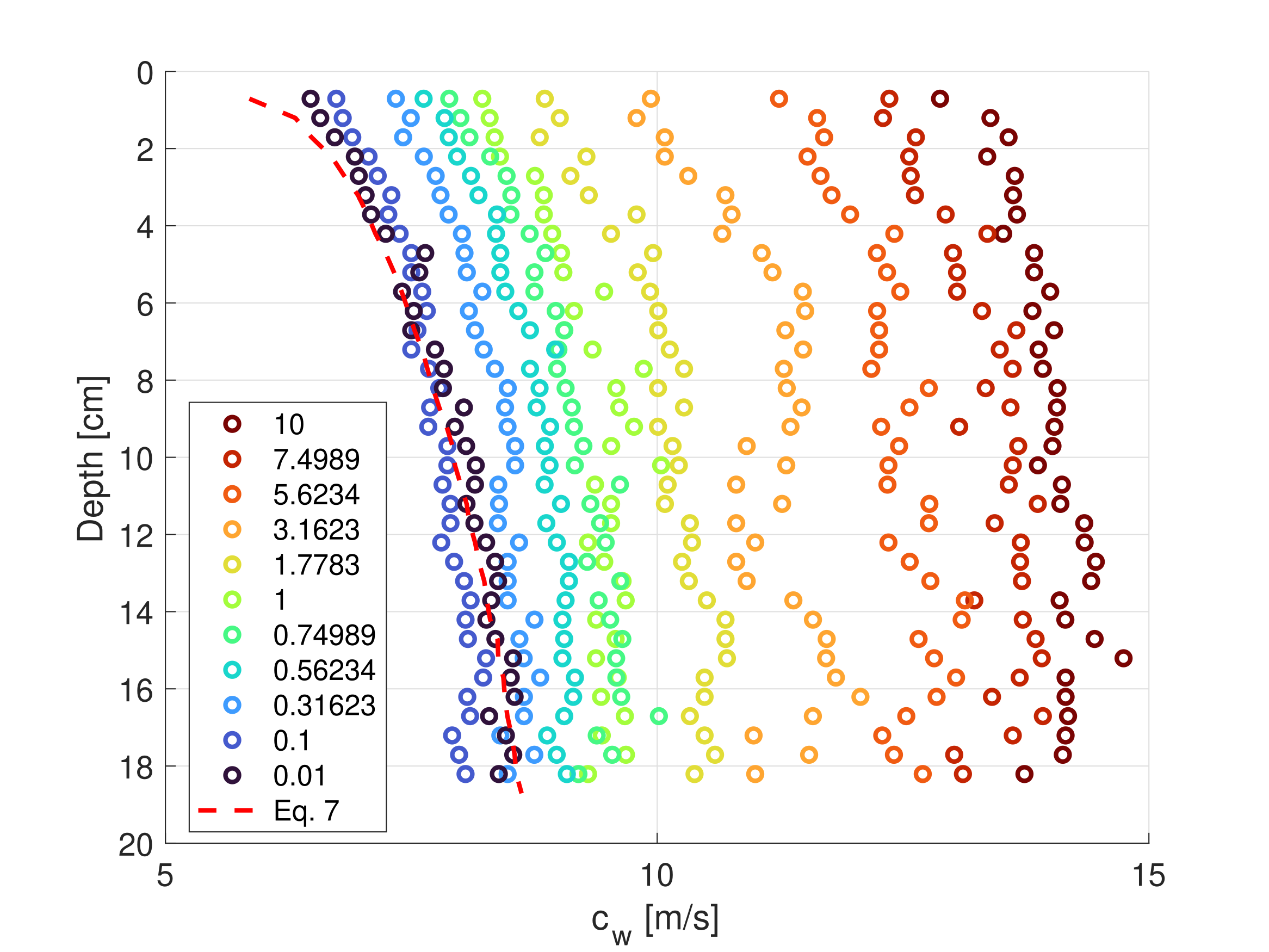}
\end{center}
\caption{\textbf{Depth vs $c_w$ for varied $v_p$.} Using the slope found from Figure \ref{fig:measurementfits} we show $c_w$ and its depth dependence for varied $v_p$ in a 2 $m$ channel filled 20 $cm$ deep with particles (compact packing). Acoustics waves are grouped together and produce roughly the same $c_w$, while shocks show an increase in $c_w$ with $v_p$. The red dashed trend line shows that the predicted sound speed, $(\sqrt{3}/2) c_0$  (eq. \ref{eq:sound_speed_nesterenko}) agrees with our numerical sound speed (bluest data points, corresponding to low $v_p$) .}
\label{fig:wave_speed_depth_slice}
\end{figure}

This process is then repeated at every depth, the results of which we show (for shocks) in Figure \ref{fig:wave_speed_depth_slice}. As expected, $c_w$ increases with depth (\cite{hostler2005pressure}). We can also predict the sound speed using equation \ref{eq:sound_speed_nesterenko} (\cite{nesterenko2013dynamics}, eq. 1.7) along with the calculated $\delta_0$ (Fig. \ref{fig:initialdelta}). Equation \ref{eq:sound_speed_nesterenko} is derived for a `strongly compressed' (which is the case for sound waves) 1D particle chain and does not include effects from some of the physical properties in our model (friction, cohesion) so it slightly over predicted $c_0$. Applying a scaling factor of $\sqrt{3}/2$ yields the prediction of $c_0$ in Fig. \ref{fig:wave_speed_depth_slice} which agrees well with our numerical sound speeds.

\begin{equation}\label{eq:sound_speed_nesterenko}
    c_0^2 = \frac{E(2R)^{1/2}}{3(1-\nu^2)m} \delta_0^{1/2}6R^2
\end{equation}

\subsubsection{Particles in the wavefront}
\begin{figure}[h]
\begin{center}
\includegraphics[scale = .4]{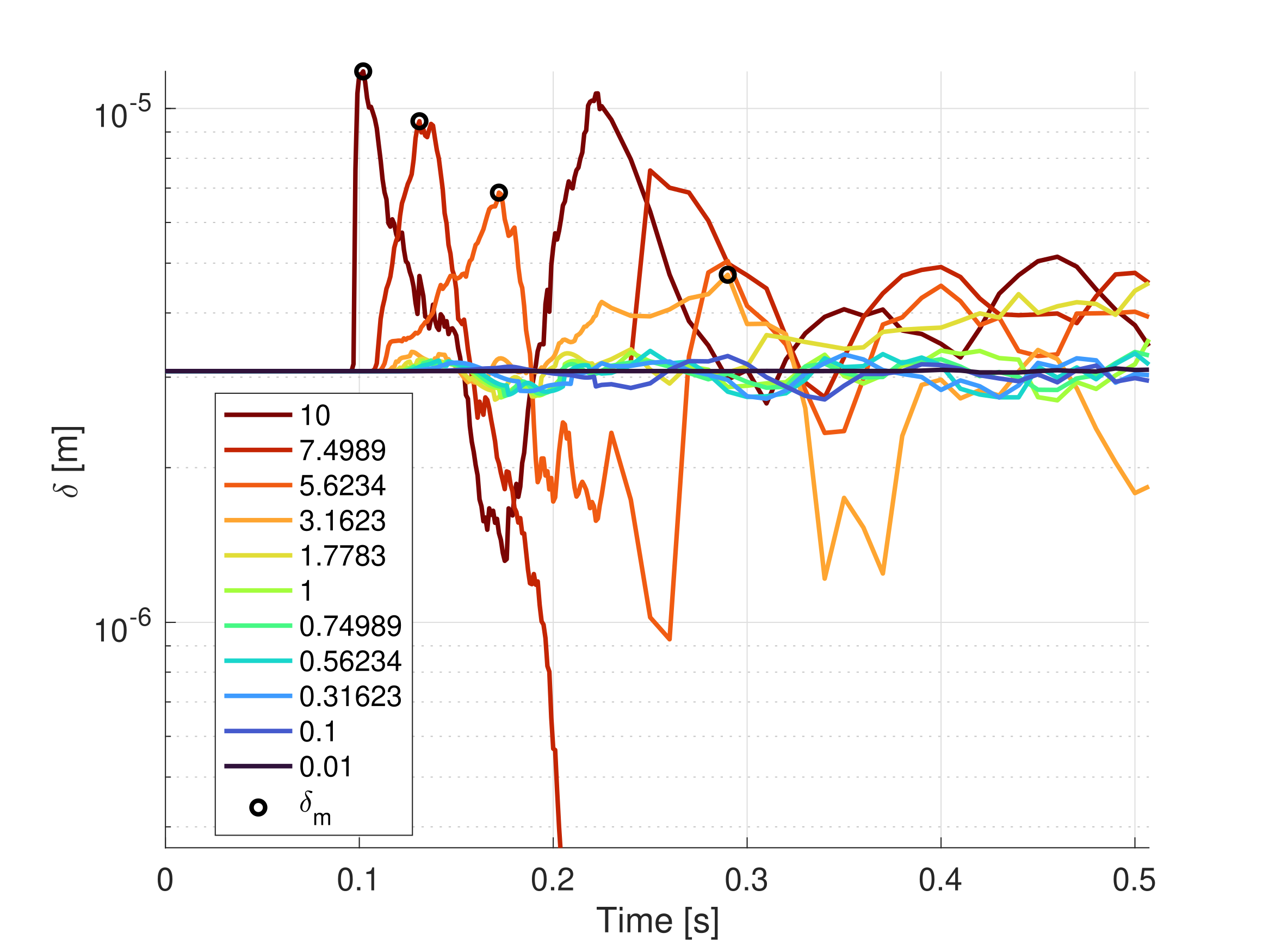}
\end{center}
\caption{\textbf{Average $\delta$ vs time.} $\delta$ evolution vs time for P waves created at varied $v_p$ from a sensor at 7.5 $cm$ depth and radial position 1 $m$ in a 2 $m$ channel filled 20 $cm$ deep with particles (compact filling). Lines have been added for clarity. The legend colors correspond to $v_p$ (m/s). The overlap the particles experience as a result of the shock, $\delta_m$, corresponds to the maximum. The red line ($v_p \sim 7.5$ $m/s$) is discontinuous between 0.2-0.25 $s$ while the particles in the cell are completely detached. $\delta_m$ is the maximum overlap experienced in the sensor as the result of the wave which is not necessarily the maximum overlap in the wave front. }
\label{fig:overlap_vs_time}
\end{figure}

To asses the strength and quality of the generated waves we will collect the maximum velocity ($v_m$) and maximum overlap ($\delta_m$) experienced by the particles in the wave front. To do this, we find the maximum $\delta$ over time in each virtual sensor.  Figure \ref{fig:overlap_vs_time} shows the average overlap vs time in a single sensor for varying wave speeds. We use the average of the maximum sensor overlaps over all depths when reporting a single value for the overlap induced by the wave front ($\delta_m$), see, for example, Fig. \ref{fig:velsweep_overlap}. Note that $\delta_m$ does not correspond with the peak force (Fig. \ref{fig:force_vs_time}) except in the case of the largest $v_p$. $\delta$ continues to grow following the impulse as the a result of collisions behind the wave front, so $\delta_m$ should be considered the maximum overlap experienced by a force sensor following the wave's passage (as opposed to the maximum overlap in the wave front). Notice that particles experiencing wave fronts induced by the largest $v_p$ ($\geq 5.6$ $m/s$) undergo a period of unloading directly following $\delta_m$. This occurs as surface particles are ejected following passage of the wave front. At times they are even completely detached from their neighbors as particles within the sensor experience lofting. The maximum velocity in the collision is collected using the same procedure, though a separate plot is not shown here since finding $v_m$ is more straightforward (coincides with time of peak force).

\subsubsection{Particle stress}
To gain a general understanding of the mechanics behind any surface dilation we observe, visualizing particle pressures over time is a valuable tool. We use LIGGGHTS' computed stresses to visualize shock characteristics (sec. \ref{sec:shock-mechanism}) which are computed as the virial stress. For our granular system, this reduces to a kinetic energy contribution and a pairwise energy contribution from neighboring particles, see eq. \ref{eq:stress}:

\begin{equation}\label{eq:stress}
    S_{ab} = -\left[mv_av_b + \frac{1}{2} \sum^N_{n=1} (r1_aF1_b + r2_aF2_b) \right]   \\
\end{equation}

The stress (S) elements a and b take values x, y, and z to form all 6 elements of a symmetric tensor (xx, yy, zz, xy, xz, yz). The stress occurs when particles (located at relative position r1 and r2) moving at relative velocity (v) experience forces (F1 and F2) during a collision. Units are reported in pressure*volume ($N*m$) so that tracking instantaneous deformable particle volume is not necessary. Though \cite{subramaniyan2008continuum} concluded that the virial stress is equivalent to the Cauchy stress, this has been debated in the literature. Commenting on this debate is outside the scope if our work. We use the LIGGGHTS provided calculation for simplicity and use the results only to guide a qualitative analysis of the surface dilation mechanism.

\subsection{Shock validation} \label{sec:3dvalidation}
\begin{figure}[h]
\begin{center}
\includegraphics[scale = .4]{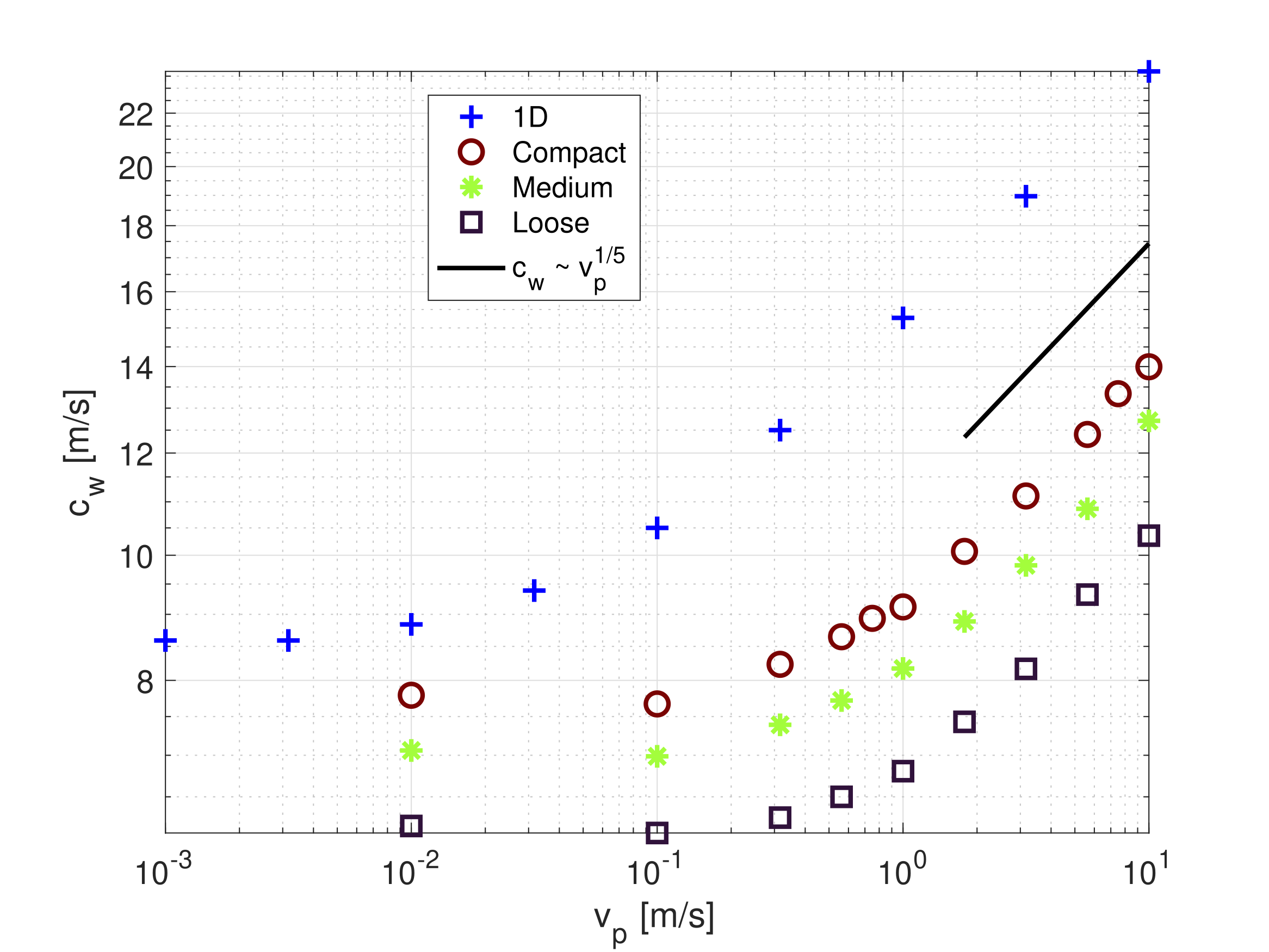}
\end{center}
\caption{ \textbf{ Depth averaged $c_w$ vs $v_p$ for varied geometries}. We show wave speed vs incident speed for the 1D case as well as three different $\phi_0$ for the 3D bed corresponding to the loose, medium, and compact levels of packing from Table \ref{tab:fill_param}.}
\label{fig:validation_3D_comparison}
\end{figure}

We used a 2 $m$ channel filled 20 $cm$ deep with particles to verify the expected weak/strong shock behavior. In a 3D randomly packed assembly, we should obtain the same power law relationship from Section \ref{sec:model_validation} as has been seen in experiments (\cite{van2013shock}, \cite{tell2020acoustic}) and in SSDEM (\cite{sanchez2022transmission}). Averaging across all depths we find the average induced $c_w$ and $c_p$ for a given $v_p$ and this is plotted in Figure \ref{fig:validation_3D_comparison}. We clearly see the expected weak/strong regions. The waves produced by low $v_p$ produce acoustic waves of the same speed (about $7$ $m/s$). $c_0$ is taken as the average of the $c_w$ of the acoustic waves ($v_p \leq 0.1 m/s$) and the results for the different beds are summarized in Table \ref{tab:soundspeed_vsphi}. In the same table, we estimate the bulk modulus by rearranging $c_0 = \sqrt{E_{bulk}/\rho_{bulk}}$ (\cite{campbell2005stress})  and neglecting cohesion.  The largest $v_p$ produce $c_w$ that increase with increasing $v_p$ and we obtain the same $v_p^{1/5}$ power law as in \cite{gomez2012shocks}.

\begin{table}[h]
\begin{center}
\begin{minipage}{174pt}
\caption{Sound speed and bulk modulus in beds of various $\phi_0$. $\rho_{bulk}$ is computed using the particle density from Table \ref{tab:BaseSimParam} with mean $\phi_0$ from Table \ref{tab:fill_param} corresponding to the Loose, Medium, and Compact channels in the 2 $m$ long, 20 $cm$ deep configuration. }\label{tab:soundspeed_vsphi}
\begin{tabular}{@{}cccc@{}}
\toprule
\begin{tabular}{c} $\phi_0$ \\ $  (\%) $ \end{tabular} & \begin{tabular}{c} $\rho_{bulk} $ \\ ($g/cm$)  \end{tabular} & \begin{tabular}{c} $c_0$ \\\ ($m/s$) \end{tabular} & \begin{tabular}{c} $E_{bulk}$ \\ ($kPa$)  \end{tabular}  \\
\midrule
54.72 & 1.37 & $6.55$   & $58.78$    \\
58.65 & 1.42 & $7.01$   & $69.78$ \\
61.96 & 1.55 & $7.51$  &  $87.42$  \\
\bottomrule
\end{tabular}
\end{minipage}
\end{center}
\end{table}

We report the P shock wave speed in Figure \ref{fig:validation_3D_comparison} so that we can compare results from the 1D assembly from Figure \ref{fig:1Dsonicv} in addition to varied $\phi_0$ in the 3D beds. Both the 1D and 3D cases follow the expected 1/5 power law. The magnitude of speeds are comparable between 1D and 3D in our simulations which is due to the slower speeds in a 1D assembly as compared to 3D (\cite{el2008acoustic}) but a larger particle size in the 1D experiment ($\sim10$ $mm$ for 1D, $1.25$ $mm$ for 3D) and larger particles lead to larger sound speed (eq. \ref{eq:sound_speed_nesterenko}) . The reduction in bulk modulus as compared to the 1D case reduces $c_w$ for the same $v_p$, while the reduced bulk density shifts the transition from acoustic waves to shocks to larger $v_p$. Wave speed in the 3D bed increases with increasing $\phi_0$ since a denser particle bed has more contacts and force can be transmitted more quickly. Given our agreement with expectations, we conclude that the acoustic and shock waves simulated in our assembly are physical.  We note here that the averaging obscures some evidence that the near surface waves increase in speed faster, closer to $v_p^{1/4}$. That $c_w$ increases more quickly than expected at the surface is an expected result - anomalous behavior in granular assemblies at low confinement pressure remains an open topic (\cite{goddard1990nonlinear}, \cite{makse1999effective}, \cite{makse2004granular}, \cite{tournat2010acoustics}, \cite{tell2020acoustic}). However, given the noise in our current $c_w$ identification method (due to both fit method and data resolution) a detailed investigation of this result is not conducted.

\begin{figure}[h]
\begin{center}
\includegraphics[scale = .4]{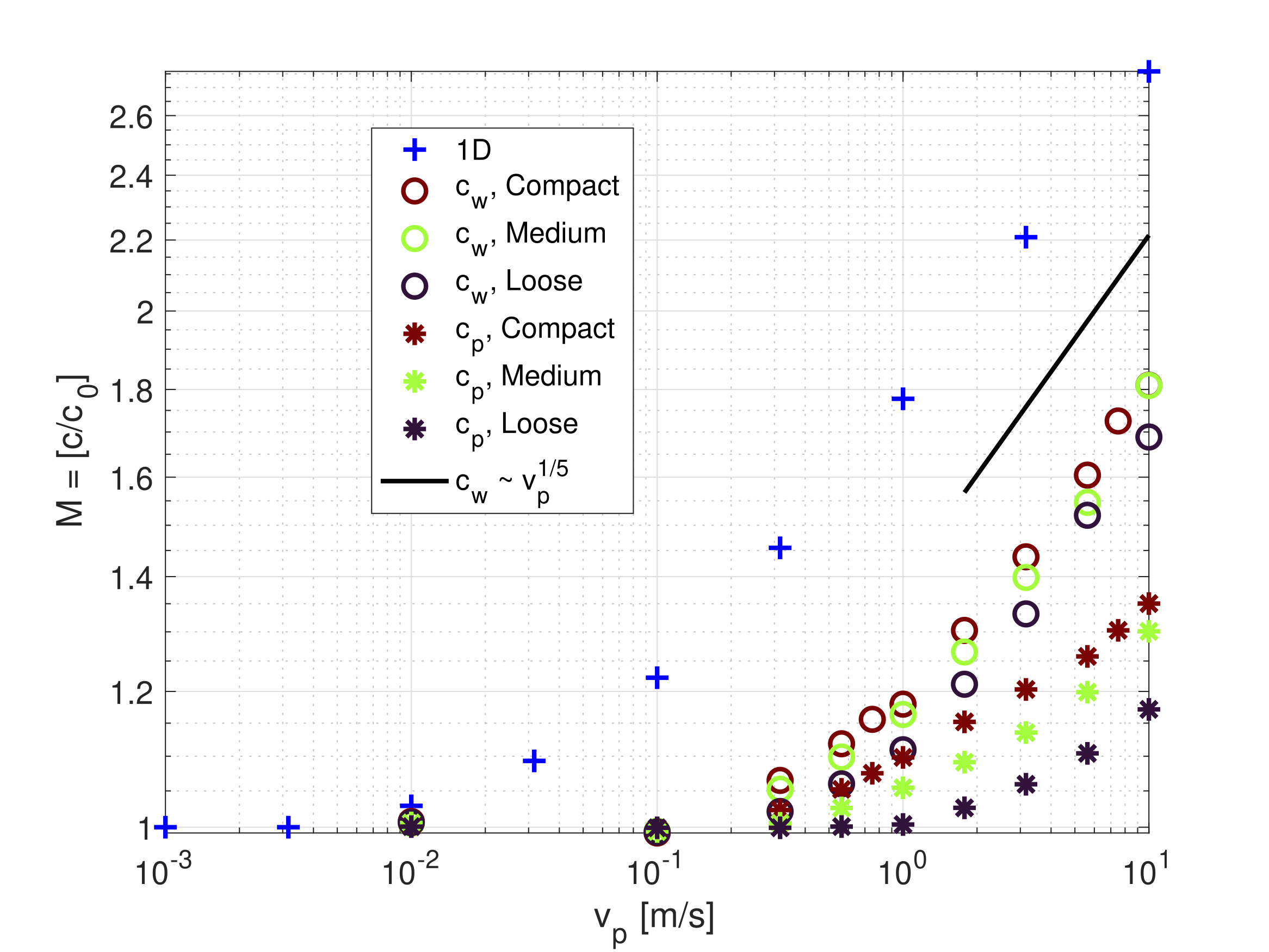}
\end{center}
\caption{ \textbf{ Mach number vs $v_p$ for varied geometries}. Wave speeds for different $v_p$ resulting in shocks in the 1D (plus marker) and 3D shocks (circles) are normalized against $c_0$. The solitary wave speed $c_p$ is corresponds to the star markers.}
\label{fig:validation_3D_normalized}
\end{figure}

At this point we can continue the model validity discussion of section \ref{sec:model_validation}. Figure \ref{fig:validation_3D_normalized} now includes the solitary wave speed $c_p$ and normalizes all wave speeds against $c_0$ to retrieve the Mach number ($M = c/c_0$). While we see that the largest $v_p$ produce shocks that follow the expected power law, these $v_p$ are an appreciable fraction of $c_m$. In reality, the produced $c_w$ would be affected by plasticity near the contact site or even fragmentation for the largest $v_p$. Despite our particles undergoing contacts with velocities beyond the elastic limit as the shock is initially created, the shocks produced in our assembly travel at speeds $< 2c_0$ which is in line with the assembly shock speeds explored in \cite{abd2010force}. We are also more interested in the wave as it travels at its solitary wave speed ($c_p$), since it is in this state when consistent dilation is initiated across the channel. Our $c_p$ do not exceed $1.4c_0$ and so the effects of impact-site-plasticity and fragmentation would be less significant for particles in the channel that are not directly adjacent to the impact site. In table \ref{tab:speed_waveDetails}, we show that the particle velocities experienced within the solitary wave front don't exceed 200 $mm/s$ (with the elastic limit around 100 $mm/s$).

\subsection{Test plan} \label{sec:planned_tests}
Having confirmed the validity of our method, we can now examine various cases to determine if and when surface dilation will occur due to generated waves.  We characterize this surface dilation with respect to wave speed, wave type, packing configuration, and boundary effects (particle fill height, length of channel). The channel length and fill height tests determine if the geometry of the container has an impact on our results. Finally, varied wave speeds will reveal if surface dilation can occur for sound waves or shocks. The wave type test will determine the dependence of dilation on P and S waves. Varied $\phi_0$ will determine the dilation response sensitivity to initial conditions. We also briefly consider the case of two colliding waves. The results of these tests are shared in the following section.

\section{Results and discussion}  \label{sec:Results}
We find that in all cases, a single wave is created which can persist across the entire length of the channel, depending on the input $v_p$. When $v_p$ is large enough to generate a shock, the wave undergoes a rapid initial decay, but then propagates at reduced speed, although still supersonic, across the channel. The propagating wave appears to be a solitary wave as it travels with minimal dissipation and maintains its shape. Even for barely supersonic solitary waves, surface dilation occurs as long as the wave is maintained. However, sustained dilation only occurs as the result of a compressive (P) wave. We first provide a qualitative description of grain motion during the induced dilation before quantifying the resultant bulk volume change and discussing implications.

\subsection{Impulse-induced granular dilation} \label{sec:shock-mechanism}

\begin{figure*}[h]
\begin{center}
\includegraphics[scale = .85]{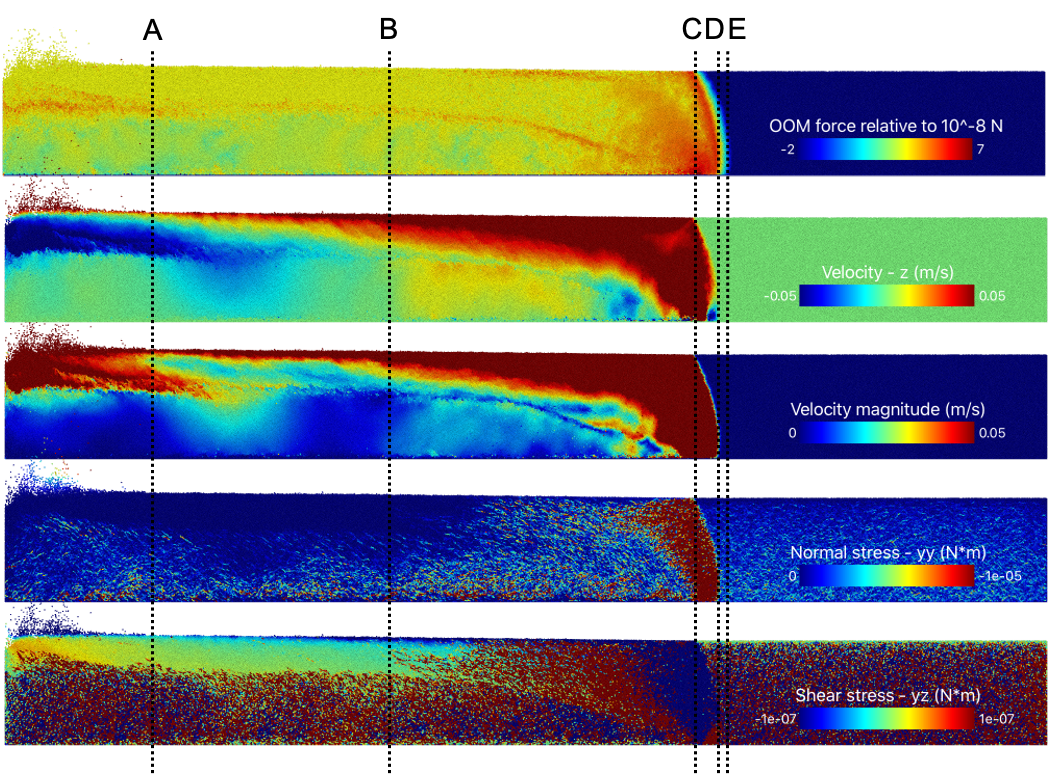}
\end{center}
\caption{ \textbf{P wave visualization.} Various parameter visualizations of a 2 $m$ long channel filled 20 $cm$ deep with particles (compact filling) at t=0.133 $s$ after P shock initiation ($v_p = 10$ $m/s$). In order of top to bottom, the panels color the particles by their force magnitude, z-velocity, velocity magnitude, normal stress (yy), and shear stress (yz). We show the legend scale and units inlaid on the right of the channel in each panel. Dashed lines are guides corresponding to labels which are described in the text. The initial shock decays to steady solitary wave and is maintained across the entirety of the channel.}
\label{fig:shock_vis_Pwave_combo}
\end{figure*}

Our laterally propagating P shock waves initially experience a rapid exponential decay in force before reaching a steady value that is maintained for a distance, in some cases across the entire channel. Despite the decrease in speed from the initial $c_w$, the solitary wave speed $c_p$ is still supersonic. As the solitary wave propagates along the channel, particles in the near surface are mobilized after the wave front passes by, with the top most layers detaching and following a parabolic trajectory (seen in Fig. \ref{fig:dilation_vs_time}). The sensor elements in the top row of Figure \ref{fig:channelcartoon}, box A confirm this behavior. Looking closely at the surface grains and using the top box boundary as a guide, we see that the grid elements on the right (just before and after wave front passage) still have a flat uniform surface that is in line with the initial height of the bed while those on the left do not, as the bulk of the assembly has dilated to fill the sensor element with particles. In the wake of the solitary wave front, a secondary wave is reflected off the floor and travels vertically upward. The reflected floor wave reaches the surface particles while they are in an inertial state. We refer the reader to frames c-f of Fig. \ref{fig:shockframes} and Fig. \ref{fig:collision} to see the floor wave. Ultimately, the solitary wave reflects off the end wall but disperses rapidly and does not travel back down the channel. The reflected end-wall wave only affects surface modification at the very end of the channel ($y = [1.5, 2]$ $m$ in Fig. \ref{fig:velsweep_dilation}, for example). The final settled height of the particles is increased across the bed, except very near the wave generation site and at the end wall (seen in section \ref{sec:surface_dilation}). In the appendix, we provide frames from various times during the simulation to show how the force profile of the wave evolves (Fig. \ref{fig:shockframes}).

\begin{figure*}[h]
\begin{center}
\includegraphics[scale = .85]{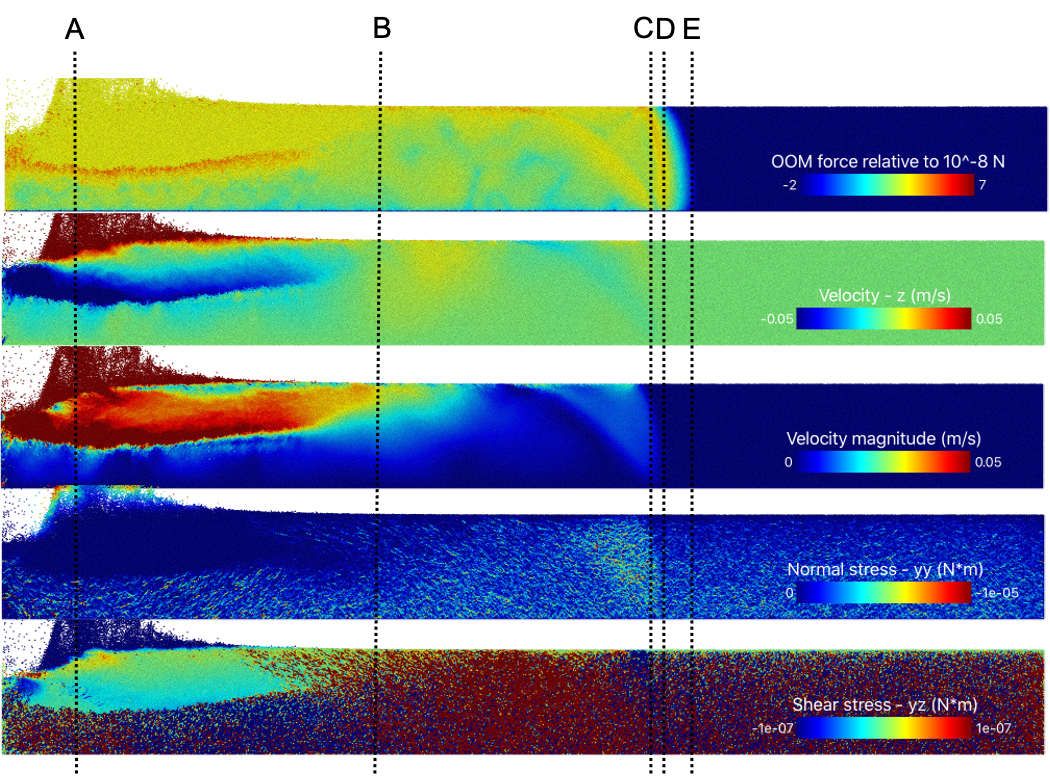}
\end{center}
\caption{ \textbf{S wave visualization.} Various parameter visualizations of a 2 $m$ long channel filled 20 $cm$ deep with particles (compact filling) at t=0.133 $s$ after S shock initiation ($A = 3.1831x10^{-5}$ $m$). In order of top to bottom, the panels color the particles by their force magnitude, z-velocity, velocity magnitude, normal stress (yy), and shear stress (yz). We show the legend scale and units inlaid on the right of the channel in each panel. Dashed lines are guides corresponding to labels which are described in the text.}
\label{fig:shock_vis_Swave_combo}
\end{figure*}

The details of the surface dilation mechanism are most easily seen by evaluating the response of the assembly across several physical parameters at the same time. Figure \ref{fig:shock_vis_Pwave_combo} and Figure \ref{fig:shock_vis_Swave_combo} visualize force, velocity, normal stress (yy), and shear stress (yz) of the same assembly at a single time step after being subjected to P and S shocks (respectively). Surface dilation occurs for a P wave but not for an S wave of comparable strength. Section \ref{sec:surface_dilation} will quantify this result, but we characterize the differences between the two waves here using the alphabetical labels in Figures \ref{fig:shock_vis_Pwave_combo} and \ref{fig:shock_vis_Swave_combo} as a guide. Consider a single radial position along the channel. As the P solitary wave arrives in Figure \ref{fig:shock_vis_Pwave_combo}, the particles at this radial position first experience an increase in force (E - force panel). Rapidly thereafter they experience the peak force which is aligned with the arrival of the compressive wave front (D - force and normal stress panels). Particles attain a velocity at this point too, though it is initially only a lateral velocity and the particles are pushed in the radial direction into their neighbors (D, z-velocity and velocity magnitude panels). However, this y-velocity is quickly converted into an upwards z-velocity looking just to the left of D in the z-velocity panel. The compressive front triggers particle ejection through a frictional collision with neighboring particles. Following this, the shear front arrives, seen as a rapid transition from negative to positive yz-shear stress seen at C in the shear stress panel. At this point, the region between A and B is completely detached (zero normal and shear stress as seen in those respective panels) and particles experience a ballistic trajectory. Left of A, particles are falling back down and the initial stress profile starts to reestablish. Note the presence of a vortex immediately following the compressive front (D, z-velocity and normal stress panels) of the particle-floor boundary. The highest forces present as a result of the shock are at the shear wave front and floor interface (C, force and shear stress panels).  The shear front following the compressive wave sustains the wave and drives dilation to occur across the entire channel.

\begin{figure*}[h]
\begin{center}
\includegraphics[scale = .85]{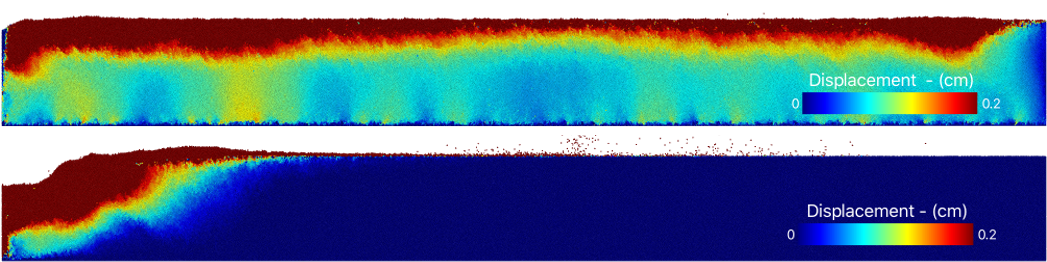}
\end{center}
\caption{\textbf{Dilation visualization.} Particles are colored by their displacement in height as compared to their initial position for the compact 2 $m$ channel filled 20 $cm$ deep with particles. The top channel experienced a P shock wave while the bottom channel experienced an S shock wave. }
\label{fig:PvsS_dilation_vis}
\end{figure*}

Without the strong compressive front, the initial S shock (Fig. \ref{fig:shock_vis_Swave_combo}) does not generate a solitary wave and is unable to trigger distant dilation. The strength of the wave has decayed by the same time the P solitary wave has reached an equivalent radial distance (E, force panel). The bender element does not completely remove P wave effects (as expected) and there is a weaker compressive wave front visible (D, normal stress panel). The particles acquire a small velocity from the compression, but it is never converted into a z-velocity (B-D, z-velocity panel). There is no high force region at the particle-floor interface maintaining the solitary wave as with the P wave and this is due to the difference in shear stress profile. The scale of the shear stress direction flip (C, shear stress panel) is greatly reduced in the S wave case and cannot sustain the wave. There is a region of detachment which is confined to be close to the initiation site (0 stress between A and B in the normal and shear stress panels), whereas the P wave detachment region formed a band of lofted particles that persisted across the entire channel. We show in section \ref{sec:results_wavetype} that dilation from the S wave is confined to the near initiation region while the P-wave-resultant dilation spans the channel in a near surface band that resembles the detachment band in Figure \ref{fig:shock_vis_Pwave_combo}. Given that dilation only occurs over long distances for the compressive wave, the remainder of the results section focuses on our P wave simulations.

\subsection{Surface dilation quantification} \label{sec:surface_dilation}
This section focuses on quantifying height change (and therefore dilation) away from the initiation site. We find the wave-induced height change for a given test by taking the average at-rest bed height from a measurement region in between red dashed lines in the following plots that provides some margin from the walls. We exclude the region nearest to the left wall where the initial shock is active since surface particles are ejected due to the larger impact speeds experienced within the shock front. As is the case in blast-loaded 1D particle granular systems (those that undergo a short impulse), the decay region of our shocks is a function of path length. As seen in Fig. \ref{fig:velsweep_overlap}, the $v_p$ that produce shocks all decay to their stationary strength by approximately the same point in the channel. The decay region length depends on the number of particles in the path of the force chain. In our simulations, the decay region ends after roughly 30 $cm$ which is longer than the 5 $cm$ that would be the case in 1D chain ($\sim 10$ particle diameters,\cite{nesterenko2013dynamics}). To avoid the region perturbed by the shock (and similarly, the region in which the solitary wave decays after reflection off the end wall) with some extra margin, we ignore the height change of particles within 50 $cm$ of the initiation (left) and end (right) walls in the final bed height change calculation.

\subsubsection{Wave type effects} \label{sec:results_wavetype}

We first subjected the same compact bed to the strongest shocks, $v_p = 10$ $m/s$ for the P wave and corresponding bender displacement amplitude $A = 3.1831x10^{-5}$ $m$ for the S wave. Figure \ref{fig:PvsS_dilation_vis} shows particle height change as a result of the shock waves. We see that volume change is only affected beyond the initiation region by P waves and a solitary wave is not maintained after S wave generation. The force profile in the P front is in the same direction of propagation while the S front is transverse to propagation (\cite{li2002particle}). As discussed in the prior section, the lateral compressive force in the P wave triggers a frictional collision with neighboring particles, allowing particles in the wake of the shock to attain a positive z velocity.  Since there is no surface dilation occurring as the result of even the strongest S wave shock, we consider only P waves going forward. Note that the region of most extreme dilation (the dark red band along the surface of the P wave height change in Figure \ref{fig:PvsS_dilation_vis}) corresponds with the detachment region as seen in the normal and shear stress shock profile in Figure \ref{fig:shock_vis_Pwave_combo}.

\subsubsection{Random seeding effects} \label{sec:seed_effects}

\begin{figure}[h]
\begin{center}
\includegraphics[scale = .4]{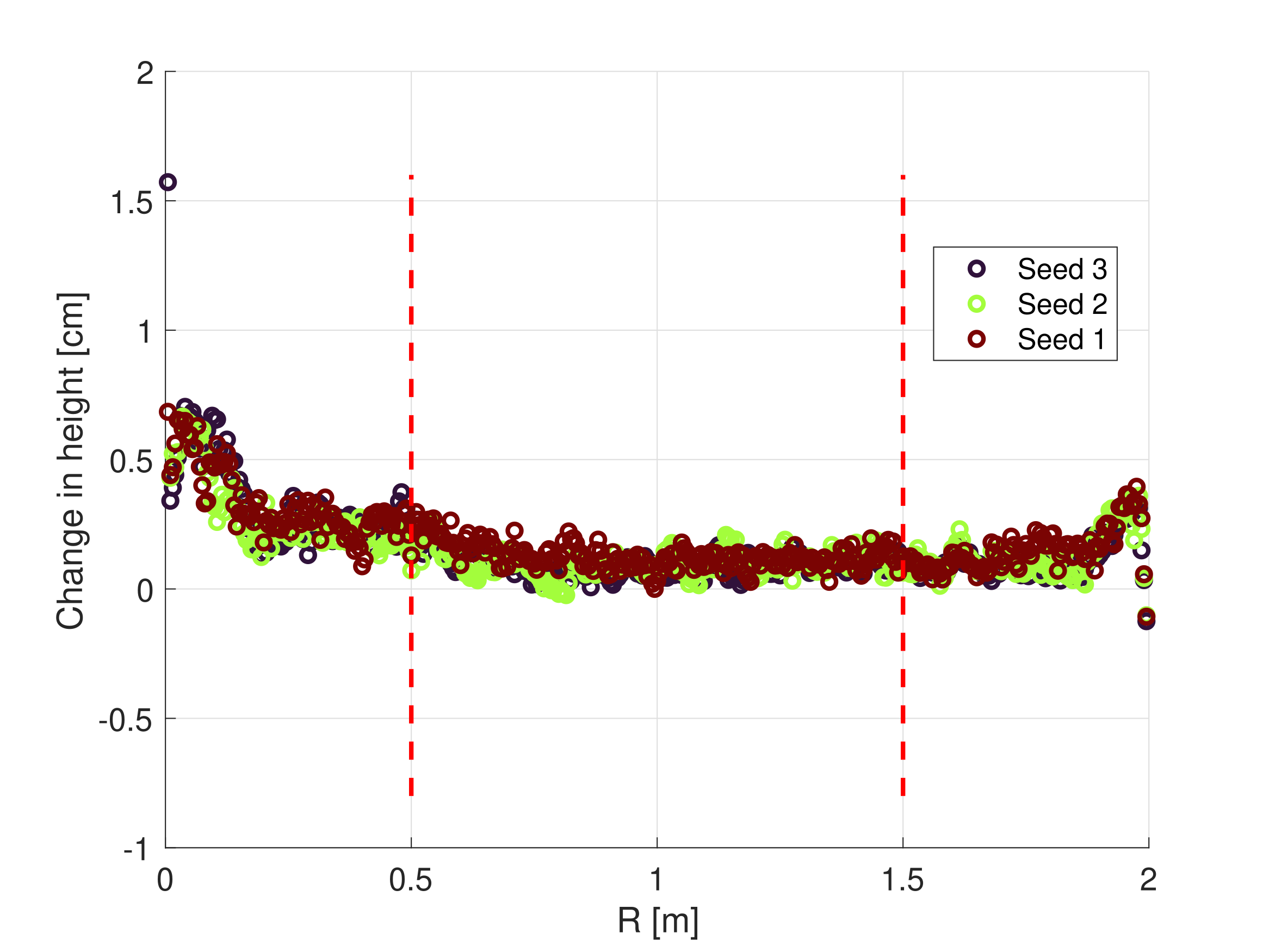}
\end{center}
\caption{ \textbf{ $\Delta z$ vs Radial position.} Color legend corresponds to seed number configuration and red dashed lines indicate the measurement region. We generated shocks ($v_p = 10$ $m/s$) in loose beds 2 $m$ in length filled 20 $cm$ deep with particles. The randomness of the packing has little effect on post shock dilation.}
\label{fig:random_seed_effects}
\end{figure}

Next we determine dilation error bars by evaluating the deviation in post-impulse bed height change due to randomness of the initial packing. Changing the seed numbers passed to the particle creation and insertion algorithm changes the initial packing of the bed. We perform this test by shocking ($v_p = 10$ $m/s$) the loosely poured bed configuration in a 2 $m$ channel filled 20 $cm$ deep. We prepare three identical beds and only vary the prime seeds passed for each case.  The results from each test are quite similar as can be seen in Figure \ref{fig:random_seed_effects}. There is very little effect of the randomness of packing on the dilation response. We quantify this variability by finding the deviation in post-impulse height change. We compute the mean ($\mu\Delta z$) and deviation ($\sigma\Delta z$) across all radial positions within the measurement range and the results are provided in Table \ref{tab:seed_summary}. $\sigma\Delta z$ is a result of the randomness in the packing along the channel within a single test and all three seed groups are comparable. Since we want to know how comparing results between two assemblies is affected by the preparation history, we take the random seed error (RSE) to be the deviation of $\mu\Delta z$.  This results in an RSE of 0.0172 $cm$ or approximately 0.2 $mm$ (or 0.1 \% bulk volume change in beds filled 20 $cm$ deep with particles). 

\begin{table}[h]
\begin{center}
\begin{minipage}{174pt}
\caption{Bed height change for different seed groups. Mean ($\mu$) and deviation ($\sigma$) are taken from the height change in the measurement region for each seed group in Fig. \ref{fig:random_seed_effects}. }\label{tab:seed_summary}
\begin{tabular}{@{}cccc@{}}
\toprule
\begin{tabular}{c} Group \end{tabular} & \begin{tabular}{c} $\mu \Delta z$ ($cm$) \end{tabular} & \begin{tabular}{c} $\sigma \Delta z$ ($cm$) \end{tabular}  \\
\midrule
Seed 1 & $0.0964$   & $0.0432$    \\
Seed 2 & $0.1017$   & $0.0500$ \\
Seed 3 & $0.1284$  &  $0.0501$  \\
\bottomrule
\end{tabular}
\end{minipage}
\end{center}
\end{table}

\subsubsection{Wave speed effect} \label{sec:wave_speed_effect}
\begin{figure}[h]
\begin{center}
\includegraphics[scale = .4]{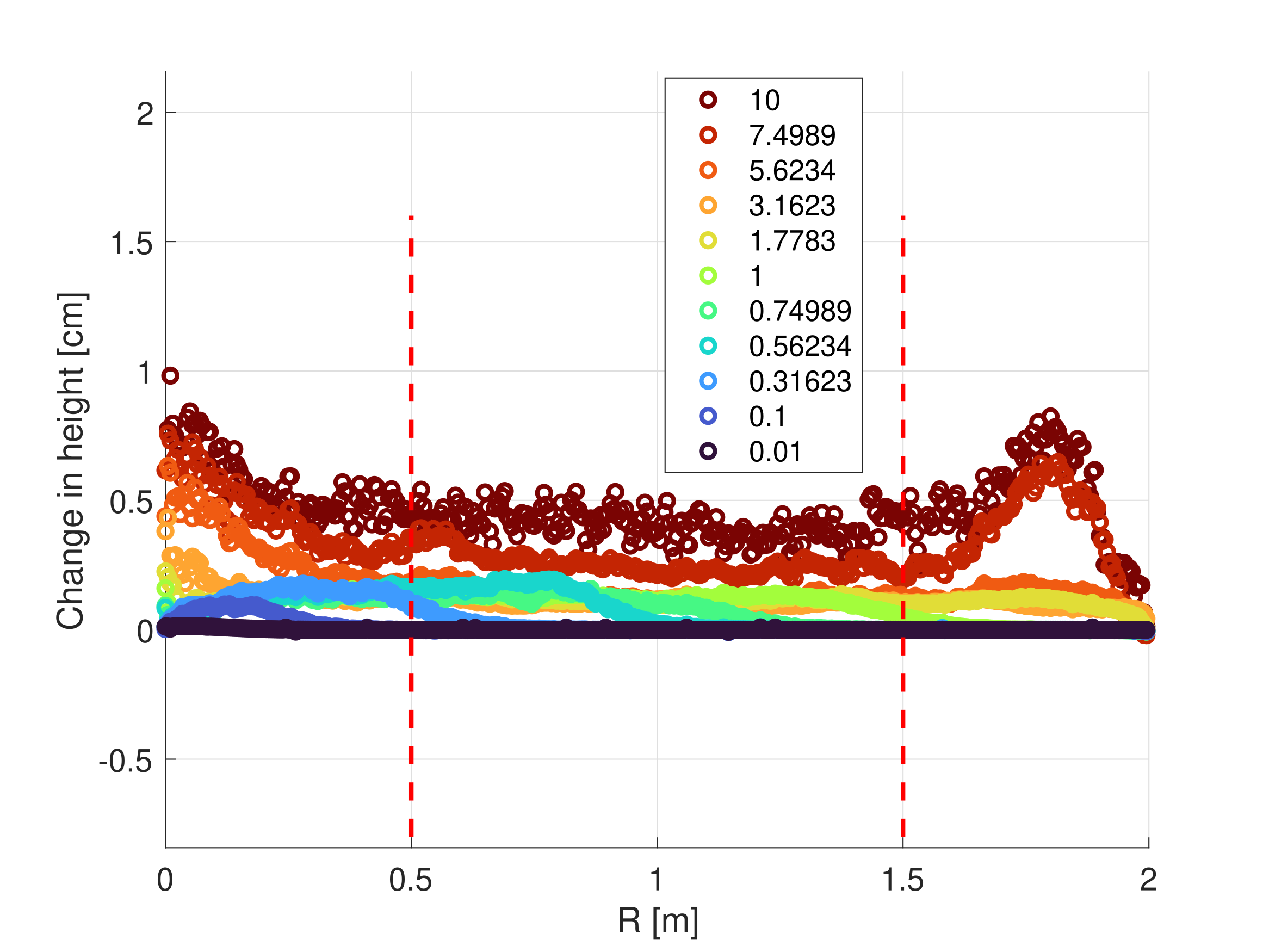}
\end{center}
\caption{\textbf{$\Delta z$ vs radial position for varied $v_p$.} The color legend shows $v_p$ in $m/s$ for waves initiated in a compact bed 2 $m$ in length filled 20 $cm$ deep with particles. Acoustic waves ($v_p \leq 0.1$ $m/s$) show no bed height changes. Waves in the transition region (0.1 $< v_p$ $\leq 1$ $m/s$) begin to show some dilation, but it does not persist across the channel. The remainder of the $v_p$ investigated induce shocks, which see uniform dilation across the channel.}
\label{fig:velsweep_dilation}
\end{figure}

We perform a velocity sweep over three decades in $v_p$ (0.01 to 10 $m/s$) for P waves an all three packing configurations (loose, medium, compact) to evaluate wave speed influence on impulse-induced dilation. In this section, all tests were run in 2 $m$ channels filled 20 $cm$ deep. The details for all the tests are given in Table \ref{tab:speed_waveDetails} which reports on the properties of the generated waves, the state of particles within the wave front, and the resultant dilation induced by the wave (as seen through height change). In all three cases, we see that that low $v_p$ generate acoustic waves which travel at the sound speed and do not induce dilation. Waves in the transition region begin to induce some post-impulse bed height change, but only the solitary waves (which decay from the strongest shocks) trigger dilation across the entire channel (shown for the compact case in Fig. \ref{fig:velsweep_dilation}). Even for these strongest shocks, the speed of particles within the front are much less that the wave speed (small $v_m/c_p$) and $v_m$ barely exceeds the Hertz elastic collision limit ($0.1$ $m/s$). While the compact and medium packed assemblies show dilation for all cases $v_p > 1$ $m/s$, the loose packed channel showed either slight dilation or slight compaction. We believe this variation indicates that the loose packed channel ($\phi_0 = 54.72 \%$) is at or near the crossover threshold between initial conditions that lead to compaction instead of dilation. This is less than the crossover $\phi_0 = 58 \%$ in \cite{royer2011role}, but that makes sense given our reduced gravity conditions and addition of cohesion. Although all $v_p$ investigated for the loose $\phi_0$ were performed in the same bed, the RSE ($\sim0.2$ $mm$) is only fully exceeded for the case at $v_p = 10$ $m/s$. This may indicate that randomness of the packing allows for variable volume change around the compaction-dilation threshold. We should also point out that there is a discrepancy in the acoustic wave speed reported for the loose bed between the $c_w$ and $c_p$ reported speeds in Table \ref{tab:speed_waveDetails} . The two speeds should be approximately the same for acoustic waves as is the case with the compact and medium packed channels. This discrepancy highlights that our wave tracking method (peak force \ref{sec:waveterminationandspeed}) is better suited for tracking stronger waves with greater $\phi_0$ because the increase in peak force is more abrupt in these cases and the peak force coincides with the end of an exponential force increase. As the peaks become less sharp, the determination of when the shock passes through the sensor becomes more challenging. Using a first arrival wave speed based on $\delta$ would likely resolve this issue.

\begin{table*}[h]
\begin{minipage}{450pt}
\caption{Wave characteristics for the velocity sweeps in the compact (C), medium (M) and loose (L) packed beds. The average initial overlap correspond to $\delta_0$ = 4.82, 4.81 and 4.23 $\mu m$ (L,M,C respectively), see sec. \ref{sec:bed_prep} for details. The columns correspond to packing fraction, piston velocity, initial (shock) wave front speed, propagating (solitary) wave front speed, particle velocity within the solitary wave front, maximum overlap experienced by particles following the passage of the solitary wave front, the ratio of particle to wave front speeds, and the at rest height change (from the initial 20 $cm$ bed height) following the wave's passage.}\label{tab:speed_waveDetails}
\begin{tabular}{@{}cccccccc@{}}
\toprule
\begin{tabular}{c} $\phi_0$ \end{tabular} & \begin{tabular}{c} $v_p$=$10^n$ \\ ($m/s$) \end{tabular} &\begin{tabular}{c} $c_w$ ($m/s$) \\ $\mu \pm \sigma$ \end{tabular} & \begin{tabular}{c} $c_p$ ($m/s$) \\ $\mu \pm \sigma$ \end{tabular}  & \begin{tabular}{c} $v_m$ ($mm/s$) \\ $\mu \pm \sigma$ \end{tabular}  & \begin{tabular}{c} $\delta_m$ ($\mu m$) \\ $\mu \pm \sigma$ \end{tabular}  & \begin{tabular}{c} $\sim\frac{v_m}{c_p}$ \\ x$10^{-3}$  \end{tabular}   &    \begin{tabular}{c} $\Delta z$ ($mm$) \\  $\mu \pm \sigma$ \end{tabular}  \\
\midrule
C & $-2$ & $7.79\pm0.52$ & $7.53 \pm 0.17$ & $0.64 \pm 0.15$  & $4.16 \pm 0.09$  & 0.09  & $ 0.00 \pm 0.02 $ \\
C & $-1$  & $7.67\pm0.33$ & $7.50 \pm 0.18$ & $5.65 \pm 0.93$  & $4.67 \pm 0.14$  & 0.75  & $ 0.00 \pm 0.02 $ \\
C & $-1/2$  & $8.23\pm0.34$ & $7.69 \pm 0.16$ & $7.47\pm 5.23$  & $4.84 \pm 0.57$  & 0.97  & $ 0.00 \pm 0.02 $ \\
C & $-1/4$ & $8.65\pm0.36$ & $7.90 \pm 0.16$ & $22.33 \pm 15.75$  & $ 6.47 \pm 1.86$  & 2.83  & $ 0.04 \pm 0.06 $ \\
C & $-1/8$ &  $8.94\pm0.41$ & $8.07 \pm 0.17$ & $28.71\pm 13.41$  &7.21  $ \pm 1.50$  & 3.56  & $ 0.35 \pm 0.37 $ \\
C & $0$  & $9.12\pm0.37$ & $8.25 \pm 0.17$ & $34.19 \pm 5.43$ & $7.86 \pm 0.64 $  & 4.15  & $ 1.14 \pm 0.22$ \\
C & $1/4$  & $10.07\pm0.49$ & $8.65 \pm 0.20$ & $32.77 \pm 6.01$  & $7.34 \pm 0.49 $  & 3.79  & 1.12 $ \pm0.11 $ \\
C & $1/2$  & $11.12\pm0.52$ & $9.04 \pm 0.23$ & $42.30\pm 10.37$  & $7.41 \pm 0.85$  & 4.68 & $ 1.07 \pm 0.11$ \\
C & $3/4$  & $12.40\pm0.41$ & $9.45 \pm 0.22$ & $78.96\pm 10.32$  & $9.90 \pm 0.79$  & 8.36 & $ 1.16 \pm 0.14$ \\
C & $7/8$ & $13.34\pm0.43$  & $9.78 \pm 0.23$ & $117.11\pm 7.87$  & $12.77 \pm 0.53 $  & 11.97 & $ 2.31 \pm 0.22 $ \\
C & $1$ & $14.00\pm0.31$ & $10.14 \pm 0.20$ & $181.17 \pm 7.17$  & $16.94 \pm 0.55 $  & 17.87 & $ 3.84 \pm 0.56$ \\
\hline
M & $-2$  & $7.06\pm0.57$ & $7.05 \pm 0.15$ & $0.13 \pm 0.06$  & $4.61 \pm 0.19$  & 0.02 & $ 0.00 \pm 0.02$ \\
M & $-1$  & $6.99\pm0.49$  & $6.97 \pm 0.17$ & $2.69 \pm 0.53$  & $4.90 \pm 0.20$  & 0.39 & $ 0.00 \pm 0.02$ \\
M & $-1/2$  &  $7.39\pm0.50$ & $7.04 \pm 0.17$ & $3.90 \pm 3.32$  & $5.04 \pm 0.43$  & 0.55 & $ 0.02 \pm 0.06$ \\
M & $-1/4$  & $7.72\pm0.48$  & $7.19 \pm 0.18$ & $10.69 \pm 7.92 $  & $5.84 \pm 0.95$  & 1.49 & $ 0.20 \pm 0.24$ \\
M & $0$  & $8.17\pm0.53$  & $7.39\pm 0.19$ & $20.57 \pm 3.91$  & $6.90 \pm 0.43$  & 2.78 & $ 0.59 \pm 0.12$ \\
M & $1/4$  & $8.89\pm0.52$  & $7.65 \pm 0.21$ & $20.21 \pm 3.97$  & $6.65 \pm 0.45$  & 2.64 & 0.62 $ \pm 0.07 $ \\
M & $1/2$  & $9.82\pm0.62$  & $7.96 \pm 0.23$ & $30.88 \pm 6.42$  & $7.01 \pm 0.56$  & 3.88 & $ 0.56 \pm 0.08 $ \\
M & $3/4$  & $10.86\pm0.53$  & $8.40 \pm 0.25$ & $65.38 \pm 8.99$  & $9.75 \pm 0.71$  & 7.78 & $ 0.81 \pm 0.13$ \\
M & $1$  & $12.71\pm0.50$  & $9.12 \pm 0.30$ & $151.46 \pm 9.23$  & $15.75 \pm 0.77$  & 16.61 & $ 3.07 \pm 0.65 $ \\
\hline
L & $-2$  & $6.17\pm0.93$  & $7.10 \pm 0.46$ & $0.34 \pm 0.12$  & $4.49 \pm 0.24$  & 0.05 & $-0.07 \pm 0.08 $ \\
L & $-1$  & $6.09\pm0.97$  & $7.09 \pm 0.42$ & $1.65 \pm 0.61$  & $4.66 \pm 0.26$  & 0.23 & $-0.08 \pm 0.08$ \\
L & $-1/2$  & $6.26\pm0.91$  & $7.09 \pm 0.42$ & $2.91 \pm 2.87$  & $4.83 \pm 0.44$  & 0.41 & $-0.07 \pm 0.08$ \\
L & $-1/4$  & $6.50\pm0.90$  & $7.10 \pm 0.44$ & $9.12 \pm 8.51$  & $5.58 \pm 1.06$  & 1.28  & $0.00 \pm 0.12$ \\
L & $0$  & $6.80\pm0.95$  & $7.12 \pm 0.51$ & $19.11 \pm 4.37$  & $6.67 \pm 0.48$  & 2.68 & $0.16 \pm 0.11 $ \\
L & $1/4$  & $7.43\pm0.97$  & $7.28 \pm 0.48$ & $19.51 \pm 4.96$  & $6.48 \pm 0.49$  & 2.68 & $0.19 \pm 0.10$ \\
L & $1/2$  & $8.17\pm0.91$  & $7.51 \pm 0.56$ & $31.39 \pm 6.85$  & $7.03 \pm 0.55$  & 4.18 & $-0.09 \pm 0.12$ \\
L & $3/4$  & $9.32\pm0.85$  & $7.83 \pm 0.65$ & $75.40 \pm 9.36$  & $10.10 \pm 0.77$  & 9.63 & $-0.30 \pm 0.18 $ \\
L & $1$  & $10.36\pm0.54$  & $8.31 \pm 0.77$ & $141.24 \pm 5.69$  & $15.27 \pm 0.56$  & 17.00 & $0.96 \pm 0.43 $ \\
\bottomrule
\end{tabular}
\end{minipage}
\end{table*}

\begin{figure}[h]
\begin{center}
\includegraphics[scale = .4]{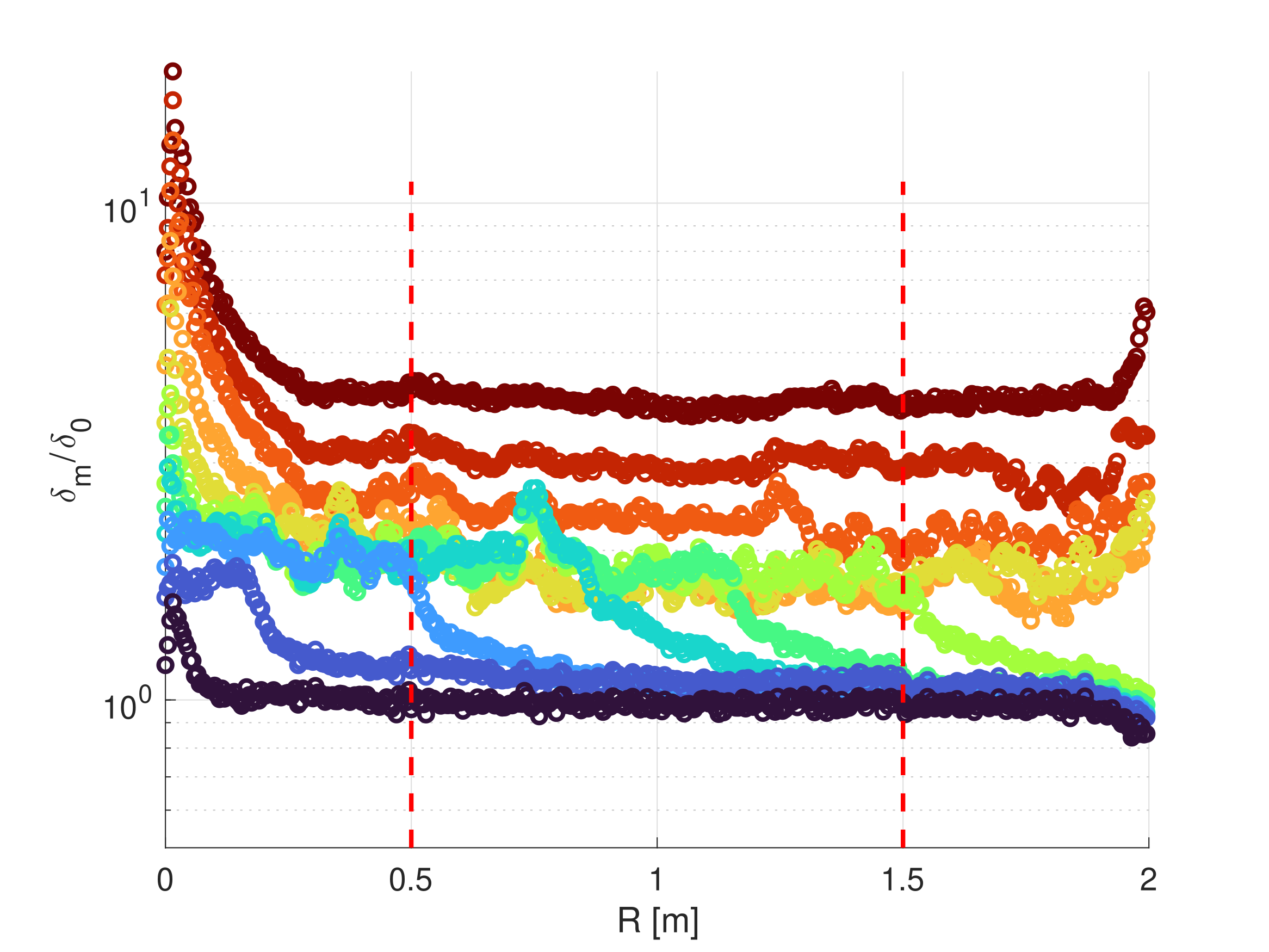}
\end{center}
\caption{\textbf{Average $\delta_m / \delta_0$ vs radial position for varied $v_p$.} Normalized maximum particle overlap over the length of a 2 $m$ channel filled 20 $cm$ deep with particles (compact filling) is shown for the same waves created in Figure \ref{fig:velsweep_dilation}. While all induced waves experience a region of initial rapid decay in $\delta_m$, only the shocks decay to solitary waves which propagate with constant $\delta_m$ (until they terminate). The strongest waves are maintained across the entire channel. The colors correspond to the same $v_p$ legend as in Fig. \ref{fig:velsweep_dilation}, but is omitted here so as not to obscure the trends.}
\label{fig:velsweep_overlap}
\end{figure}

In terms of the strength of the wave, the strongest solitary wave ($v_p = 10$, dark red) exhibits a constant $\delta_m$ across the entire channel (see fig. \ref{fig:velsweep_overlap}), i.e., there is minimal dissipation after the shock initially decays. Solitary waves corresponding to shocks initiated at $1 < v_p \leq 10^{0.875}$ $m/s$ (orange - red) maintain a constant $\delta_m$ for some distance across the channel and then begin to decay, but maintain $\delta_m > \delta_0$. For those waves initiated by shocks at $10^{-0.5} < v_p \leq 1$ $m/s$ (light blue - green), $\delta_m$ decays back to $\delta_0$ after the solitary wave terminates. The rest of the waves ($v_p \leq 10^{-0.5}$, black - dark blue) are purely acoustic and never attain a constant $\delta_m$ before returning to $\delta_0$. Be aware that taking the average over the measurement region causes larger deviations and tends to under report $\delta_m$ for the intermediate $v_p$ cases since we average over a region in which the solitary wave is no longer active. The remainder of our tests are conducted using the strongest solitary wave (from $v_p = 10$ $m/s$) since it induces the greatest dilation response.

\begin{figure}[h]
\begin{center}
\includegraphics[scale = .4]{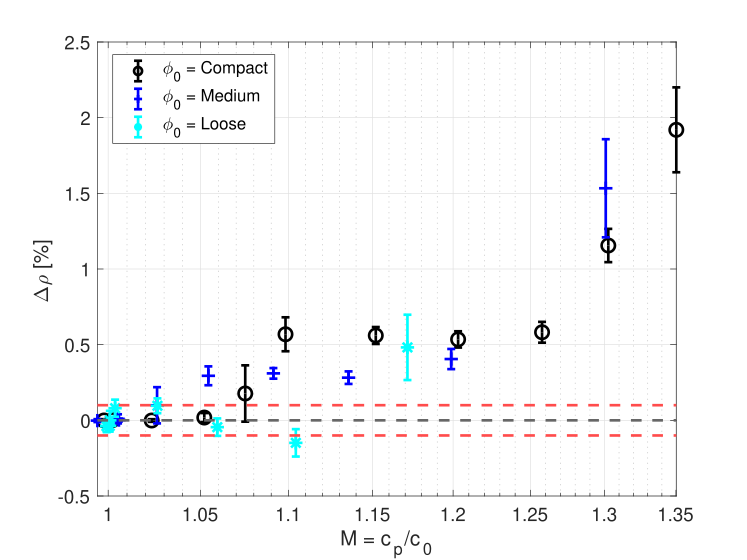}
\end{center}
\caption{\textbf{$\Delta\rho$ vs $M = c_p/c_0$.} Dilation vs Mach speed for the compact, medium, and loosely packed beds. The black dashed line at 0 is the compact-dilation threshold and the red dashed line  corresponds to the RSE.}
\label{fig:dz_vs_mach}
\end{figure}

To understand the strength of the waves inducing dilation we show Figure \ref{fig:dz_vs_mach} which plots resultant dilation vs Mach number in channels of different $\phi_0$. We find the Mach number $M = c_p/c_0$ using the solitary wave speeds from Table \ref{tab:speed_waveDetails} and the sound speed from Table \ref{tab:soundspeed_vsphi}. $\Delta\rho$ is the percent change in bulk density taken as the impulse-induced height change from Table \ref{tab:speed_waveDetails} divided by the bed height (for example, $\Delta\rho = 100$x$\Delta z / 20$  $cm$ in a channel filled to 20 $cm$ with particles). The acoustic waves from all tests ($M=1$) are all clustered around $\Delta\rho = 0$, as expected. However, once the waves are no longer acoustic (i.e., barely supersonic), dilation begins to occur. Disregarding the loose case (since it barely exceeds the RSE), the intermediate strength solitary waves in the medium ($1.02 < M < 1.2$) and compact ($1.07 < M < 1.26$) exhibit roughly constant height change. The strongest solitary wave sees more than a doubling in $\Delta\rho$ over the constant height change region, however the $v_m$ experienced by these particles do exceed the static limit (though, less than 2x) so they may be more influenced by plastic deformation in reality. Nevertheless, these results show that barely supersonic solitary waves can excite surface dilation.

\subsubsection{Boundary effects: channel length}
\begin{figure}[h]
\begin{center}
\includegraphics[scale = .4]{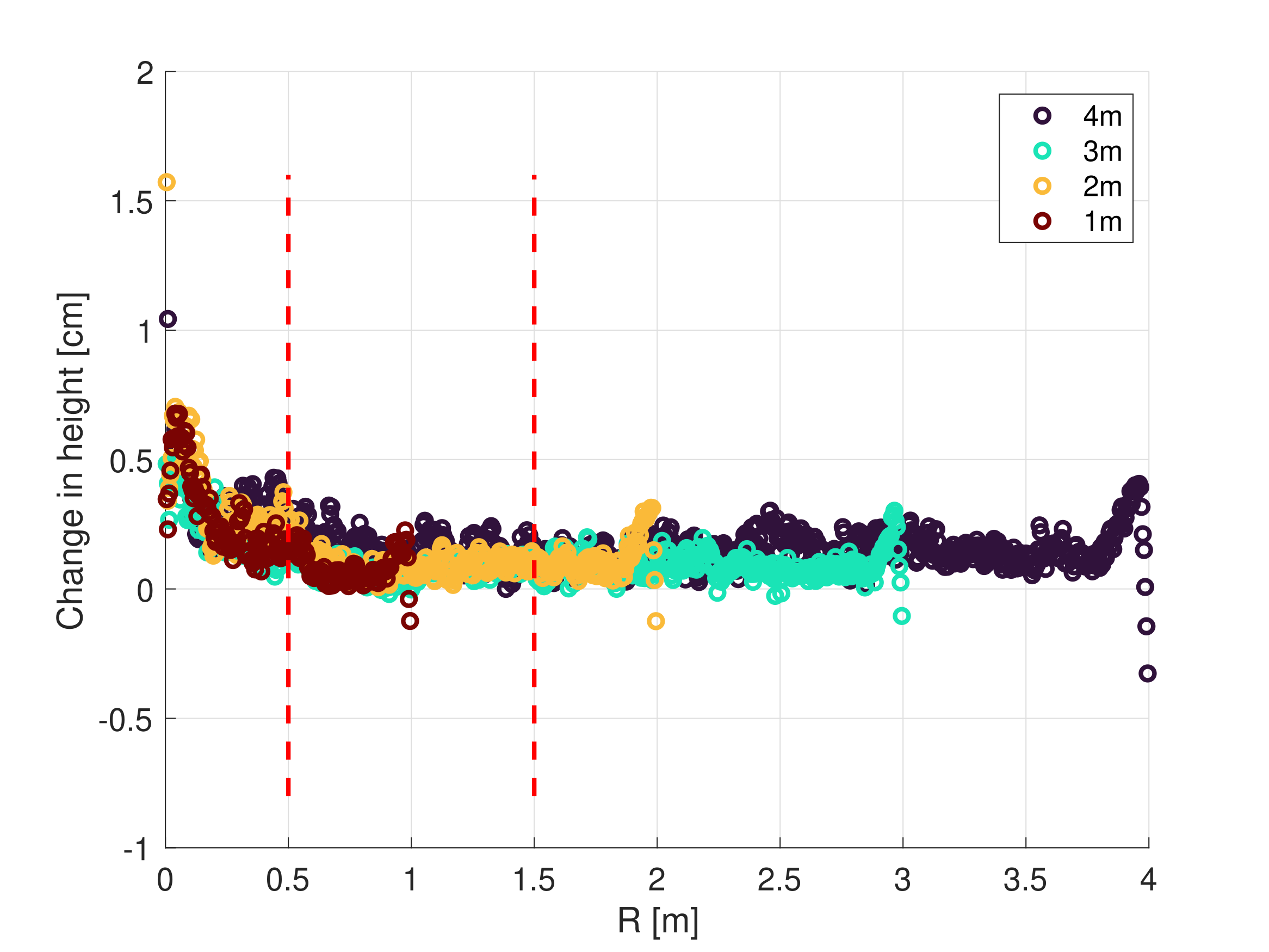}
\end{center}
\caption{\textbf{$\Delta z$ vs radial position for varied channel lengths.} Change in bed height is tracked in different beds filled 20 $cm$ deep with the loose configuration of particles. A uniform, though small, amount of dilation occurs across all channels. Color legend corresponds to channel length. }
\label{fig:channel_length_effects}
\end{figure}

To test the effect of channel length, we prepared channels of varied lengths (1 to 4 $m$) at the same loose packing to reduce computation time. Each bed is subjected to a single shock at $v_p = 10$ $m/s$. The resultant solitary wave induces uniform dilation across the channel that is approximately the same regardless of the length of the assembly (Fig. \ref{fig:channel_length_effects}). Note that the value for $\Delta\rho$ in the 4 $m$ channel differs from the 2 and 3 $m$ channels by more than the RSE (Table \ref{tab:length_heights}). We attribute this to the slightly larger $c_p$ attained in the 4 $m$ channel. Given that a wall can influence results for up to half a meter (at the start or end of the channel), the 1 $m$ channel is not long enough to include an uncorrupted measurement region. There is therefore a minimum required channel length, but otherwise we find that the length of the assembly does not have an effect on our results, which agrees with \cite{mouraille2006sound}. 

\begin{figure}[h]
\begin{center}
\includegraphics[scale = .4]{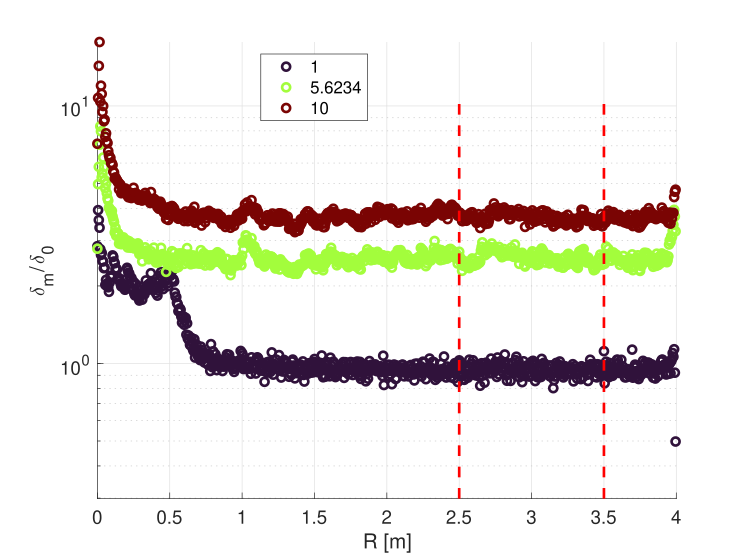}
\end{center}
\caption{\textbf{Average $\delta_m$ vs R.} We show normalized $\delta_m$ across the 4 $m$ channel for three different $v_p$ cases. The solitary waves generated with the two largest $v_p$ sustain a roughly constant $\delta_m$ over the entire length.}
\label{fig:FvsR_4m}
\end{figure}

To evaluate how the solitary waves evolve in the longer channel, we show the wave profile in Figure \ref{fig:FvsR_4m}. Figure \ref{fig:FvsR_4m} shows the normalized overlap experienced along the length of the channel. The strongest solitary wave which decayed from a shock initiated at $v_p$ persisted across the entire channel. The solitary wave generated using $v_p = 1$ $m/s$ terminated earlier (around 1 $m$) than the same $v_p$ generated wave in the compact case (Fig. \ref{fig:velsweep_overlap}, around 1.5 m). This might suggest that a more densely packed assembly can maintain the same strength wave over a longer distance. We show the dilation resulting from these three tests in Table \ref{tab:length_heights} and we see similar behavior (both dilation and compaction) as the loose bed in the prior section.

\begin{table}[h]
\begin{center}
\begin{minipage}{174pt}
\caption{$\Delta z$ in beds of various length. $c_p$ is the solitary wave speed and $\Delta\rho$ is computed as a percent difference from a bed initially filled 20 $cm$ deep with particles. }\label{tab:length_heights}
\begin{tabular}{@{}ccccc@{}}
\toprule
\begin{tabular}{c} Length \\  ($m$) \end{tabular} & \begin{tabular}{c} $v_p$ \\ ($m/s$) \end{tabular} & \begin{tabular}{c} $c_p$ ($m/s$) \\ $\mu \pm \sigma$ \end{tabular}  &  \begin{tabular}{c} $\Delta \rho$ \\ ($\%$) \end{tabular}  \\
\midrule
2 & $10.0$   & $8.29 \pm 0.55$   & 0.482 \\
3 & $10.0$   & $8.20 \pm 0.36$ & 0.379 \\
4 & $10.0$  &  $8.36 \pm 0.38$ & 0.771  \\
\hline
4 & $5.62$   & $7.41 \pm 0.52$ & -0.033   \\
4 & $1.0$   & $6.66 \pm 0.73$  & 0.001    \\
\bottomrule
\end{tabular}
\end{minipage}
\end{center}
\end{table}

\subsubsection{Boundary effects: channel depth} \label{sec:result_height}
\begin{figure}[h]
\begin{center}
\includegraphics[scale = .4]{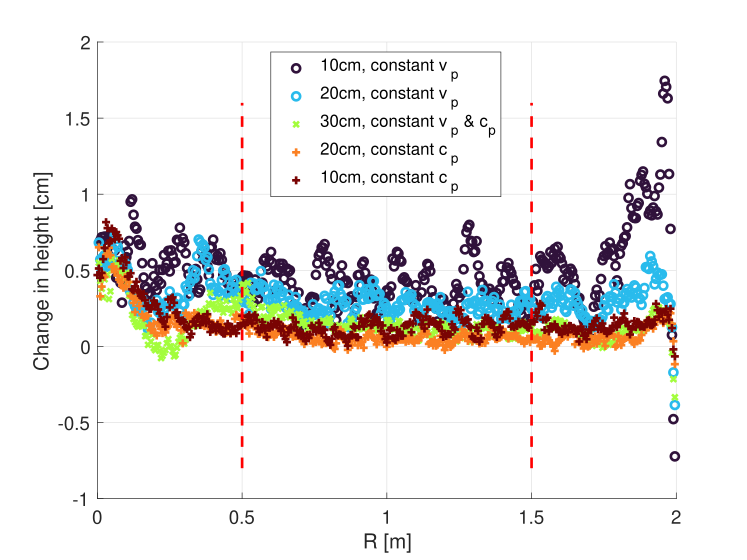}
\end{center}
\caption{\textbf{$\Delta z$ vs radial position for varied fill height.} There is a slight increase in the post impulse height change with decreasing bed height (2 $m$ long beds filled at the loose packing configuration). }
\label{fig:dilation_height_effects}
\end{figure}

With a 2 $m$ length channel we filled assemblies to depths of 10, 20, and 30 $cm$ (`fill height') to investigate any height effects on the surface dilation. We again use the loose packing to reduce preparation time. There is a slight increase in height change with a decrease in bed height (Fig. \ref{fig:dilation_height_effects}). 

\begin{figure}[h]
\begin{center}
\includegraphics[scale = .1]{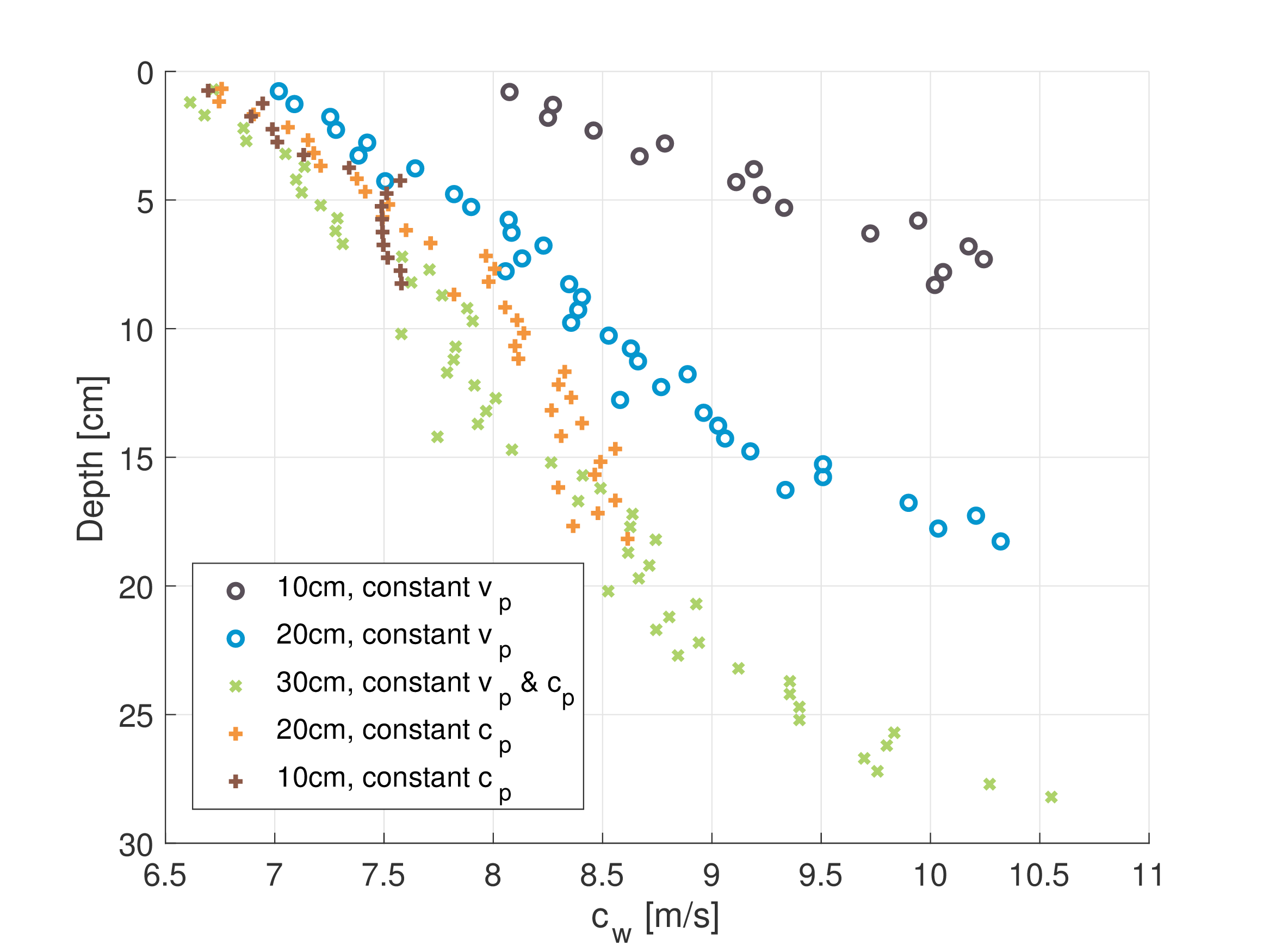}
\end{center}
\caption{\textbf{Depth vs $c_p$.} While the initially induced $c_w$ were the same, $c_p$ is greatest for the smallest channel fill height. Constant $v_p$ cases are marked with circles and constant $c_p$ with a plus sign (the case for the 30 cm channel marked with an x marker is both constant $v_p$ and $c_p$). }
\label{fig:height_effects_steady_wavespeed}
\end{figure}

The change in bed height dependence on channel depth is a result of an increased propagation speed in the bed as the fill height decreases, as was also observed in \cite{somfai2005elastic}. In our simulations we see that the maximum speed in the solitary wave is essentially the $v_p$ used to initiate the wave. This is evident in Figure \ref{fig:height_effects_steady_wavespeed} where we show how $c_p$ varies with depth. We first initiated shocks within each of the three channels at $v_p = 10$ $m/s$ (constant $v_p$) and $c_p$ is just over $10$ $m/s$ at the floor of the assembly in each of the constant $v_p$ cases. As a result of the wave speed dependence on $\delta$ and the $\delta$ dependence on the overburden force, the speed of the generated wave must be greatest at the bottom floor of the channel. Surface particles in a taller assembly therefore will experience reduced speeds as compared to a shorter assembly since, for the same floor speed, they will undergo a greater reduction in speed due a larger gravity gradient. 

\begin{figure}[h]
\begin{center}
\includegraphics[scale = .1]{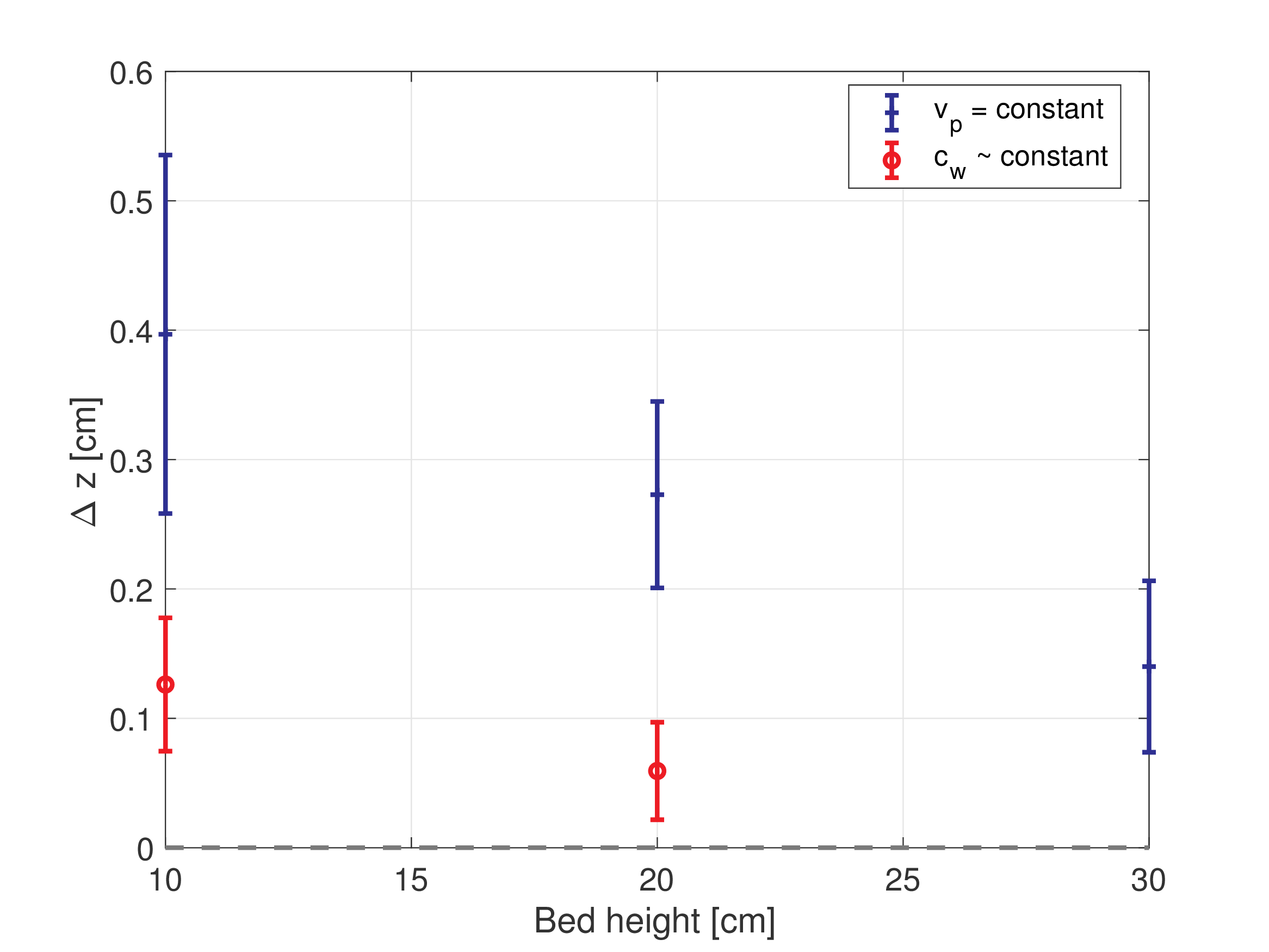}
\end{center}
\caption{\textbf{$\Delta z$ vs channel height.} Shows cases for constant $v_p$ and for roughly constant propagating $c_w$ . $\Delta z$ is given instead of $\Delta\rho$ since the different initial bed heights influences the percent difference calculation. The single data point at 30 cm can be considered as both constant $v_p$ and constant $c_p$.}
\label{fig:dz_vs_channel_height}
\end{figure}

To verify that it is the wave speed change that leads to greater bed height change (as opposed to some geometrical effect of a shorter channel) we reduced the input $v_p$ in the 10 and 20 $cm$ deep channels. Reducing $v_p$ generates solitary waves with speeds in the wavefront that are roughly the same as those in 30 $cm$ channel with $v_p = 10$ $m/s$. The input $v_p$ and resultant $c_p$ are given in table \ref{tab:propagating_cw_heights} and compared to the constant $v_p$ height changes in Figure \ref{fig:dz_vs_channel_height}.  The height change for the constant $v_p$ cases appears to be linearly decreasing as bed height increases while the height change for the constant $c_w$ is much closer to constant.

\begin{table}[h]
\begin{center}
\begin{minipage}{174pt}
\caption{$c_p$ and $v_p$ in beds of various height. We show the input $v_p$ used to generate solitary waves of different $c_p$. }\label{tab:propagating_cw_heights}
\begin{tabular}{@{}cccc@{}}
\toprule
\begin{tabular}{c} Height \\  ($cm$) \end{tabular} & \begin{tabular}{c} $v_p$ \\ ($m/s$) \end{tabular} & \begin{tabular}{c} $c_p$ ($m/s$) \\ $\mu \pm \sigma$ \end{tabular}  \\
\midrule
10 & $10.0$   & $9.54 \pm 0.55$    \\
20 & $10.0$   & $7.72 \pm 0.36$ \\
30 & $10.0$  &  $7.01 \pm 0.30$  \\
\hline
20 & $8.8$   & $7.32 \pm 0.36$    \\
10 & $7.5$   & $7.09 \pm 0.28$    \\
\bottomrule
\end{tabular}
\end{minipage}
\end{center}
\end{table}

\subsubsection{Initial packing effects} \label{sec:packing_effects}
\begin{figure}[h]
\begin{center}
\includegraphics[scale = .4]{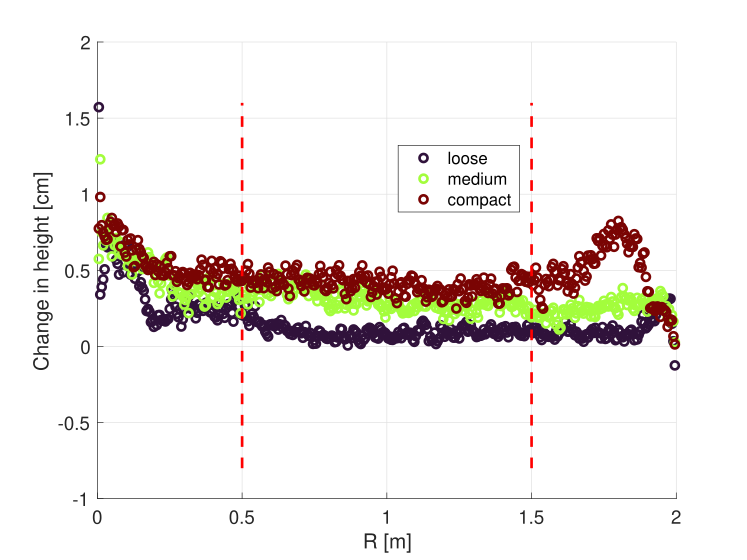}
\end{center}
\caption{\textbf{$\Delta z$ vs radial position for different $\phi_0$.} We examine the dilation sensitivity to $\phi_0$ by initiating shocks in beds (2 $m$ in length filled 20 $cm$ deep) corresponding to the loose, medium, and compact $\phi_0$ from Table \ref{tab:fill_param}. In each case, the height change is uniform across the bed, besides the initiation region and the end wall region. The heap near the end wall is the result of particles ejected as the wave reflects off the end wall. Red dashed lines indicate the measurement region.}
\label{fig:dilation_initial_packing}
\end{figure}

\begin{figure}[h]
\begin{center}
\includegraphics[scale = .4]{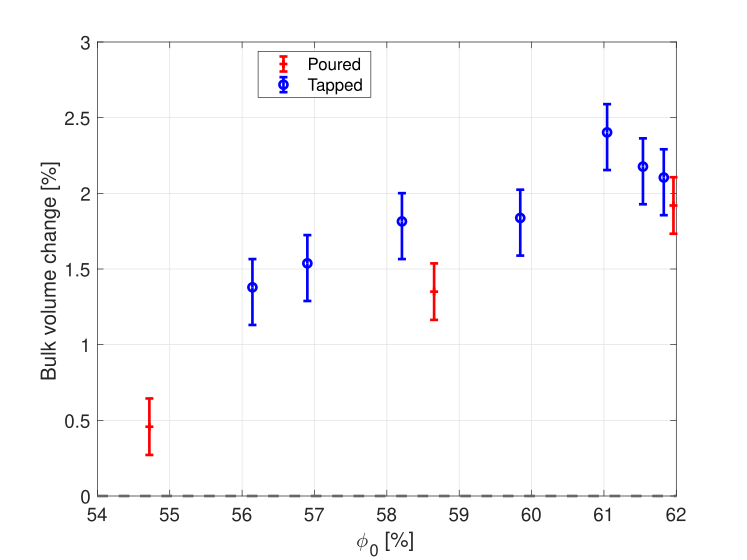}
\end{center}
\caption{\textbf{Bulk density change vs $\phi_0$.} Percentage change in volume fraction is computed as an average of height changes the measurement region from Figure \ref{fig:dilation_initial_packing} assuming the entire vertical extent of the bed  has undergone dilation. Since the dilation band is a fraction of the total bed height, these values are lower bounds. Results are included for both poured and tapped beds.}
\label{fig:initial_packing_effects_comp}
\end{figure}

We excite shocks at $v_p = 10$ $m/s$ in the loose, medium, and compact beds to evaluate the influence of $\phi_0$ on impulse-induced bed height change. Similar to the work from \cite{royer2011role}, we expect that granular dilation will be the most pronounced for the initially compacted bed and decreases as the packing becomes looser. Figure \ref{fig:dilation_initial_packing} confirms this trend. Quantifying the percent change in height, Figure \ref{fig:initial_packing_effects_comp} shows that dilation occurs even for the loosest bed and considering the random seeding error. The crossover initial state between a compaction and dilation result is $\phi_0 < $ 0.55, lower than the 0.58 crossover seen in \cite{royer2011role}.  The reduced threshold is attributable to the reduction in gravity and increase in cohesion of our assembly. We also tested whether preparation history had an effect on resultant dilation by tapping the loosest bed to the $\phi_0$ reported in Table \ref{tab:tapping_durations} and then initiating shocks each bed (see sec. \ref{sec:Tapping}). Figure \ref{fig:initial_packing_effects_comp} includes the results of the poured and tapped beds. Similar results are produced regardless of the bed preparation method, as we see the same linear trend in the tapped beds as the poured beds. The slight increase in the percent dilation for the tapped bed over the poured beds is the result of the reduced bed height (15 $cm$ deep tapped beds, see sec. \ref{sec:Tapping}) as determined in section \ref{sec:result_height}. The method of bed preparation does not influence the dilation response beyond the randomness of the packing.

\begin{figure*}[h]
\begin{center}
\includegraphics[scale = .85]{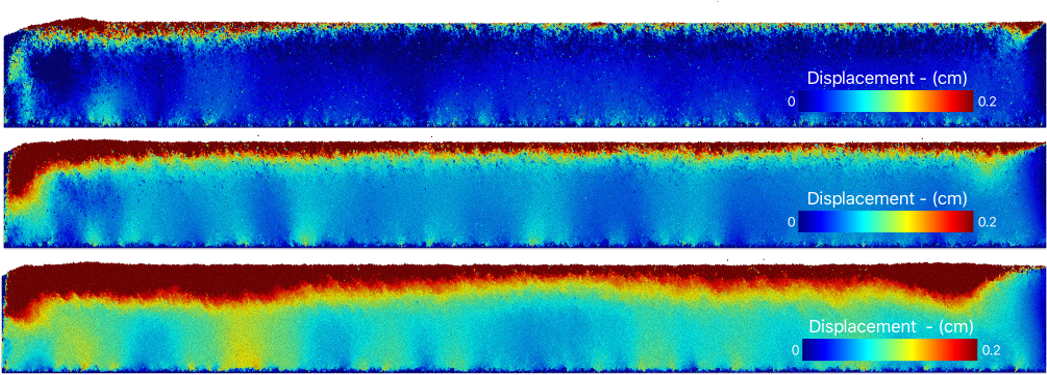}
\end{center}
\caption{\textbf{Particle height change.} Each particle is color coded to show it's change in height. We show results for loose (top), medium (middle), and compact (bottom) packing of 2 $m$ beds filled 20 $cm$ deep with particles after a shock was initiated at $v_p = 10$ $m/s$. The extent of dilation increases with $\phi_0$ and, in the case of the compact bed, a consistent portion of the near surface (to about 5 $cm$ in depth) experienced more than 0.2 $cm$ increase in height.}
\label{fig:height_dilation_bed}
\end{figure*}

Figure \ref{fig:height_dilation_bed} shows a visualization of the resultant impulse induced dilation for the three poured $\phi_0$. This visualization confirms that our data analysis method accurately captures surface dilation and it shows the depth dependence of particle mobility. The band in which particles expand upwards increases in lateral extent as $\phi_0$ increases. For the compact case, the particles are most mobile in depths down to about one third of the total channel height ($\sim 8$ $cm$), although the majority of the channel has experienced upward particle motion (dilation) as the result of a solitary wave.

\subsubsection{Error analysis}
\begin{figure}[h]
\begin{center}
\includegraphics[scale = .4]{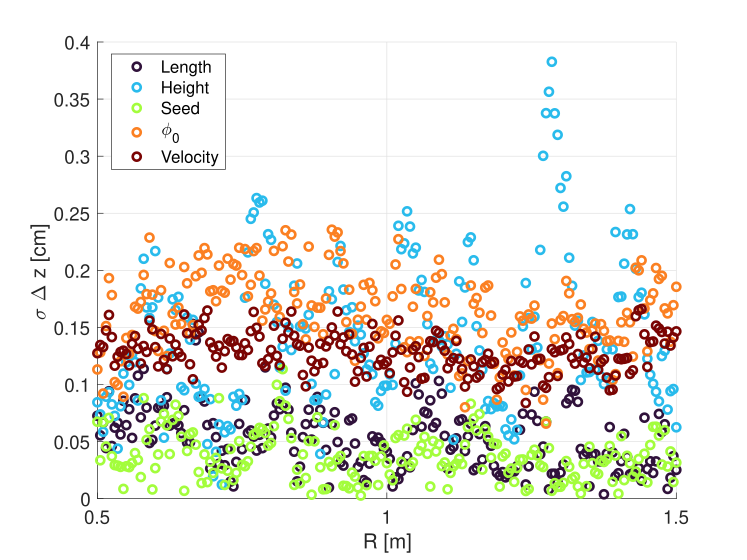}
\end{center}
\caption{\textbf{$\sigma \Delta z$ vs radial position.} Deviation is computed for the radial positions in common across the parameters investigated. The legend shows the test from which the deviation is computed.}
\label{fig:variance_height}
\end{figure}

\begin{figure}[h]
\begin{center}
\includegraphics[scale = .4]{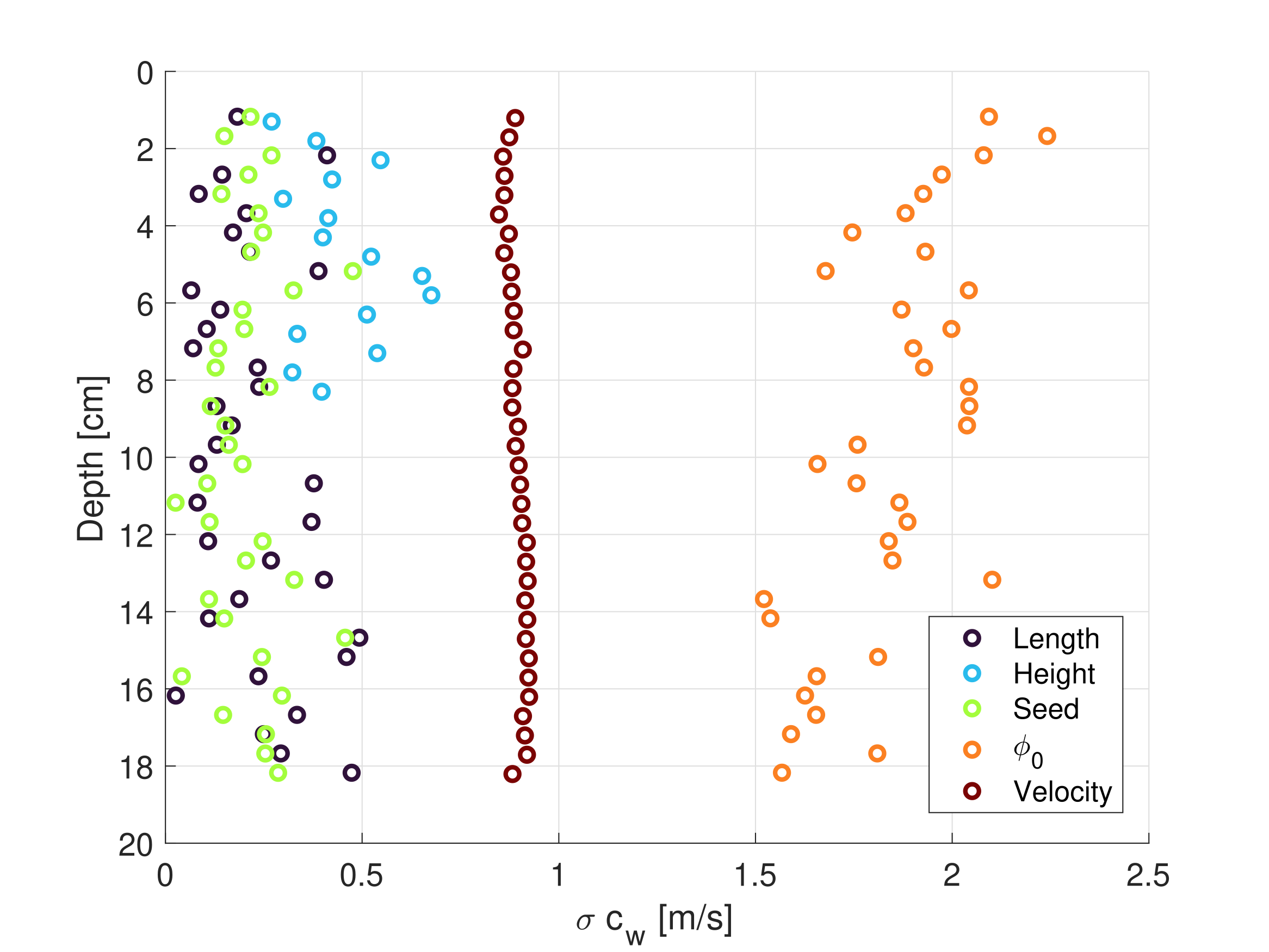}
\end{center}
\caption{\textbf{Depth vs $\sigma c_w$.} Deviation is computed for the depths in common across the parameters investigated. The legend shows the test from which the deviation is computed.}
\label{fig:variance_wavespeed}
\end{figure}

To compare results across the different parameter spaces, we find the deviation in wave speed and dilation response for each case evaluated. Since deviation of the height change (Fig. \ref{fig:variance_height}) is taken from the measurement region (0.5 $m$ away from the end walls) the 1 $m$ channel is neglected in computing the deviation for the channel length cases. The deviation in the wave speed (Fig. \ref{fig:variance_wavespeed}) is taken from depths in common across the cases and the range is [0,20] $cm$ for all cases except the height analysis, which only has depths up to 10 $cm$ in common across the three heights investigated.

In both the case of dilation and wave speed, the deviation across the beds of different lengths is equivalent to the seed deviation. The length of the channel does not influence our results any more than the randomness of the packing. Velocity, bed height, and $\phi_0$ all influence the magnitude of dilation that occurs. Wave speed is also affected by the piston velocity and packing fraction, but bed height does not impact the initiated $c_w$ more than the random packing (though it does affect $c_p$, see sec. \ref{sec:result_height}).

\subsection{Effect of wave initiation region}  \label{sec:initiationRegion}
We varied the wave initiation region within the 2 $m$ long 20 $cm$ deep compact bed to provide further bounds on when dilation can occur. We modified the P wave initiation process (sec. \ref{sec:wavegen}) in three ways by reducing the region of particles that make up the virtual piston. For the first test (thin strip), the piston still spanned the height of the channel, but only particles within one particle diameter of the origin (x dimension) were given a y velocity. This test evaluates how surface dilation is influenced by greater initial dispersion. For the second and third test, the piston spanned the width of the channel, but we modified the z dimension of particles included. Only particles within 10 $cm$ of the surface are included for the `top half' test and particles within 10 $cm$ of the floor are included for the `bottom half' test. This test allows us to evaluate how waves generated at an angle or reflected floor waves might influence surface dilation. 

\begin{figure*}[h]
\begin{center}
\includegraphics[scale = .85]{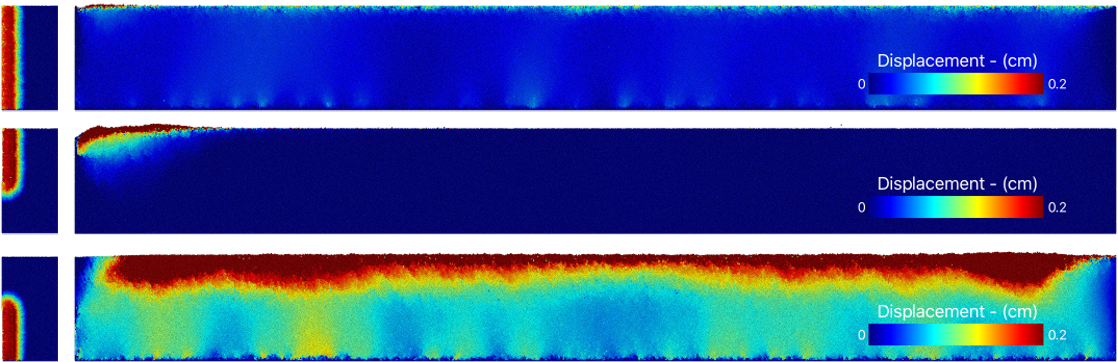}
\end{center}
\caption{\textbf{Shock profile and particle height change for various test cases.} Each frame shows the profile of the initiated shock region and the resulting bed height change. The left images show the particle force (same legend as in Fig. \ref{fig:channelcartoon}) and the right image colors particles by their z displacement with the embedded color legend in units of $m/s$. The channels correspond to the thin strip (top), top half (middle), bottom half (bottom) test cases.}
\label{fig:try_test_results}
\end{figure*}

Figure \ref{fig:try_test_results} shows that for the thin strip case a solitary wave still forms and is able to propagate down the entire channel and affect uniform dilation, albeit at a reduced level. In the top half case there is almost no surface modification, though the bottom half case still produces dilation similar to the full plane piston. Additionally, we found that the results of the `top half' test were the same as when the full plane piston was used, but no particles were fixed to the floor of the assembly. Therefore, the region of initiation must include a stronger floor for the force in the shock to be radially sustained.

\begin{figure*}[h]
\begin{center}
\includegraphics[scale = .7]{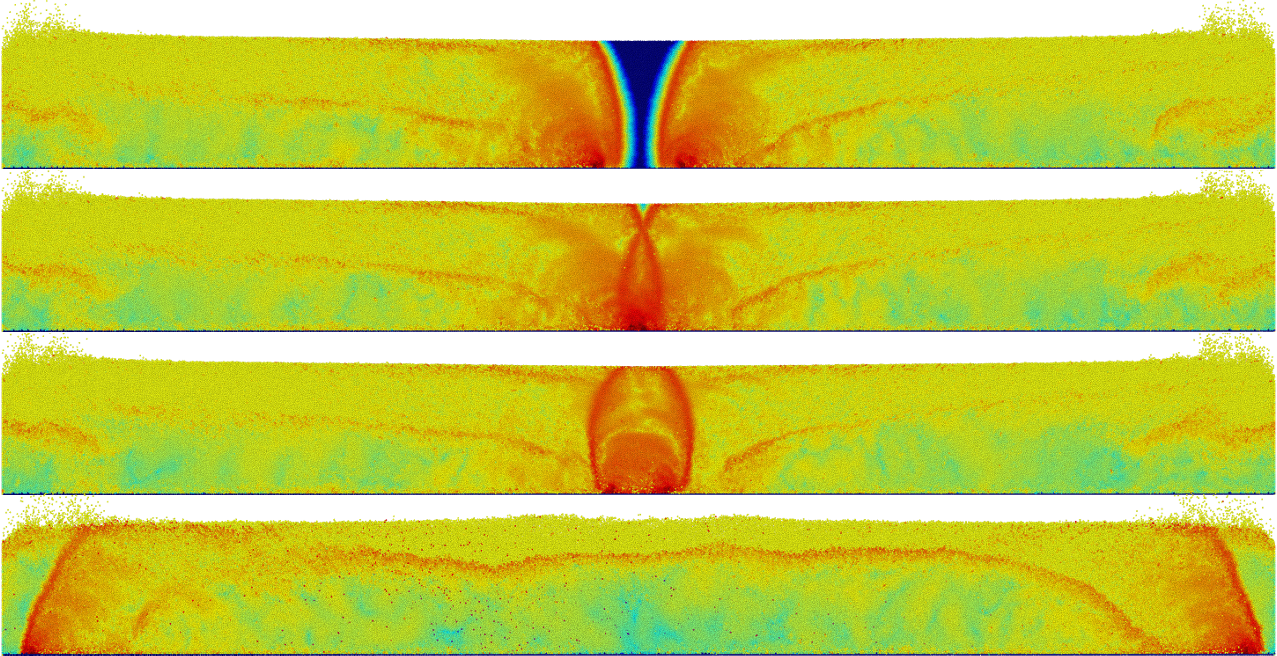}
\end{center}
\caption{\textbf{Interacting waves.} Solitary waves with equal strength ($v_p =$ $10$ $m/s$) are generated in opposite directions from each end of a 2 $m$ long channel filled $20$ $cm$ deep with particles ($\phi_0 =$ compact). From top to bottom the frames correspond to $t = $ 0.093, 0.099, 0.104 and 0.195 s. Particles are colored corresponding to the order of magnitude of force experienced (same scale as in Fig. \ref{fig:channelcartoon}). The strength of each wave is maintained during and after the collision. The shape of the wavefront is largely maintained, though there is less curvature to the wavefront near the floor of the channel as the waves approach their terminal points (bottom frame). }
\label{fig:collision}
\end{figure*}

Finally, we generated a shock at each end of a channel to see how the resultant solitary waves interact with each other. Figure \ref{fig:collision} shows several frames from this interaction performed in the 2 $m$ long channel filled 20 $cm$ deep ($\phi_0$ = compact) and a piston velocity $v_p = $ 10 $m/s$. Qualitatively, both the strength and shape of the wave are preserved through the collision, the hallmark of a soliton (\cite{sen1998solitonlike}, \cite{nesterenko2013dynamics} - chapter 1.6.4). 

\subsection{Discussion of solitary waves}  
In our simulations, we have been able to easily create solitary waves as the decay product of an initiated shock. A single, solitary wave front emerges from a lateral impulse in which a constant $\delta_m$ compression (Figs. \ref{fig:velsweep_overlap}, \ref{fig:FvsR_4m}) is maintained. The shape of the wave front and its spatial size (width) are preserved (see the wavefront in Fig. \ref{fig:channelcartoon}, top frames from Figs. \ref{fig:shock_vis_Pwave_combo} and \ref{fig:collision}, Fig. \ref{fig:shockframes} frame d, or video OL1). The elastic limit for impact velocities (0.1 $m/s$) is only appreciably exceeded (less than doubled) by particles in the wave front of the strongest solitary waves ($v_m$ for $v_p = 10$ $m/s$ in Table \ref{tab:speed_waveDetails}) which supports impulse-induced dilation as a physical result. The spatial size of the wave front is much smaller than the distance it travels and we saw the strongest solitary waves to be sustained over the entire 4 $m$ channel (Fig. \ref{fig:FvsR_4m}), only terminating upon contact with the end wall. Here we have neither quantified nor predicted the termination distance of the solitary waves we generate, understanding the criteria for lateral solitary wave generation and sustainment in a 3D randomly packed granular assembly are important topics for future investigations.

Another subject which deserves further study is the soliton-like behavior of the solitary waves we have generated. In addition to maintaining strength and shape and propagating for much longer distances than the wavefront is wide, our solitary waves also qualitatively preserve shape and strength through a collision (Fig.  \ref{fig:collision}). We briefly evaluate our numerically obtained solitary wave speeds against expected soliton speed. Equation 1.43 from \cite{nesterenko2013dynamics} (equation \ref{eq:solitonspeed}) gives the soliton speed (in a 1D chain) based on particle velocity in the wavefront ($v_m$, Table \ref{tab:speed_waveDetails}). Equation \ref{eq:solitonspeed} is valid in the sonic vacuum regime which is the case when the granular assembly is weakly compressed. An assembly can be considered weakly compressed when the maximum overlap in the wave front is greater than (not on the same order) as the initial overlap, $\delta_m /  \delta_0 \gtrsim 2$ (\cite{nesterenko2013dynamics} ch. 1.3). 

\begin{align}
     V_s = \left(\frac{16}{25}\right)^{1/5}\left(  \frac{2E}{\pi \rho_p(1-\nu^2)} \right)^{2/5}v_m^{1/5}  \label{eq:solitonspeed}
\end{align}

\begin{figure}[h]
\begin{center}
\includegraphics[scale = .4]{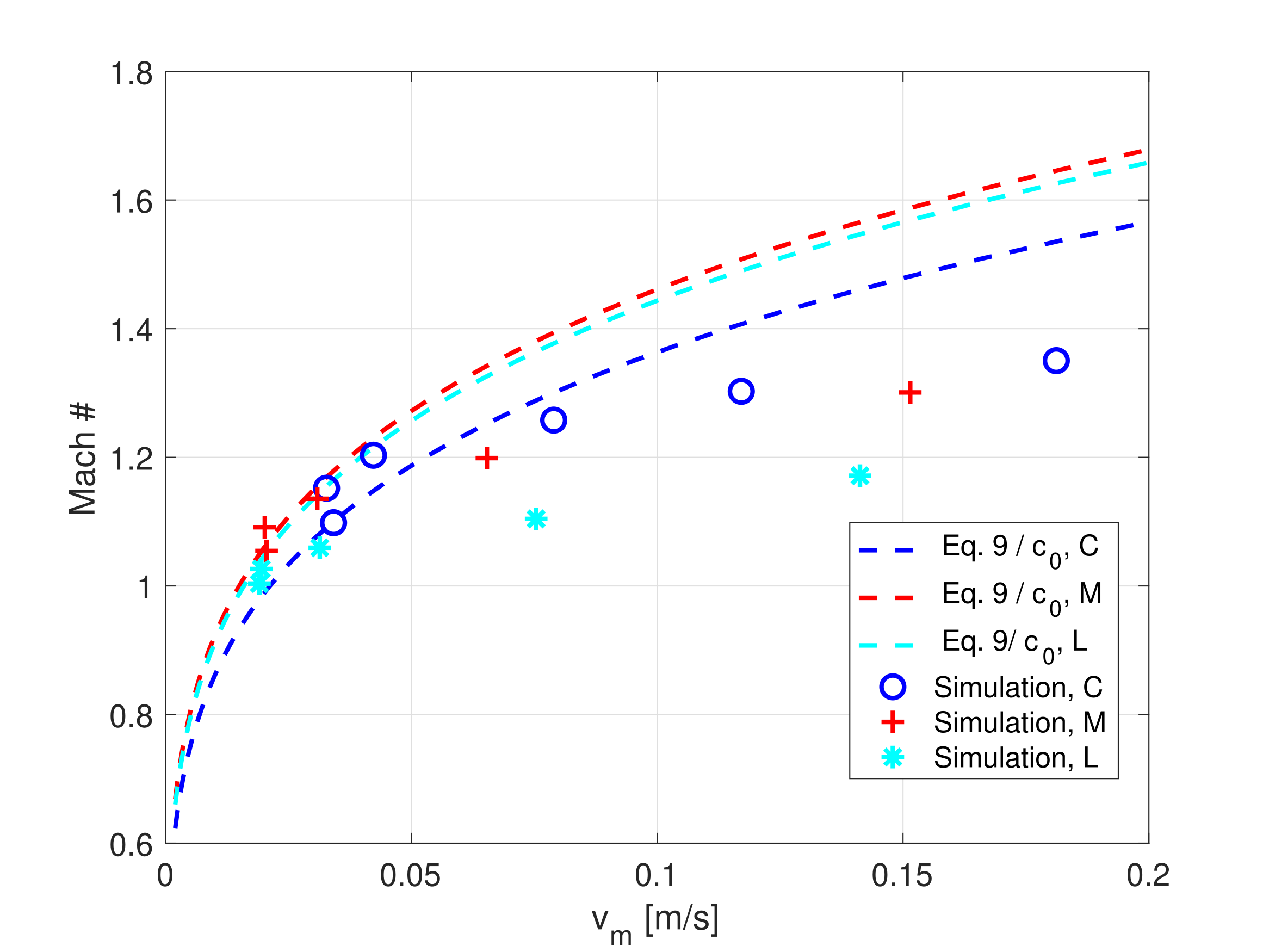}
\end{center}
\caption{\textbf{Mach \# vs $v_m$.} Equation \ref{eq:solitonspeed} is divided by $c_0$ corresponding to the loose (L), medium (M) and compact (C) packed assemblies to produce the predicted soliton speed (dashed lines) for a range of $v_m$. For the corresponding simulated results (various markers) we find $M = c_p/c_0$ and plot against $v_m$ from table \ref{tab:speed_waveDetails}. Simulated results are only plotted for tests at $v_p \geq 1$.}
\label{fig:machVSdelta}
\end{figure}

Our strongest solitary waves fall into the weakly compressed regime (Table \ref{tab:speed_waveDetails}), though we include points from any test that had an initial shock decay to a solitary wave in this analysis. We divide equation \ref{eq:solitonspeed} by $c_0$ of each bed (Table \ref{tab:soundspeed_vsphi}) to create Figure \ref{fig:machVSdelta} which shows the analytically and numerically calculated Mach number vs $v_m$. Points are plotted only for waves that were initiated as shocks. All three cases show a similar trend (faster initial increase in Mach at lower $v_m$), though the loose bed shows reduced Mach number compared to the medium and compact. We believe this is due to some inaccuracies in the wave speed calculation in the loose bed (discussed in sec. \ref{sec:wave_speed_effect}). Simulated results from the medium and compact cases agree with the soliton speed at lower velocities, but diverge as $v_m$ increases.  While our simulated solitary waves do not follow the trend exactly, they are in the range of what is expected. Note that eq. \ref{eq:solitonspeed} is derived for 1D particles and does not include any modifications to reflect the material properties implemented in our model (friction, cohesion) nor geometrical effects of the 3D packing. Also recall that our numerical results are averaged over particles within a virtual cell which is 0.5 $cm$ in width which may obscure the trend. We have a grid resolution of about 8 sensors in the wave front ($\sim$15 particles, $\sim$4 $cm$), but augmenting the results with single designated particle sensors would help to refine the results. Predicting the behavior of laterally propagating solitary waves at the surface of a vacuum interface should be a focus of future work.

\subsection{Dilation and scaling discussion}
We've shown that in the absence of atmospheric confinement pressure and under low gravity, a band of grains at the surface of a granular assembly can experience inertial dilation as the result of a laterally propagating solitary wave. First, a granular assembly must be in an initially compact state for a lateral impulse to induce dilation. Surface dilation is then triggered for solitary waves that are barely supersonic ($M > 1.05$, Figure \ref{fig:dz_vs_mach}) and can be sustained over large distances (up to 4 $m$, so far) as long as the wave is active. The extent of dilation is primarily controlled by $c_p$ since the increases in dilation observed as a result of increasing $\phi_0$ or decreasing bed height were fundamentally a result of increased solitary wave speed. This new particle lofting mechanism is capable of affecting surface bulk dilation in our simulations, but how does it scale to other environments?

We do not yet have a method for predicting the final resting height of particles (and thus, dilation) based on the solitary wave speed, but we can roughly predict the depth to which particles will loft. Our simulations show that a sheet of particles became completely detached following the passage of the solitary wave front. There is an easily discernible cutoff depth for this detached sheet at around 8 $cm$ in depth visible in Figure \ref{fig:shock_vis_Pwave_combo} corresponding to the demarcation between particles experiencing large and then zero normal and shear stresses. For the particles to be in a lofted state, they must have acquired some upward velocity, so any particle down to that cutoff depth will have experienced an upward force. The depth dependence on lofting suggests that an imbalance between the forces in the wave front and the overburden is the source of the upward velocity.  We define $\theta$ as the angle of the wave front which determines the amount of force experienced by a particle and is redirected in the z direction. $\theta$ describes the angle between horizontal (y) and the vector normal to the wave front. This vector is normal to the curvature in the force profile of the wave front and the curvature can be seen in Figs. \ref{fig:channelcartoon}, top frame of \ref{fig:shock_vis_Pwave_combo} and frame d of \ref{fig:shockframes}. Near the surface, the direction normal to the shock front ($\theta$) is greater than 0, while at depth theta is approximately 0 (an x-plane wave). In our simulation, the lofting depth is roughly equivalent to the depth at which the wave front becomes planar so we assume that the cutoff angle for lofting occurs when $\theta$ is small (we assume  $1^{\circ}$). The wave front shape (and thus $\theta$) can be derived analytically, but we leave that as an exercise for our future work. We then use the constant $\theta$ to find the gravitational forces (eq. \ref{eq:fgrav}) in the vertical direction and balance them with the wavefront forces (eq. \ref{eq:fwf}): $F_{grav} = sin(\theta) F_{WF}$. 

\begin{equation}
	F_{grav} = \left(\phi(\pi R^2)z +  \frac{4}{3} \pi R^3\right) \rho_p g   \label{eq:fgrav}
\end{equation}

\begin{equation}
	F_{WF} = \mid - k_n (\delta_m-\delta_0)^{3/2} - \gamma_nv_n + k_cA_c  \mid \label{eq:fwf}
\end{equation}

Solving the balance for depth then gives the loft depth, $z_{loft}$, in equation \ref{eq:loft_solve}:

\begin{align}
      z_{loft}  =  \frac{sin(\theta) F_{WF} - \frac{4}{3}\pi R^3 \rho_p g}{\phi \pi \rho_p g R^2}   \label{eq:loft_solve}
\end{align}

\begin{figure}[h]
\begin{center}
\includegraphics[scale = .4]{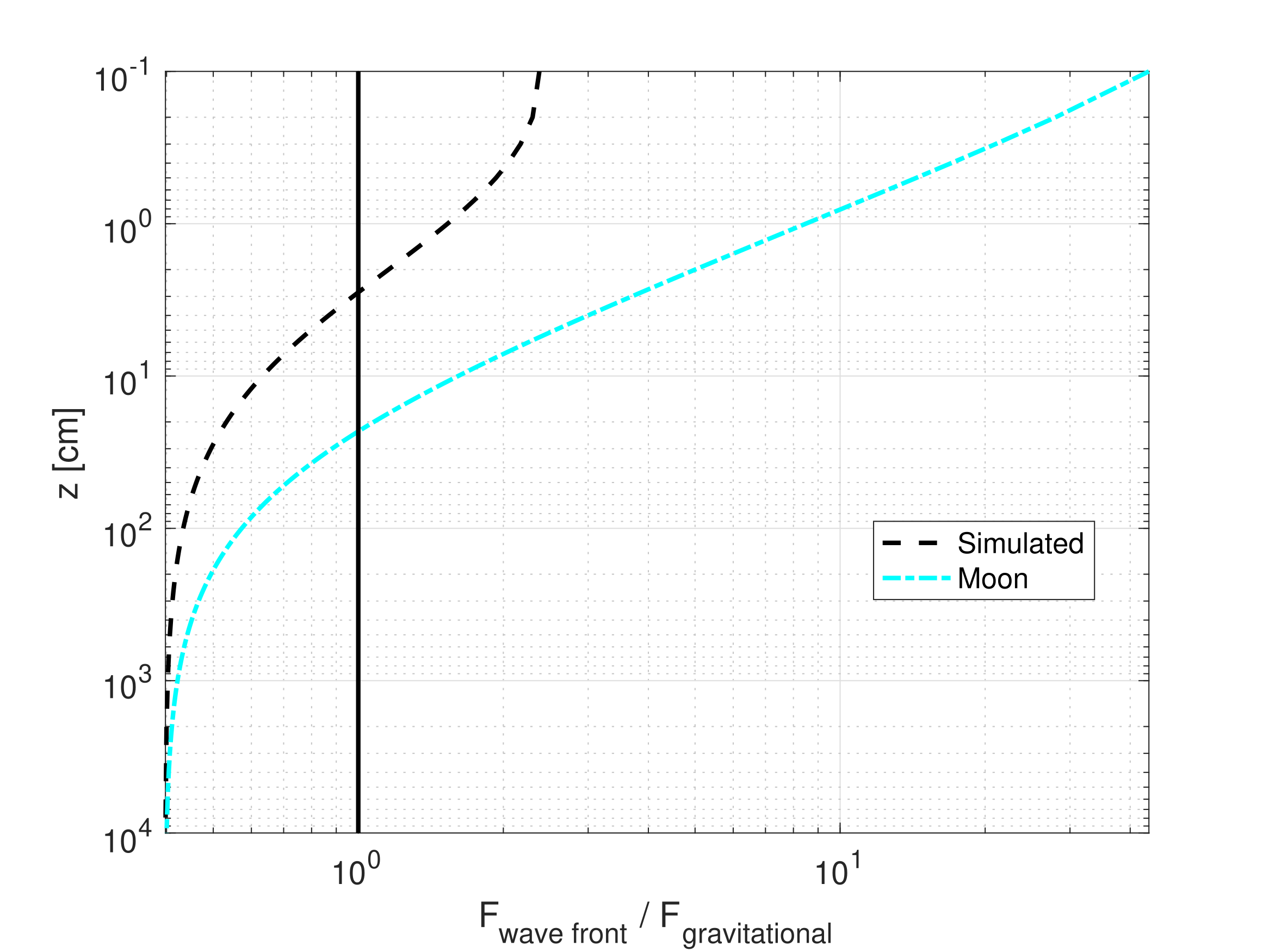}
\end{center}
\caption{\textbf{$z$ vs force balance.} The lofting depth $z_{loft}$ occurs when $F_{gravitational}$ and $F_{wave front}$ are equal (denoted by the solid black line). The simulated and Lunar trends are produced using the material parameters described in the text along with eq. \ref{eq:loft_solve}. When the fraction is greater than 1, wave front forces in the vertical direction exceed gravitational forces and particles will loft above the depth $z_{loft} = z(F_{wf} / F_g =1)$. }
\label{fig:zVsFraction}
\end{figure}

Solving for the value of depth that satisfies this equality yields the lofting depth, $z_{loft}$. The spring force and cohesion depend on the strength of the solitary wave ($\delta_m$) and the damping component depends on impact velocity, which is the maximum velocity of particles in the wave front ($v_m$). Note that this formulation neglects the effect of friction. We make a few additional simplifying assumptions to use this equation. To solve for the damping force (eq. \ref{eq:normal_damp}), we'll use a constant $v_m = 0.1$ $m/s$ (the static limit) which is in the range of what is experienced in our simulations, but really is a function of the wave front strength ($\delta_m$). Also note that this formulation is based on the particle-particle material properties, not bulk properties (particle properties can be orders of magnitude larger than bulk - \cite{zhang2018rotational}).  We show the procedure to solve for $z_{loft}$ in Figure \ref{fig:zVsFraction}, with $\delta_m / \delta_0 = 4$ (maximum experienced by our solitary waves). The ratio of the maximum to initial overlap can be though of as describing the strength of the wave, though keep in mind that it is not the same as Mach number.

The depth at which the force balance (x axis) equals unity is the lofting depth, which for the case pictured resulted in $z_{loft} \sim 3$ $cm$ for our simulation.  While we expect a larger lofting depth, $3$ $cm$ is on the order of the $8$ $cm$ we observed in Fig. \ref{fig:shock_vis_Pwave_combo}. Given the numerous assumptions made while constructing equation \ref{eq:loft_solve}, it is encouraging that the results are within a factor of three. Using equation \ref{eq:loft_solve} to consider how the loft depth would change in different conditions also makes intuitive sense. The loft depth will increase for denser particles (increased wave speed) and reduced gravity (decrease of $\rho_p$ and $g$) and increase for harder particles (increase in $E$, which is embedded in $k_n$) while the dependence on particle size ($R$) is more complicated. Atmospheric pressure would be an additional term on the gravitational side of the equation and would decrease $z_{loft}$.

\subsection{Relevance to LCS}
As an initial assessment of lateral impulse-induced dilation's candidacy as the LCS formation mechanism, the results we have seen here are encouraging. The solitary waves that induce dilation can be minimally disruptive which is important since the LCS show no visible signs of surface modification. The solitary waves inducing bulk dilation in our simulations occur at low speeds ($M\gtrsim1.05$) and have wave fronts in which the maximum compression isn't much more than 1\% of the particle radius (1.4\% for our strongest solitary wave with $\delta_m = 16.94$ $\mu m$, see Table \ref{tab:speed_waveDetails}). The solitary waves inducing the largest dilation response were also able to travel long distances and were only constrained by the size of our channel, although the $4$ $m$ propagation distance we observed is much less than is needed for LCS formation ($\sim100$ km). To understand if a laterally propagating solitary wave could be the LCS formation mechanism, more work must be done to predict the strength of the initiated solitary wave and its termination distance. In our simulations, the initiated shock wave needed to be in contact with the hard subsurface floor in order to be maintained and this is likely the case on the Moon as well. Surface regolith undergoes a rapid increase in density in the first 10s of $cm$ of the surface (\cite{carrier1991physical}) and below that there are varying layers of regolith composition that can have abrupt transitions (\cite{fa2015regolith}).

Although the greatest bulk dilation we observed in any of our tests was $\Delta\rho \sim 0.5\%$ which is an order of magnitude less than the expected LCS dilation ($\sim 4\%$)), recall that that our idealized particles (monodisperse) have properties (size, modulus) that differ by orders of magnitude as compared to the Lunar surface regolith grains. We expect an assembly with material properties that more closely resemble the Lunar environment to show greater dilation than seen here. In the Lunar regolith, wave speed (which is dominant in controlling impulse-induced dilation) will increase due to harder particles and dilation will be enhanced by the smaller particles, irregular particle shapes, and polydispersity (\cite{jia1999ultrasound}). Based on shear band thickness dependence on particle size as a result of planar flow ($\propto D^{-1/3}$, \cite{mohan2002frictional}), we expect at least a doubling in the vertical extent of the dilation band as compared to the results presented here.

\begin{figure}[h]
\begin{center}
\includegraphics[scale = .4]{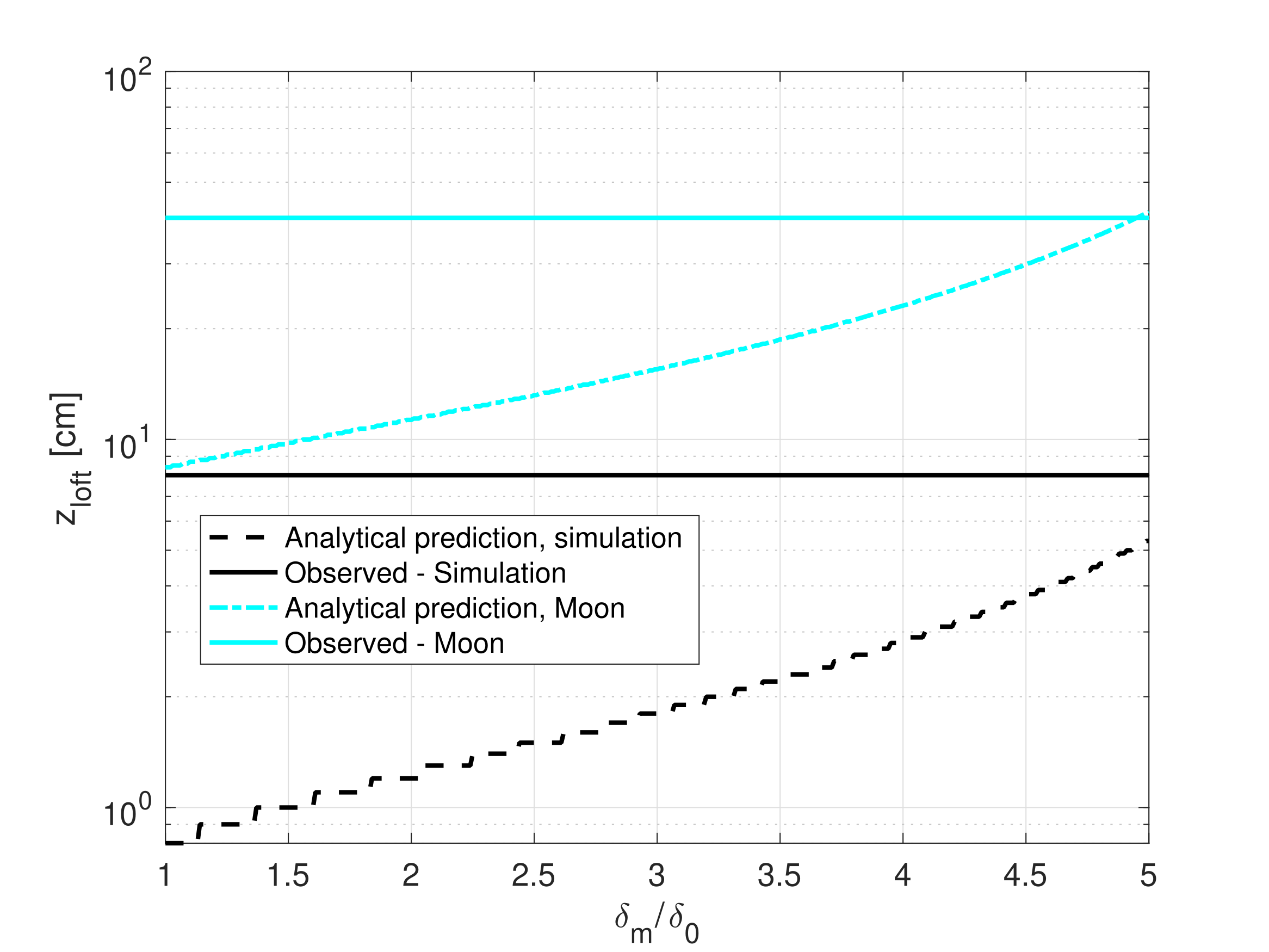}
\end{center}
\caption{\textbf{$z_loft$ vs shock strength.} We solve equation \ref{eq:loft_solve} for $z_{loft}$ at varying shock strengths ($\delta_m/\delta_0$). The expected lofting depths are approximately achieved for the strongest solitary waves ($\delta_m/\delta_0 = 5$) for both the Moon (cyan) and our simulated (black) conditions. There is a consistent nearly order of magnitude scaling of $z_{loft}$ between our simulated case and the Lunar environment.}
\label{fig:zloftVsStrength}
\end{figure}

We use equation \ref{eq:loft_solve} to evaluate how our simulated results might scale to the Lunar environment. We conduct a sweep over the space $0< \delta_m / \delta_0 < 5$ as a proxy for solitary wave strength. We first need to define a few particle properties for an idealized Lunar surface grain. In our simulation, we had a particle elastic modulus of $E = 5$ $MPa$ which resulted in a bulk modulus around $75$ $kPa$ (Table \ref{tab:soundspeed_vsphi}). If the ratio of bulk-to-particle is the same on the Moon, then this leads to Lunar surface regolith particle with modulus $500$ $MPa$.  For the cohesion, we assume a particle-particle value that is also two orders of magnitude greater than the bulk value (\cite{zhang2018rotational}), so $k_c = 100$ $kJ/m^3$ (scaling from Table \ref{tab:BaseSimParam}). Using the respective values for our simulated and Lunar particles yields Figure \ref{fig:zloftVsStrength}.

In Figure \ref{fig:zloftVsStrength}, the solid lines represent the observed $z_{loft}$ in our simulation ($8$ $cm$, black) and hypothesized $z_{loft}$ on the Moon ($40$ $cm$ \cite{bandfield2014lunarCS}, cyan). For stronger solitary waves, particles are lofted down to approximately the depths we expect. The loft depth of $40$ cm is surpassed for the Lunar grains around $\delta_m/\delta_0 = 5$ and approaches our observed $8$ $cm$ simulated lofting depth for the same strength. More importantly, there is a roughly constant scaling ratio in $z_{loft}$ of nearly an order of magnitude ($\sim8.5$) for the Lunar surface particles compared to simulated. This is exciting because, even with our assumptions and the omission of some material parameters (like friction), we approximately capture the scaling between assembles in different environments. Given the scaling of Equation \ref{eq:loft_solve} discussed in the prior section, the impulse-induced surface dilation seen here may also be active on other airless and low gravity bodies. Developing a complete model (which predicts final bed height change) and understanding the sensitivity of impulse-induced-dilation to various physical and mechanical parameters is the focus of our ongoing research.

\section{Summary and conclusions}
We have constructed and validated an open source tool for investigating inertial dilation and near surface wave propagation in a granular medium exposed to vacuum and microgravity. In our simulations, lateral shock waves were generated via impulsive piston impact and quickly decayed to a solitary wave. We have provided the first characterization and explanation for the particle lofting mechanism that occurs as long as the solitary wave is maintained. The primary results of our work are:

\begin{itemize}
    \item vertical surface dilation will occur as the result of particle lofting induced by a single laterally propagating granular wave in microgravity and vacuum
    
    \item an impulse-induced lateral shock in a randomly packed 3D assembly can decay to a solitary wave which propagates (and affects dilation) over long distances if the shock is generated in contact with a hard subsurface floor
    
    \item  impulse-induced granular dilation is confined to a narrow region at the grain-vacuum interface and increases with wave speed (which increases with $\phi_0$ and height of the assembly's surface above a soft-to-hard particle interface)
    
    \item the expression of dilation does not depend on the length of a granular assembly (for the distances inspected in this paper) and would occur in a short channel to the same extent as it would in a longer channel with the same $\phi_0$ and depth

\end{itemize}
We conclude that impulse-induced granular dilation is a plausible LCS formation mechanism and deserves further study. We hypothesize that initial Lunar impact-generated shocks decay to 3D solitary waves which are maintained over large distances, dilating the surface regolith as they travel.  Regardless of the outcome, understanding small airless body response to surface waves has applications for understanding the evolutionary history of the solar system as well as future human exploration and construction on the Moon.

\backmatter

\bmhead{Supplementary information}
We include supplementary videos showing P wave evolution in the compact bed using the same filters as the panels in Figure \ref{fig:shock_vis_Pwave_combo}. Please see the online version of this article to access the videos (Online Resources 1-5).

\bmhead{Acknowledgements}
Comments from the two anonymous reviewers were invaluable in improving this work. The authors thank the A. James \& Alice B. Clark Foundation for their support. We also thank C. Callejon for providing an initial simulation template and T. Leps and Y. Zhang for their granular dynamics guidance. Thanks as well to the University of Maryland High Performance Computing and Division of Information Technology for their support. This material is based upon work supported by the National Science Foundation Graduate Research Fellowship Program under Grant No. DGE 1840340. Any opinions, findings, and conclusions or recommendations expressed in this material are those of the author(s) and do not necessarily reflect the views of the National Science Foundation.

 \section*{Declarations}
 
\begin{itemize}
\item Funding: This material is based upon work supported by the National Science Foundation Graduate Research Fellowship Program under Grant No. DGE 1840340. Any opinions, findings, and conclusions or recommendations expressed in this material are those of the author(s) and do not necessarily reflect the views of the National Science Foundation.
\item Competing interests: The authors have no competing interests to declare that are relevant to the content of this article.
\item Ethics approval: Not applicable.
\item Consent to participate: Not applicable.
\item Consent for publication: Not applicable.
\item Availability of data and materials: Given the size of our datasets we provide a subset of data used in this paper. The restart files and scripts needed to launch shock simulations along with said restart files are provided in \cite{frizzell2023data}. We also provide all the output from a single shock run. Any reader could use the provided files with open source software (see Code availability below) to easily recreate our work and validate against the provided single shock run.
\item Code availability: We use open source code for simulation and visualization. The files used to implement our model and run the simulations are provided in our public repository (\cite{frizzell2023code}).
\item Authors' contributions: ESF implemented, carried out, and analyzed the simulation results in this paper and was also responsible for original manuscript preparation. CMH conceptualized this study, provided manuscript editing and review, and also analyzed simulation results.
\end{itemize}

\noindent

\begin{appendices}

\section{Contact model} \label{sec:appendix_contact}
Our granular assembly is a collection of spheres subject to Hertz contact law (\cite{mindlin1953elastic}) which accurately captures wave propagation in granular media (\cite{coste1999validity}). We also include static and rolling friction (\cite{ai2011assessment}) and cohesion (\cite{johnson1971surface}) in our force model. Here we provide an overview of our model considering two monodisperse particles in contact. When two particles of radius $R$ come into contact they begin to experience a nonlinear restorative elastic force proportional to the amount of overlap ($\delta$), $F \propto \delta^{3/2}$. Normal and tangential directions are with respect to the particle contact. When the overlap is purely normal ($\delta_n$) the restorative elastic force ($F_{n,e}$, eq. \ref{eq:normal_elastic}) comes from the sphere's stiffness, $k_n$.

\begin{align}
     F_{n,e} = - k_n \delta_n^{3/2} \nonumber \\   k_n = \frac{4}{3}E_{eq}\sqrt{R_{eq}} \label{eq:normal_elastic}
\end{align}

Equivalent radius ($R_{eq}$) and mass ($m_{eq}$) are given in eq. \ref{eq:equivalents_RM}. Equivalent compressive ($E_{eq}$) and shear ($G_{eq}$) strengths are provided in eq. \ref{eq:equivalents_EG}.

\begin{align}
    R_{eq} = \frac{R_1R_2}{R_1 + R_2} \nonumber \\  m_{eq} = \left [ \frac{1}{m_1} + \frac{1}{m_2} \right ]^{-1}  \label{eq:equivalents_RM}
\end{align}

\begin{align}
    E_{eq} = \left[\frac{1-\nu_1^2}{E_1} + \frac{1-\nu_2^2}{E_2}\right]^{-1} \nonumber \\  G_{eq} = \cdots \nonumber \\ \left [ \frac{2(2+\nu_1)(1-\nu_1)}{E_1} + \frac{2(2+\nu_2)(1-\nu_2)}{E_2} \right ]^{-1} \ \label{eq:equivalents_EG}
\end{align}

There is also a normal damping force ($F_{n,d}$, eq. \ref{eq:normal_damp}) which depends on particle normal velocity ($v_n$) and the normal damping coefficient ($\gamma_n$) which depends coefficient of restitution ($e$), in addition to material properties already defined. Intermediate quantities in eq. \ref{eq:normal_damp} are defined in eq. \ref{eq:normal_damp_sup}.

\begin{align}
    F_{n,d} = -\gamma_nv_n \nonumber \\  \gamma_n = -2\sqrt{\frac{5}{6}} \beta \sqrt{S_nm_{eq}} \label{eq:normal_damp} 
\end{align}

\begin{align}
      \beta = \frac{ln(e)}{\sqrt{ln(e)^2 + \pi^2}} \nonumber \\  S_n = 2E_{eq} \sqrt{R_{eq} \delta_n} \label{eq:normal_damp_sup} 
\end{align}

Finally, the normal force also captures the \cite{johnson1971surface} cohesive force ($F_c$, eq. \ref{eq:normal_cohesion}). This depends on the particle's cohesion energy density, $k_c$, and particle contact area, $A_c$ (\cite{kloss2012models}), shown here in eq. \ref{eq:normal_cohesion_area}).

\begin{equation}\label{eq:normal_cohesion}
    \textbf{F}_c = k_c A_c 
\end{equation}

\begin{align}
     A_c = -\frac{\pi}{4} \frac{1}{c_{12}^2} (c_{12}-R_1-R_2)(c_{12}+R_1-R_2) \cdots \nonumber \\ (c_{12}-R1+R_2)(c_{12}+R_1+R_2)  \nonumber \\  c_{12} = \delta_n - R_1 - R_2 \label{eq:normal_cohesion_area}
\end{align}

The total normal contact force (eq. \ref{eq:normal_force_total}) is then $F_n = F_{n,e} + F_{n,d} + F_c$.

\begin{equation}\label{eq:normal_force_total}
    \textbf{F}_n = - k_n \delta_n^{3/2} - \gamma_nv_n + k_cA_c  \\
\end{equation}

The tangential contact forces are similar to the normal forces, but with a constraint to appropriately capture Coulomb sliding friction. The tangential elastic force ($F_{t,e}$, eq. \ref{eq:tangential_elastic}) depends on the particle's tangential stiffness, $k_t$, which depends on the tangential overlap, $\delta_t$. 

\begin{align}
  F_{t,e} = -k_t\delta_t \nonumber \\  k_{t} = 8G_{eq}\sqrt{R_{eq}\delta_n} \label{eq:tangential_elastic}
\end{align}

As before, there is a tangential damping force ($F_{t,d}$ eq. \ref{eq:tangential_damp}) which depends on particle tangential velocity ($v_t$) and the tangential damping coefficient ($\gamma_t$). Intermediate quantities in eq. \ref{eq:tangential_damp} are defined in eq. \ref{eq:tangential_damp_sup}.

\begin{align}
     F_{t,d} = -\gamma_tv_t \nonumber \\  \gamma_t = -2\sqrt{\frac{5}{6}} \beta \sqrt{S_tm_{eq}} \label{eq:tangential_damp}
\end{align}

\begin{align}
    \beta = \frac{ln(e)}{\sqrt{ln(e)^2 + \pi^2}} \nonumber \\  S_t = 8G_{eq}\sqrt{R_{eq}\delta_n} \label{eq:tangential_damp_sup}
\end{align}

The total tangential contact force (eq. \ref{eq:tangential_force_total}) is then $F_t = F_{t_e} + F_{t,d}$. However, $F_t$ is re-scaled and $\delta_t$ becomes `historical' when the contact is in the Coulomb sliding regime ($\mu_s$ - sliding friction coefficient). Including contact history is important for capturing shear behavior of an assembly (\cite{fleischmann2016importance}).

\begin{align}
    \textbf{F}_t = - k_t \delta_t^{3/2} - \gamma_tv_t \nonumber \\  \mid F_t \mid \leq \mu_s \mid F_n \mid \label{eq:tangential_force_total}
\end{align}

Rolling friction  is included to provide packing support and mimic aspherical particles. Rolling (\cite{mohamed2010comprehensive}, \cite{jiang2015novel}) and twisting (\cite{jiang2015novel}) friction are needed to accurately capture shear response. See \cite{ai2011assessment} for a comprehensive overview of the model and a helpful diagram (their Figure 3).  We provide a brief overview here. We model rolling friction (eq. \ref{eq:rolling_friction}) as an elastic-plastic-spring-dashpot (EPSD) which includes a spring torque ($M_{r,k}$) and a viscous damping torque ($M_{r,d}$).

\begin{equation}\label{eq:rolling_friction}
    \textbf{M}_r = \textbf{M}_{r,k} + \textbf{M}_{r,d}
\end{equation}

The spring torque is computed incrementally (eq. \ref{eq:torque_spring}) using the relative rotation between the particles ($\Delta\boldsymbol\theta$) and the rolling spring stiffness, $k_r$. The spring torque depends on the rolling friction parameter ($\mu_r$) and is limited to the torque achieved at full mobilization ($M_{r,k}^m = \mu_rR_{eq}F_n$) which allows for cyclic rolling. The relative rotation implementation of rolling friction also captures twisting (torsional) friction (\cite{sanchez2016disruption}).

\begin{align}
     \textbf{M}_{r,k}^{t+\Delta t} = \textbf{M}_{r,k}^t + \Delta \textbf{M}_{r,k} \nonumber \\
     \Delta \textbf{M}_{r,k} = -k_r \Delta \boldsymbol\theta \nonumber \\
     \mid M_{r,t+\Delta t}^k \mid \leq M_{r,k}^m \nonumber \\ k_r = 2.55k_n \mu_r^2R_{eq}^2 \label{eq:torque_spring}
\end{align}

The damping torque (eq. \ref{eq:torque_damp}) depends on the relative rolling angular velocity ($\boldsymbol{\dot{\theta}}$) and is dependent on the rolling damping parameter, $\gamma_r$. The full model from \cite{ai2011assessment} allows for different damping regimes but our implementation considers the case where viscous damping torque is active only before the particle has become fully mobilized as was done in \cite{sanchez2016disruption}.
 
\begin{align}
    \textbf{M}_{r,d}^{t+\Delta t} = -\gamma_r \boldsymbol{\dot{\theta}} \text{  while  }  \mid M_{r,t+\Delta t}^k \mid \leq M_{r,k}^m \nonumber \\ \text{else } = 0 \label{eq:torque_damp}
\end{align}

In equation \ref{eq:governing_eq} we give the governing equations which are integrated at each step. Torque, normal and tangential contact forces, and gravity are included. Gravitational force is $F_g = m\boldsymbol{g}$, with $g$ the gravitational acceleration vector. $\textbf{x}_i$ is the particle state, $\textbf{r}_{c,i}$ the vector connecting the centers of the colliding particles, $I_i$ the equivalent moment of inertia of the contact, and $\boldsymbol\omega_i$ the angular velocity:

\begin{align}
    m_i \ddot{\textbf{x}}_i = \textbf{F}_{n,i} + \textbf{F}_{t,i} + \textbf{F}_{g,i} \nonumber  \\ I_i \frac{d \boldsymbol\omega_i}{dt} = \textbf{r}_{c,i} \times \textbf{F}_{t,i} + \textbf{M}_{r,i} \label{eq:governing_eq} 
\end{align}

We use $\Delta t < 0.01\tau_c$ (eq. \ref{eq:contact_time}, \cite{abd2010force}), with $V_{in}$ the incident particle velocity and $M$ and $R$ the mass and radius of the incident particle. 

\begin{equation}\label{eq:contact_time}
    \tau_c = 2.94 \left (\frac{m_{eq}^2}{\left (\frac{16}{15} \right )^2 E_{eq}^2 R V_{in}} \right )^{1/5}
\end{equation}

\section{Wave Visualization}
\begin{figure*}[h]
\begin{center}
\includegraphics[scale = .85]{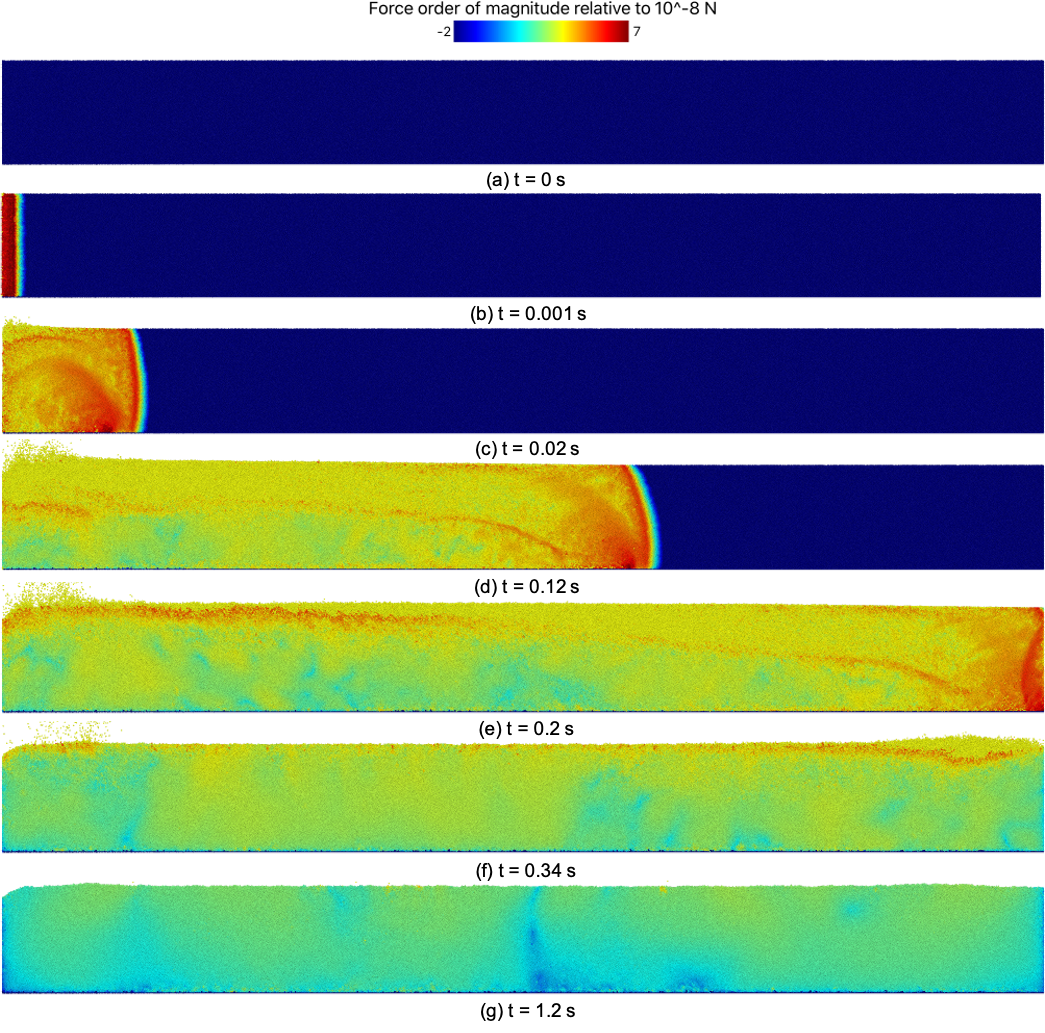}
\end{center}
\caption{\textbf{Wave Visualization.} Wave propagation seen through force visualization. We performed the shocking procedure on a $2$ $m$ channel filled $20$ $cm$ deep with grains (compact packing). Each particle is colored based on a scale that represents the order of magnitude force experienced, relative to $10^{-8}$ $N$. Frame (a) is the settled particle bed. In Frame (b), the shock has been initiated and the wavefront is uniform. In Frame (c) the wave is propagating as a solitary wave and the front has acquired curvature due to slower surface speeds than those at depth. Frame (d) shows the beginning of the reflected floor wave and mobilization of surface particles in the first 1m of the channel is already evident. In Frame (e) the shock has just reached the end wall and Frame (f) shows that the reflected wave terminates before it can travel back down the channel. The reflected floor wave reaches the surface grains, though note the thin strip of particles that do not experience a force from this reflection. These are the fully detached particles. Frame (g) show  that the surface modification remains while the bed begins to settle. The small high force region at the surface in the middle of the channel is the result of an ejected particle returning to the surface, which does not further modify bed height.}
\label{fig:shockframes}
\end{figure*}

We provide a visualization of the laterally propagating wave as it decays from the initially induced shock and propagates through our granular assembly as a solitary wave. Figure \ref{fig:shockframes} shows the force evolution of particles in the bed over time.

\end{appendices}



\bibliographystyle{unsrtnat} 
\bibliography{manuscript.bib}

\begin{thebibliography}{127}
\providecommand{\natexlab}[1]{#1}
\providecommand{\url}[1]{\texttt{#1}}
\expandafter\ifx\csname urlstyle\endcsname\relax
  \providecommand{\doi}[1]{doi: #1}\else
  \providecommand{\doi}{doi: \begingroup \urlstyle{rm}\Url}\fi

\bibitem[Royer et~al.(2011)Royer, Conyers, Corwin, Eng, and Jaeger]{royer2011role}
J.R. Royer, B.~Conyers, E.I. Corwin, P.J. Eng, and H.M. Jaeger.
\newblock The role of interstitial gas in determining the impact response of granular beds.
\newblock \emph{Europhys. Lett.}, 93\penalty0 (2):\penalty0 28008, 2011.
\newblock URL \url{https://doi.org/10.1209/0295-5075/93/28008}.

\bibitem[Reynolds(1885)]{reynolds1885lvii}
O.~Reynolds.
\newblock {LVII}. {O}n the dilatancy of media composed of rigid particles in contact. {W}ith experimental illustrations.
\newblock \emph{The Lond., Edinb., and Dublin Philosoph. Mag. and J. of Sci.}, 20\penalty0 (127):\penalty0 469--481, 1885.
\newblock URL \url{https://doi.org/10.1080/14786448508627791}.

\bibitem[Tillemans and Herrmann(1995)]{tillemans1995simulating}
H.-J. Tillemans and H.J. Herrmann.
\newblock Simulating deformations of granular solids under shear.
\newblock \emph{Physica A: Stat. Mech. and its Appl.}, 217\penalty0 (3-4):\penalty0 261--288, 1995.
\newblock URL \url{https://doi.org/10.1016/0378-4371(95)00111-J}.

\bibitem[Wensrich(2002)]{wensrich2002experimental}
C.~Wensrich.
\newblock Experimental behaviour of quaking in tall silos.
\newblock \emph{Powder Technol.}, 127\penalty0 (1):\penalty0 87--94, 2002.
\newblock URL \url{https://doi.org/10.1016/S0032-5910(02)00105-5}.

\bibitem[Philippe and Bideau(2002)]{philippe2002compaction}
P.~Philippe and D.~Bideau.
\newblock Compaction dynamics of a granular medium under vertical tapping.
\newblock \emph{Europhys. Lett.}, 60\penalty0 (5):\penalty0 677, 2002.
\newblock URL \url{https://doi.org/10.1209/epl/i2002-00362-7}.

\bibitem[Sture et~al.(1998)Sture, Costes, Batiste, Lankton, AlShibli, Jeremic, Swanson, and Frank]{sture1998mechanics}
S.~Sture, N.C. Costes, S.N. Batiste, M.R. Lankton, K.A. AlShibli, B.~Jeremic, R.A. Swanson, and M.~Frank.
\newblock Mechanics of granular materials at low effective stresses.
\newblock \emph{J. of Aero. Eng.}, 11\penalty0 (3):\penalty0 67--72, 1998.
\newblock URL \url{https://doi.org/10.1061/(ASCE)0893-1321(1998)11:3(67)}.

\bibitem[Tournat et~al.(2005)Tournat, Zaitsev, Nazarov, Gusev, and Castagn{\`e}de]{tournat2005experimental}
V.~Tournat, V.Y. Zaitsev, V.E. Nazarov, V.E. Gusev, and B.~Castagn{\`e}de.
\newblock Experimental study of nonlinear acoustic effects in a granular medium.
\newblock \emph{Acoustical Phys.}, 51\penalty0 (5):\penalty0 543--553, 2005.
\newblock URL \url{https://doi.org/10.1134/1.2042573}.

\bibitem[Van~der Elst et~al.(2012)Van~der Elst, Brodsky, Le~Bas, and Johnson]{van2012auto}
N.J. Van~der Elst, E.E. Brodsky, P.-Y. Le~Bas, and P.A. Johnson.
\newblock Auto-acoustic compaction in steady shear flows: Experimental evidence for suppression of shear dilatancy by internal acoustic vibration.
\newblock \emph{J. of Geophys. Res.: Solid Earth}, 117\penalty0 (B9), 2012.
\newblock URL \url{https://doi.org/10.1029/2011JB008897}.

\bibitem[Campbell(2005)]{campbell2005stress}
C.S. Campbell.
\newblock Stress-controlled elastic granular shear flows.
\newblock \emph{J. of Fluid Mech.}, 539:\penalty0 273--297, 2005.
\newblock URL \url{https://doi.org/10.1017/S0022112005005616}.

\bibitem[Daniels(2013)]{daniels2013rubble}
K.E. Daniels.
\newblock Rubble-pile near earth objects: {I}nsights from granular physics.
\newblock \emph{Asteroids: {P}rospective Energy and Mat. Resour.}, pages 271--286, 2013.
\newblock URL \url{https://doi.org/10.1007/978-3-642-39244-3\_11}.

\bibitem[Walsh et~al.(2022)Walsh, Ballouz, Jawin, Avdellidou, Barnouin, Bennett, Bierhaus, Bos, Cambioni, Connolly~Jr, et~al.]{walsh2022near}
K.J. Walsh, R.-L. Ballouz, E.R. Jawin, C.~Avdellidou, O.S. Barnouin, C.A. Bennett, E.B. Bierhaus, B.J. Bos, S.~Cambioni, H.C. Connolly~Jr, et~al.
\newblock Near-zero cohesion and loose packing of {B}ennu’s near subsurface revealed by spacecraft contact.
\newblock \emph{Sci. Adv.}, 8\penalty0 (27):\penalty0 eabm6629, 2022.
\newblock URL \url{https://doi.org/10.1126/sciadv.abm6229}.

\bibitem[Murdoch et~al.(2013)Murdoch, Rozitis, Nordstrom, Green, Michel, de~Lophem, and Losert]{murdoch2013convection}
N.~Murdoch, B.~Rozitis, K.~Nordstrom, S.F. Green, P.~Michel, T.-L. de~Lophem, and W.~Losert.
\newblock Granular convection in microgravity.
\newblock \emph{Phys. Rev. Lett.}, 110:\penalty0 018307, Jan 2013.
\newblock URL \url{https://link.aps.org/doi/10.1103/PhysRevLett.110.018307}.

\bibitem[Shinbrot et~al.(2017)Shinbrot, Sabuwala, Siu, Lazo, and Chakraborty]{shinbrot2017size}
T.~Shinbrot, T.~Sabuwala, T.~Siu, M.V. Lazo, and P.~Chakraborty.
\newblock Size sorting on the rubble-pile asteroid {I}tokawa.
\newblock \emph{Phys. Rev. Lett.}, 118\penalty0 (11):\penalty0 111101, 2017.
\newblock URL \url{https://doi.org/10.1103/PhysRevLett.118.111101}.

\bibitem[Wright et~al.(2020)Wright, Quillen, South, Nelson, Sanchez, Martini, Schwartz, Nakajima, and Asphaug]{wright2020boulder}
E.~Wright, A.C. Quillen, J.~South, R.C. Nelson, P.~Sanchez, L.~Martini, S.R. Schwartz, M.~Nakajima, and E.~Asphaug.
\newblock Boulder stranding in ejecta launched by an impact generated seismic pulse.
\newblock \emph{Icarus}, 337:\penalty0 113424, 2020.
\newblock URL \url{https://doi.org/10.1016/j.icarus.2019.113424}.

\bibitem[Farr(1986)]{farr1986loading}
J.V. Farr.
\newblock \emph{Loading rate effects on the one-dimensional compressibility of four partially saturated soils}.
\newblock University of Michigan, 1986.

\bibitem[Holsapple(1993)]{holsapple1993scaling}
K.A. Holsapple.
\newblock The scaling of impact processes in planetary sciences.
\newblock \emph{Ann. Rev. of Earth and Planet. Sci.}, 21\penalty0 (1):\penalty0 333--373, 1993.
\newblock URL \url{https://doi.org/10.1146/annurev.ea.21.050193.002001}.

\bibitem[Aoki and Akiyama(1995)]{aoki1995simulation}
K.M. Aoki and T.~Akiyama.
\newblock Simulation studies of pressure and density wave propagations in vertically vibrated beds of granules.
\newblock \emph{Phys. Rev. E}, 52\penalty0 (3):\penalty0 3288, 1995.
\newblock URL \url{https://doi.org/10.1103/PhysRevE.52.3288}.

\bibitem[Gusev et~al.(2008)Gusev, Aleshin, and Tournat]{gusev2008reflection}
V.~Gusev, V.~Aleshin, and V.~Tournat.
\newblock Reflection of nonlinear acoustic waves from the mechanically free surface of an unconsolidated granular medium.
\newblock \emph{Acta Acustica United w. Acustica}, 94\penalty0 (2):\penalty0 215--228, 2008.
\newblock URL \url{https://doi.org/10.3813/AAA.918025}.

\bibitem[S{\'a}nchez et~al.(2022)S{\'a}nchez, Scheeres, and Quillen]{sanchez2022transmission}
P.~S{\'a}nchez, D.J. Scheeres, and A.C. Quillen.
\newblock Transmission of a seismic wave generated by impacts on granular asteroids.
\newblock \emph{The Planet. Sci. J.}, 3\penalty0 (10):\penalty0 245, 2022.
\newblock URL \url{https://doi.org/10.3847/PSJ/ac960c}.

\bibitem[Tancredi et~al.(2022)Tancredi, Liu, Campo-Bagatin, Moreno, and Dominguez]{tancredi2022lofting}
G.~Tancredi, P-Y Liu, A.~Campo-Bagatin, F.~Moreno, and B~Dominguez.
\newblock Lofting of low speed ejecta produced in the {DART} experiment and production of a dust cloud.
\newblock \emph{Mon. Not. of the R. Astro. Soc.}, 2022.
\newblock URL \url{https://doi.org/10.1093/mnras/stac3258}.

\bibitem[Goldshtein et~al.(1996)Goldshtein, Shapiro, and Gutfinger]{goldshtein1996mechanics}
A.~Goldshtein, M.~Shapiro, and C.~Gutfinger.
\newblock Mechanics of collisional motion of granular materials. {P}art 4. {E}xpansion wave.
\newblock \emph{J. of Fluid Mech.}, 327:\penalty0 117--138, 1996.
\newblock URL \url{https://doi.org/10.1017/S0022112096008488}.

\bibitem[Omidvar et~al.(2012)Omidvar, Iskander, and Bless]{omidvar2012stress}
M.~Omidvar, M.~Iskander, and S.~Bless.
\newblock Stress-strain behavior of sand at high strain rates.
\newblock \emph{Int. J. of Impact Eng.}, 49:\penalty0 192--213, 2012.
\newblock URL \url{https://doi.org/10.1016/j.ijimpeng.2012.03.004}.

\bibitem[Lu and Fall(2018)]{lu2018state}
G.~Lu and M.~Fall.
\newblock State-of-the-art modelling of soil behaviour under blast loading.
\newblock \emph{Geotech. and Geolog. Eng.}, 36:\penalty0 3331--3355, 2018.
\newblock URL \url{https://doi.org/10.1007/s10706-018-0560-5}.

\bibitem[Collins(2014)]{collins2014numerical}
G.S. Collins.
\newblock Numerical simulations of impact crater formation with dilatancy.
\newblock \emph{J. of Geophys. Res.: Planets}, 119\penalty0 (12):\penalty0 2600--2619, 2014.
\newblock URL \url{https://doi.org/10.1002/2014JE004708}.

\bibitem[Gowd and Rummel(1980)]{gowd1980effect}
T.N. Gowd and F.~Rummel.
\newblock Effect of confining pressure on the fracture behaviour of a porous rock.
\newblock In \emph{Int. J. of Rock Mech. and Mining Sci. \& Geomech. Abstracts}, volume~17, pages 225--229. Elsevier, 1980.
\newblock URL \url{https://doi.org/10.1016/0148-9062(80)91089-X}.

\bibitem[Brown and Jaeger(2012)]{brown2012role}
E.~Brown and H.M. Jaeger.
\newblock The role of dilation and confining stresses in shear thickening of dense suspensions.
\newblock \emph{J. of Rheology}, 56\penalty0 (4):\penalty0 875--923, 2012.
\newblock URL \url{https://doi.org/10.1122/1.4709423}.

\bibitem[Bandfield et~al.(2014)Bandfield, Song, Hayne, Brand, Ghent, Vasavada, and Paige]{bandfield2014lunarCS}
J.L. Bandfield, E.~Song, P.O. Hayne, B.D. Brand, R.R. Ghent, A.R. Vasavada, and D.A. Paige.
\newblock Lunar cold spots: Granular flow features and extensive insulating materials surrounding young craters.
\newblock \emph{Icarus}, 231:\penalty0 221--231, 2014.
\newblock URL \url{https://doi.org/10.1016/j.icarus.2013.12.017}.

\bibitem[Williams et~al.(2018)Williams, Bandfield, Paige, Powell, Greenhagen, Taylor, Hayne, Speyerer, Ghent, and Costello]{williams2018lunarCS}
J.-P. Williams, J.L. Bandfield, D.A. Paige, T.M. Powell, B.T. Greenhagen, S.~Taylor, P.O. Hayne, E.J. Speyerer, R.R. Ghent, and E.S. Costello.
\newblock Lunar {C}old {S}pots and {C}rater {P}roduction on the {M}oon.
\newblock \emph{J. of Geophys. Res.: Planets}, 123\penalty0 (9):\penalty0 2380--2392, 2018.
\newblock URL \url{https://doi.org/10.1029/2018JE005652}.

\bibitem[Duffy and Mindlin(1957)]{duffy1957stress}
J.~Duffy and R.D. Mindlin.
\newblock Stress-strain relations and vibrations of a granular medium.
\newblock \emph{J. of Appl. Mech.}, 24\penalty0 (4):\penalty0 585--593, 1957.
\newblock URL \url{https://doi.org/10.1115/1.4011605}.

\bibitem[Nesterenko(2013)]{nesterenko2013dynamics}
V.F. Nesterenko.
\newblock \emph{Dynamics of heterogeneous materials}.
\newblock Springer Science \& Business Media, 2013.
\newblock URL \url{https://doi.org/10.1007/978-1-4757-3524-6}.

\bibitem[Nesterenko(1984)]{nesterenko1984propagation}
V.F. Nesterenko.
\newblock Propagation of nonlinear compression pulses in granular media.
\newblock \emph{J. Appl. Mech. Tech. Phys.(Engl. Transl.);(United States)}, 24\penalty0 (5), 1984.
\newblock URL \url{https://doi.org/10.1007/BF00905892}.

\bibitem[Coste et~al.(1997)Coste, Falcon, and Fauve]{coste1997solitary}
C.~Coste, E.~Falcon, and S.~Fauve.
\newblock Solitary waves in a chain of beads under {H}ertz contact.
\newblock \emph{Phys. Rev. E}, 56\penalty0 (5):\penalty0 6104, 1997.
\newblock URL \url{https://doi.org/10.1103/PhysRevE.56.6104}.

\bibitem[Sen et~al.(1998)Sen, Manciu, and Wright]{sen1998solitonlike}
S.~Sen, M.~Manciu, and J.D. Wright.
\newblock Solitonlike pulses in perturbed and driven {H}ertzian chains and their possible applications in detecting buried impurities.
\newblock \emph{Phys. Rev. E}, 57\penalty0 (2):\penalty0 2386, 1998.
\newblock URL \url{https://doi.org/10.1103/PhysRevE.57.2386}.

\bibitem[Hong et~al.(2000)Hong, Kim, and Hwang]{hong2000characterization}
J.~Hong, H.~Kim, and J.-P. Hwang.
\newblock Characterization of soliton damping in the granular chain under gravity.
\newblock \emph{Phys. Rev. E}, 61\penalty0 (1):\penalty0 964, 2000.
\newblock URL \url{https://doi.org/10.1103/PhysRevE.61.964}.

\bibitem[Shoaib and Kari(2011)]{shoaib2011discrete}
M.~Shoaib and L.~Kari.
\newblock Discrete element simulation of elastoplastic shock wave propagation in spherical particles.
\newblock \emph{Adv. in Acoust. and Vib.}, 2011, 2011.
\newblock URL \url{https://doi.org/10.1155/2011/123695}.

\bibitem[Chakravarty and Sen(2018)]{chakravarty2018possibility}
S.~Chakravarty and S.~Sen.
\newblock Possibility of useful mechanical energy from noise: the solitary wave train problem in the granular chain revisited.
\newblock \emph{Granular Matter}, 20\penalty0 (3):\penalty0 1--10, 2018.
\newblock URL \url{https://doi.org/10.1007/s10035-018-0811-4}.

\bibitem[Shukla and Damania(1987)]{shukla1987experimental}
A.~Shukla and C.~Damania.
\newblock Experimental investigation of wave velocity and dynamic contact stresses in an assembly of disks.
\newblock \emph{Exp. Mech.}, 27\penalty0 (3):\penalty0 268--281, 1987.
\newblock URL \url{https://doi.org/10.1007/BF02318093}.

\bibitem[Sen and Sinkovits(1996)]{sen1996sound}
S.~Sen and R.S. Sinkovits.
\newblock Sound propagation in impure granular columns.
\newblock \emph{Phys. Rev. E}, 54\penalty0 (6):\penalty0 6857, 1996.
\newblock URL \url{https://doi.org/10.1103/PhysRevE.54.6857}.

\bibitem[Owens and Daniels(2011)]{owens2011sound}
E.T. Owens and K.E. Daniels.
\newblock Sound propagation and force chains in granular materials.
\newblock \emph{Europhys. Lett.}, 94\penalty0 (5):\penalty0 54005, 2011.
\newblock URL \url{https://doi.org/10.1209/0295-5075/94/54005}.

\bibitem[Nishida et~al.(2009)Nishida, Tanaka, and Ishida]{nishida2009simulation}
M.~Nishida, K.~Tanaka, and T.~Ishida.
\newblock {DEM} simulation of wave propagation in two-dimensional ordered array of particles.
\newblock In \emph{Shock Waves}, pages 815--820. Springer, 2009.
\newblock URL \url{https://doi.org/10.1007/978-3-540-85181-3\_3}.

\bibitem[Awasthi et~al.(2012)Awasthi, Smith, Geubelle, and Lambros]{awasthi2012propagation}
A.P. Awasthi, K.J. Smith, P.H. Geubelle, and J.~Lambros.
\newblock Propagation of solitary waves in {2D} granular media: {A} numerical study.
\newblock \emph{Mech. of Materials}, 54:\penalty0 100--112, 2012.
\newblock URL \url{https://doi.org/10.1016/j.mechmat.2012.07.005}.

\bibitem[Leonard et~al.(2013)Leonard, Fraternali, and Daraio]{leonard2013directional}
A.~Leonard, F.~Fraternali, and C.~Daraio.
\newblock Directional wave propagation in a highly nonlinear square packing of spheres.
\newblock \emph{Exp. Mech.}, 53\penalty0 (3):\penalty0 327--337, 2013.
\newblock URL \url{https://doi.org/10.1007/s11340-011-9544-6}.

\bibitem[Pal et~al.(2014)Pal, Awasthi, and Geubelle]{pal2014characterization}
R.K. Pal, A.P. Awasthi, and P.H. Geubelle.
\newblock Characterization of wave propagation in elastic and elastoplastic granular chains.
\newblock \emph{Phys. Rev. E}, 89\penalty0 (1):\penalty0 012204, 2014.
\newblock URL \url{https://doi.org/10.1103/PhysRevE.89.012204}.

\bibitem[Waymel et~al.(2018)Waymel, Wang, Awasthi, Geubelle, and Lambros]{waymel2018propagation}
R.F. Waymel, E.~Wang, A.P. Awasthi, P.H. Geubelle, and J.~Lambros.
\newblock Propagation and dissipation of elasto-plastic stress waves in two dimensional ordered granular media.
\newblock \emph{J. of the Mech. and Phys. of Solids}, 120:\penalty0 117--131, 2018.
\newblock URL \url{https://doi.org/10.1016/j.jmps.2017.11.007}.

\bibitem[Leonard et~al.(2014)Leonard, Chong, Kevrekidis, and Daraio]{leonard2014traveling}
A.~Leonard, C.~Chong, P.G. Kevrekidis, and C.~Daraio.
\newblock Traveling waves in {2D} hexagonal granular crystal lattices.
\newblock \emph{Granular Matter}, 16\penalty0 (4):\penalty0 531--542, 2014.
\newblock URL \url{https://doi.org/10.1007/s10035-014-0487-3}.

\bibitem[Zhang et~al.(2020)Zhang, Li, Lambros, Bergman, and Vakakis]{zhang2020pulse}
Q.~Zhang, W.~Li, J.~Lambros, L.A. Bergman, and A.F. Vakakis.
\newblock Pulse transmission and acoustic non-reciprocity in a granular channel with symmetry-breaking clearances.
\newblock \emph{Granular Matter}, 22\penalty0 (1):\penalty0 1--16, 2020.
\newblock URL \url{https://doi.org/10.1007/s10035-019-0982-7}.

\bibitem[Tournat and Gusev(2010)]{tournat2010acoustics}
V.~Tournat and V.E. Gusev.
\newblock Acoustics of unconsolidated “model” granular media: An overview of recent results and several open problems.
\newblock \emph{Acta Acustica united w. Acustica}, 96\penalty0 (2):\penalty0 208--224, 2010.
\newblock URL \url{https://doi.org/10.3813/AAA.918271}.

\bibitem[Tell et~al.(2020)Tell, Drei{\ss}igacker, Tchapnda, Yu, and Sperl]{tell2020acoustic}
K.~Tell, C.~Drei{\ss}igacker, A.C. Tchapnda, P.~Yu, and M.~Sperl.
\newblock Acoustic waves in granular packings at low confinement pressure.
\newblock \emph{Rev. of Sci. Instr.}, 91\penalty0 (3):\penalty0 033906, 2020.
\newblock URL \url{https://doi.org/10.1063/1.5122848}.

\bibitem[G{\'o}mez et~al.(2012)G{\'o}mez, Turner, van Hecke, and Vitelli]{gomez2012shocks}
L.R. G{\'o}mez, A.M. Turner, M.~van Hecke, and V.~Vitelli.
\newblock Shocks near jamming.
\newblock \emph{Phys. Rev. Letters}, 108\penalty0 (5):\penalty0 058001, 2012.
\newblock URL \url{https://doi.org/10.1103/PhysRevLett.108.058001}.

\bibitem[Rogers and Don(1994)]{rogers1994location}
A.J. Rogers and C.G. Don.
\newblock Location of buried objects by an acoustic impulse technique.
\newblock \emph{Acoustics Australia}, 22:\penalty0 5--5, 1994.

\bibitem[Sen and Manciu(2001)]{sen2001solitary}
S.~Sen and M.~Manciu.
\newblock Solitary wave dynamics in generalized {H}ertz chains: An improved solution of the equation of motion.
\newblock \emph{Phys. Rev. E}, 64\penalty0 (5):\penalty0 056605, 2001.
\newblock URL \url{https://doi.org/10.1103/PhysRevE.64.056605}.

\bibitem[Sen et~al.(2005)Sen, Krishna~Mohan, Visco~Jr, Swaminathan, Sokolow, Avalos, and Nakagawa]{sen2005using}
S.~Sen, T.R. Krishna~Mohan, D.P. Visco~Jr, S.~Swaminathan, A.~Sokolow, E.~Avalos, and M.~Nakagawa.
\newblock Using mechanical energy as a probe for the detection and imaging of shallow buried inclusions in dry granular beds.
\newblock \emph{Int. J. of Mod. Phys. B}, 19\penalty0 (18):\penalty0 2951--2973, 2005.
\newblock URL \url{https://doi.org/10.1142/S0217979205031997}.

\bibitem[Sen et~al.(2017)Sen, Krishna~Mohan, and Tiwari]{sen2017impact}
S.~Sen, T.R. Krishna~Mohan, and M.~Tiwari.
\newblock Impact dispersion using 2{D} and 3{D} composite granular packing.
\newblock \emph{KONA Powder and Particle J.}, page 2017014, 2017.
\newblock URL \url{https://doi.org/10.14356/kona.2017014}.

\bibitem[Hostler and Brennen(2005)]{hostler2005pressure}
S.R. Hostler and C.E. Brennen.
\newblock Pressure wave propagation in a granular bed.
\newblock \emph{Phys. Rev. E}, 72\penalty0 (3):\penalty0 031303, 2005.
\newblock URL \url{https://doi.org/10.1103/PhysRevE.72.031303}.

\bibitem[Quillen et~al.(2022)Quillen, Neiderbach, Suo, South, Wright, Skerrett, S{\'a}nchez, C{\'u}{\~n}ez, Miklavcic, and Askari]{quillen2022propagation}
A.C. Quillen, M.~Neiderbach, B.~Suo, J.~South, E.~Wright, N.~Skerrett, P.~S{\'a}nchez, F.D. C{\'u}{\~n}ez, P.~Miklavcic, and H.~Askari.
\newblock Propagation and attenuation of pulses driven by low velocity normal impacts in granular media.
\newblock \emph{Icarus}, 386:\penalty0 115139, 2022.
\newblock URL \url{https://doi.org/10.1016/j.icarus.2022.115139}.

\bibitem[Jiao et~al.(2023)Jiao, Chen, Takato, Sen, and Huang]{jiao2023revisiting}
T.~Jiao, W.~Chen, Y.~Takato, S.~Sen, and D.~Huang.
\newblock Revisiting nesterenko’s solitary wave in the precompressed granular alignment held between fixed ends.
\newblock \emph{Granular Matter}, 25\penalty0 (2):\penalty0 17, 2023.
\newblock URL \url{https://doi.org/10.1007/s10035-023-01309-y}.

\bibitem[Sutton and Duennebier(1970)]{sutton1970elastic}
G.H. Sutton and F.K. Duennebier.
\newblock Elastic properties of the lunar surface from {S}urveyor spacecraft data.
\newblock \emph{J. of Geophys. Res.}, 75\penalty0 (35):\penalty0 7439--7444, 1970.
\newblock URL \url{https://doi.org/10.1029/JB075i035p07439}.

\bibitem[Cooper et~al.(1974)Cooper, Kovach, and Watkins]{cooper1974lunar}
M.R. Cooper, R.L. Kovach, and J.S. Watkins.
\newblock Lunar near-surface structure.
\newblock \emph{Rev. of Geophys.}, 12\penalty0 (3):\penalty0 291--308, 1974.
\newblock URL \url{https://doi.org/10.1029/RG012i003p00291}.

\bibitem[Mouraille et~al.(2009)Mouraille, Herbst, and Luding]{mouraille2009sound}
O.~Mouraille, O.~Herbst, and S.~Luding.
\newblock Sound propagation in isotropically and uni-axially compressed cohesive, frictional granular solids.
\newblock \emph{Eng. Fracture Mech.}, 76\penalty0 (6):\penalty0 781--792, 2009.
\newblock URL \url{https://doi.org/10.1016/j.engfracmech.2008.09.001}.

\bibitem[Botello et~al.(2016)Botello, Castellanos, and Tournat]{botello2016ultrasonic}
F.R. Botello, A.~Castellanos, and V.~Tournat.
\newblock Ultrasonic probing of cohesive granular media at very low consolidation.
\newblock \emph{Ultrasonics}, 69:\penalty0 193--200, 2016.
\newblock URL \url{https://doi.org/10.1016/j.ultras.2015.11.011}.

\bibitem[Agui and Creager(2018)]{agui2018high}
J.H. Agui and C.M. Creager.
\newblock High impact wave propagation studies in lunar granular systems.
\newblock In \emph{Earth and Space 2018: Eng. for Extreme Environ.}, pages 99--108. American Society of Civil Engineers Reston, VA, 2018.
\newblock URL \url{https://doi.org/10.1061/9780784481899.011}.

\bibitem[Zeng et~al.(2007)Zeng, Agui, and Nakagawa]{zeng2007wave}
X.~Zeng, J.H. Agui, and M.~Nakagawa.
\newblock Wave velocities in granular materials under microgravity.
\newblock \emph{J. of Aero. Eng.}, 20\penalty0 (2):\penalty0 116--123, 2007.
\newblock URL \url{https://doi.org/10.1061/(ASCE)0893-1321(2007)20:2(116)}.

\bibitem[El~Shourbagy et~al.(2008)El~Shourbagy, Okeda, and Matuttis]{el2008acoustic}
S.AM. El~Shourbagy, S.~Okeda, and H.-G. Matuttis.
\newblock Acoustic of sound propagation in granular materials in one, two, and three dimensions.
\newblock \emph{J. of the Phys. Soc. of Japan}, 77\penalty0 (3):\penalty0 034606--034606, 2008.
\newblock URL \url{https://doi.org/10.1143/jpsj.77.034606}.

\bibitem[Burgoyne et~al.(2015)Burgoyne, Newman, Jackson, and Daraio]{burgoyne2015guided}
H.A. Burgoyne, J.A. Newman, W.C. Jackson, and C.~Daraio.
\newblock Guided impact mitigation in 2{D} and 3{D} granular crystals.
\newblock \emph{Procedia Eng.}, 103:\penalty0 52--59, 2015.
\newblock URL \url{https://doi.org/10.1016/j.proeng.2015.04.008}.

\bibitem[Fonseka et~al.(2022{\natexlab{a}})Fonseka, Awasthi, Lambros, and Geubelle]{fonseka2022shockwaves}
R.D.JI. Fonseka, A.P. Awasthi, J.~Lambros, and P.H. Geubelle.
\newblock Shockwaves in jammed ductile granular media.
\newblock \emph{J. of Appl. Mech.}, 89\penalty0 (5):\penalty0 051003, 2022{\natexlab{a}}.
\newblock URL \url{https://doi.org/10.1115/1.4053622}.

\bibitem[Fonseka et~al.(2022{\natexlab{b}})Fonseka, Geubelle, and Lambros]{fonseka2022effect}
R.D.JI. Fonseka, P.H. Geubelle, and J.~Lambros.
\newblock Effect of confinement on the impact response of a granular array.
\newblock \emph{Exp. Mech.}, 62\penalty0 (5):\penalty0 849--862, 2022{\natexlab{b}}.
\newblock URL \url{https://doi.org/10.1007/s11340-022-00819-9}.

\bibitem[Kloss et~al.(2012)Kloss, Goniva, Hager, Amberger, and Pirker]{kloss2012models}
C.~Kloss, C.~Goniva, A.~Hager, S.~Amberger, and S.~Pirker.
\newblock Models, algorithms and validation for opensource {DEM} and {CFD--DEM}.
\newblock \emph{Prog. in Comp. Fluid Dyn., an Int. J.}, 12\penalty0 (2-3):\penalty0 140--152, 2012.
\newblock URL \url{https://doi.org/10.1504/PCFD.2012.047457}.

\bibitem[Jia et~al.(1999)Jia, Caroli, and Velicky]{jia1999ultrasound}
X.~Jia, C.~Caroli, and B.~Velicky.
\newblock Ultrasound propagation in externally stressed granular media.
\newblock \emph{Phys. Rev. Lett}, 82\penalty0 (9):\penalty0 1863, 1999.
\newblock URL \url{https://doi.org/10.1103/PhysRevLett.82.1863}.

\bibitem[Goddard and Didwania(1998)]{goddard1998computations}
J.D. Goddard and A.K. Didwania.
\newblock Computations of dilatancy and yield surfaces for assemblies of rigid frictional spheres.
\newblock \emph{The Quart. J. of Mech. and Appl. Math.}, 51\penalty0 (1):\penalty0 15--44, 1998.
\newblock URL \url{https://doi.org/10.1093/qjmam/51.1.15}.

\bibitem[Makse et~al.(1999)Makse, Gland, Johnson, and Schwartz]{makse1999effective}
H.A. Makse, N.~Gland, D.L. Johnson, and L.M. Schwartz.
\newblock Why effective medium theory fails in granular materials.
\newblock \emph{Phys. Rev. Lett.}, 83\penalty0 (24):\penalty0 5070, 1999.
\newblock URL \url{https://doi.org/10.1103/PhysRevLett.83.5070}.

\bibitem[Makse et~al.(2004)Makse, Gland, Johnson, and Schwartz]{makse2004granular}
H.A. Makse, N.~Gland, D.L. Johnson, and L.~Schwartz.
\newblock Granular packings: Nonlinear elasticity, sound propagation, and collective relaxation dynamics.
\newblock \emph{Phys. Rev. E}, 70\penalty0 (6):\penalty0 061302, 2004.
\newblock URL \url{https://doi.org/10.1103/PhysRevE.70.061302}.

\bibitem[Cundall and Strack(1979)]{cundall1979discrete}
P.A. Cundall and O.DL. Strack.
\newblock A discrete numerical model for granular assemblies.
\newblock \emph{G{\'e}otechnique}, 29\penalty0 (1):\penalty0 47--65, 1979.
\newblock URL \url{https://doi.org/10.1680/geot.1979.29.1.47}.

\bibitem[Thornton(2000)]{thornton2000numerical}
C.~Thornton.
\newblock Numerical simulations of deviatoric shear deformation of granular media.
\newblock \emph{G{\'e}otechnique}, 50\penalty0 (1):\penalty0 43--53, 2000.
\newblock URL \url{https://doi.org/10.1680/geot.2000.50.1.43}.

\bibitem[Tanaka et~al.(2002)Tanaka, Nishida, Kunimochi, and Takagi]{tanaka2002discrete}
K.~Tanaka, M.~Nishida, T.~Kunimochi, and T.~Takagi.
\newblock Discrete element simulation and experiment for dynamic response of two-dimensional granular matter to the impact of a spherical projectile.
\newblock \emph{Powder Technol.}, 124\penalty0 (1-2):\penalty0 160--173, 2002.
\newblock URL \url{https://doi.org/10.1016/S0032-5910(01)00489-2}.

\bibitem[Ning et~al.(2015)Ning, Khoubani, and Evans]{ning2015shear}
Z.~Ning, A.~Khoubani, and T.M. Evans.
\newblock Shear wave propagation in granular assemblies.
\newblock \emph{Comput. and Geotechnics}, 69:\penalty0 615--626, 2015.
\newblock URL \url{https://doi.org/10.1016/j.compgeo.2015.07.004}.

\bibitem[Schwartz et~al.(2012)Schwartz, Richardson, and Michel]{schwartz2012implementation}
S.R. Schwartz, D.C. Richardson, and P.~Michel.
\newblock An implementation of the soft-sphere discrete element method in a high-performance parallel gravity tree-code.
\newblock \emph{Granular Matter}, 14\penalty0 (3):\penalty0 363--380, 2012.
\newblock URL \url{https://doi.org/10.1007/s10035-012-0346-z}.

\bibitem[S{\'a}nchez and Scheeres(2011)]{sanchez2011simulating}
P.~S{\'a}nchez and D.J. Scheeres.
\newblock Simulating asteroid rubble piles with a self-gravitating soft-sphere distinct element method model.
\newblock \emph{The Astrophys. J.}, 727\penalty0 (2):\penalty0 120, 2011.
\newblock URL \url{https://doi.org/10.1088/0004-637X/727/2/120}.

\bibitem[Tancredi et~al.(2012)Tancredi, Maciel, Heredia, Richeri, and Nesmachnow]{tancredi2012granular}
G.~Tancredi, A.~Maciel, L.~Heredia, P.~Richeri, and S.~Nesmachnow.
\newblock Granular physics in low-gravity environments using discrete element method.
\newblock \emph{Mon. Notices of the Roy. Astronom. Soc.}, 420\penalty0 (4):\penalty0 3368--3380, 2012.
\newblock URL \url{https://doi.org/10.1111/j.1365-2966.2011.20259.x}.

\bibitem[S{\'a}nchez and Scheeres(2016)]{sanchez2016disruption}
P.~S{\'a}nchez and D.J. Scheeres.
\newblock Disruption patterns of rotating self-gravitating aggregates: {A} survey on angle of friction and tensile strength.
\newblock \emph{Icarus}, 271:\penalty0 453--471, 2016.
\newblock URL \url{https://doi.org/10.1016/j.icarus.2016.01.016}.

\bibitem[DeMartini et~al.(2019)DeMartini, Richardson, Barnouin, Schmerr, Plescia, Scheirich, and Pravec]{demartini2019using}
J.V. DeMartini, D.C. Richardson, O.S. Barnouin, N.C. Schmerr, J.B. Plescia, P.~Scheirich, and P.~Pravec.
\newblock Using a discrete element method to investigate seismic response and spin change of 99942 {A}pophis during its 2029 tidal encounter with {E}arth.
\newblock \emph{Icarus}, 328:\penalty0 93--103, 2019.
\newblock URL \url{https://doi.org/10.1016/j.icarus.2019.03.015}.

\bibitem[Zhang et~al.(2021)Zhang, Michel, Richardson, Barnouin, Agrusa, Tsiganis, Manzoni, and May]{zhang2021creep}
Y.~Zhang, P.~Michel, D.C. Richardson, O.S. Barnouin, H.F. Agrusa, K.~Tsiganis, C.~Manzoni, and B.H. May.
\newblock Creep stability of the {DART}/{H}era mission target 65803 {D}idymos {II}. the role of cohesion.
\newblock \emph{Icarus}, 362:\penalty0 114433, 2021.
\newblock URL \url{https://doi.org/10.1016/j.icarus.2021.114433}.

\bibitem[O’Donovan et~al.(2016)O’Donovan, Ibraim, O’sullivan, Hamlin, Muir~Wood, and Marketos]{o2016micromechanics}
J.~O’Donovan, E.~Ibraim, C.~O’sullivan, S.~Hamlin, D.~Muir~Wood, and G.~Marketos.
\newblock Micromechanics of seismic wave propagation in granular materials.
\newblock \emph{Granular Matter}, 18\penalty0 (3):\penalty0 1--18, 2016.
\newblock URL \url{https://doi.org/10.1007/s10035-015-0599-4}.

\bibitem[Berger and Hrenya(2017)]{berger2017predicting}
K.J. Berger and C.M. Hrenya.
\newblock Predicting regolith erosion during a lunar landing: Role of continuous size distribution.
\newblock \emph{J. of Aero. Eng.}, 30\penalty0 (5):\penalty0 04017027, 2017.
\newblock URL \url{https://doi.org/10.1061/(ASCE)AS.1943-5525.0000735}.

\bibitem[Otto et~al.(2018)Otto, Kerst, Roloff, Janiga, and Katterfeld]{otto2018cfd}
H.~Otto, K.~Kerst, C.~Roloff, G.~Janiga, and A.~Katterfeld.
\newblock {CFD--DEM} simulation and experimental investigation of the flow behavior of lunar regolith {JSC-1A}.
\newblock \emph{Particuology}, 40:\penalty0 34--43, 2018.
\newblock URL \url{https://doi.org/10.1016/j.partic.2017.12.003}.

\bibitem[Hurley and Andrade(2015)]{hurley2015friction}
R.C. Hurley and J.E. Andrade.
\newblock Friction in inertial granular flows: competition between dilation and grain-scale dissipation rates.
\newblock \emph{Granular Matter}, 17\penalty0 (3):\penalty0 287--295, 2015.
\newblock URL \url{https://doi.org/10.1007/s10035-015-0564-2}.

\bibitem[Mindlin and Deresiewicz(1953)]{mindlin1953elastic}
R.D. Mindlin and H.~Deresiewicz.
\newblock Elastic spheres in contact under varying oblique forces.
\newblock \emph{J. of Appl. Mech.}, 20\penalty0 (3):\penalty0 327--344, 1953.
\newblock URL \url{https://doi.org/10.1115/1.4010702}.

\bibitem[Coste and Gilles(1999)]{coste1999validity}
C.~Coste and B.~Gilles.
\newblock On the validity of {H}ertz contact law for granular material acoustics.
\newblock \emph{The Euro. Phys. J. B-Condens. Matter and Complex Sys.}, 7\penalty0 (1):\penalty0 155--168, 1999.
\newblock URL \url{https://doi.org/10.1007/s100510050598}.

\bibitem[Ai et~al.(2011)Ai, Chen, Rotter, and Ooi]{ai2011assessment}
J.~Ai, J.-F. Chen, J.M. Rotter, and J.Y. Ooi.
\newblock Assessment of rolling resistance models in discrete element simulations.
\newblock \emph{Powder Technol.}, 206\penalty0 (3):\penalty0 269--282, 2011.
\newblock URL \url{https://doi.org/10.1016/j.powtec.2010.09.030}.

\bibitem[Johnson et~al.(1971)Johnson, Kendall, and Roberts]{johnson1971surface}
K.L. Johnson, K.~Kendall, and A.D. Roberts.
\newblock Surface energy and the contact of elastic solids.
\newblock \emph{Proc. of the R. Soc. of Lond. {A}. Math. and Phys. Sci.}, 324\penalty0 (1558):\penalty0 301--313, 1971.
\newblock URL \url{https://doi.org/10.1098/rspa.1971.0141}.

\bibitem[G\'omez et~al.(2012)G\'omez, Turner, and Vitelli]{gomez2012uniform}
L.R. G\'omez, A.M. Turner, and V.~Vitelli.
\newblock Uniform shock waves in disordered granular matter.
\newblock \emph{Phys. Rev. E}, 86:\penalty0 041302, Oct 2012.
\newblock URL \url{https://link.aps.org/doi/10.1103/PhysRevE.86.041302}.

\bibitem[Mase et~al.(2009)Mase, Smelser, and Mase]{mase2009continuum}
G.T. Mase, R.E. Smelser, and G.E. Mase.
\newblock \emph{Continuum Mechanics for Engineers}.
\newblock CRC press, 2009.
\newblock URL \url{https://doi.org/10.1201/9780429174391}.

\bibitem[Sunday et~al.(2020)Sunday, Murdoch, Tardivel, Schwartz, and Michel]{sunday2020validating}
C.~Sunday, N.~Murdoch, S.~Tardivel, S.R. Schwartz, and P.~Michel.
\newblock Validating {N}-body code {CHRONO} for granular {DEM} simulations in reduced-gravity environments.
\newblock \emph{Mon. Notices of the Roy. Astronom. Soc.}, 498\penalty0 (1):\penalty0 1062--1079, 2020.
\newblock URL \url{https://doi.org/10.1093/mnras/staa2454}.

\bibitem[Abd-Elhady et~al.(2010)Abd-Elhady, Abd-Elhady, Rindt, and Van~Steenhoven]{abd2010force}
M.S. Abd-Elhady, S.~Abd-Elhady, C.C.M. Rindt, and A.A. Van~Steenhoven.
\newblock Force propagation speed in a bed of particles due to an incident particle impact.
\newblock \emph{Adv. Powder Technol.}, 21\penalty0 (2):\penalty0 150--164, 2010.
\newblock URL \url{https://doi.org/10.1016/j.apt.2009.11.009}.

\bibitem[Abd-Elhady et~al.(2009)Abd-Elhady, Rindt, and van Steenhoven]{abd2009contact}
M.S. Abd-Elhady, C.C.M. Rindt, and A.A. van Steenhoven.
\newblock Contact time of an incident particle hitting a 2{D} bed of particles.
\newblock \emph{Powder Technol.}, 191\penalty0 (3):\penalty0 315--326, 2009.
\newblock URL \url{https://doi.org/10.1016/j.powtec.2008.10.024}.

\bibitem[Gupta et~al.(2016)Gupta, Sun, and Ooi]{gupta2016cfd}
P.~Gupta, J.~Sun, and J.Y. Ooi.
\newblock {DEM-CFD} simulation of a dense fluidized bed: {W}all boundary and particle size effects.
\newblock \emph{Powder Technol.}, 293:\penalty0 37--47, 2016.
\newblock URL \url{https://doi.org/10.1016/j.powtec.2015.11.050}.

\bibitem[Potapov and Campbell(1996)]{potapov1996propagation}
A.V. Potapov and C.S. Campbell.
\newblock Propagation of elastic waves in deep vertically shaken particle beds.
\newblock \emph{Phys. Rev. Lett.}, 77\penalty0 (23):\penalty0 4760, 1996.
\newblock URL \url{https://doi.org/10.1103/PhysRevLett.77.4760}.

\bibitem[Stukowski(2009)]{stukowski2009visualization}
A.~Stukowski.
\newblock Visualization and analysis of atomistic simulation data with {OVITO}--the {O}pen {V}isualization {T}ool.
\newblock \emph{Modelling and Sim. in Mat. Sci. and Eng.}, 18\penalty0 (1):\penalty0 015012, 2009.
\newblock URL \url{https://doi.org/10.1088/0965-0393/18/1/015012}.

\bibitem[McKay et~al.(1991)McKay, Heiken, Basu, Blanford, Simon, Reedy, Bevan, and Papike]{mckay1991lunar}
D.S. McKay, G.~Heiken, A.~Basu, G.~Blanford, S.~Simon, R.~Reedy, M.F. Bevan, and J.~Papike.
\newblock The lunar regolith.
\newblock In Lunar {S}ourcebook: a user's guide to~the {M}oon, editor, \emph{Heiken, Grant H. and Vaniman, David T. and Bevan, M. F.}, volume~7, pages 285--356. Press Syndicate of the University of Cambridge, New {Y}ork, 1991.

\bibitem[Carrier~III et~al.(1991)Carrier~III, Olhoeft, and Mendell]{carrier1991physical}
W.D. Carrier~III, G.R. Olhoeft, and W.~Mendell.
\newblock Physical properties of the lunar surface.
\newblock In Lunar {S}ourcebook: a user's guide to~the {M}oon, editor, \emph{Heiken, G.H. and Vaniman, D.T. and Bevan, M.F.}, pages 475--594. Press Syndicate of the University of Cambridge, New {Y}ork, 1991.

\bibitem[Kovach and Watkins(1973)]{kovach1973velocity}
R.L. Kovach and J.S. Watkins.
\newblock The velocity structure of the lunar crust.
\newblock \emph{The Moon}, 7\penalty0 (1):\penalty0 63--75, 1973.
\newblock URL \url{https://doi.org/10.1007/BF00578808}.

\bibitem[Chau et~al.(2002)Chau, Wong, and Wu]{chau2002coefficient}
K.T. Chau, R.H.C. Wong, and J.J. Wu.
\newblock Coefficient of restitution and rotational motions of rockfall impacts.
\newblock \emph{Int. J. of Rock Mech. and Mining Sci.}, 39\penalty0 (1):\penalty0 69--77, 2002.
\newblock URL \url{https://doi.org/10.1016/S1365-1609(02)00016-3}.

\bibitem[Zhang et~al.(2018)Zhang, Richardson, Barnouin, Michel, Schwartz, and Ballouz]{zhang2018rotational}
Y.~Zhang, D.C. Richardson, O.S. Barnouin, P.~Michel, S.R. Schwartz, and R.-L. Ballouz.
\newblock Rotational failure of rubble-pile bodies: {I}nfluences of shear and cohesive strengths.
\newblock \emph{The Astrophys. J.}, 857\penalty0 (1):\penalty0 15, 2018.
\newblock URL \url{https://doi.org/10.3847/1538-4357/aab5b2}.

\bibitem[Wang et~al.(2020)Wang, Zhang, Feng, Yang, Wu, and Yue]{wang2020behaviors}
J.~Wang, M.~Zhang, L.~Feng, H.~Yang, Y.~Wu, and G.~Yue.
\newblock The behaviors of particle-wall collision for non-spherical particles: {E}xperimental investigation.
\newblock \emph{Powder Technol.}, 363:\penalty0 187--194, 2020.
\newblock URL \url{https://doi.org/10.1016/j.powtec.2019.12.041}.

\bibitem[Holmes et~al.(2016)Holmes, Brown, Wauters, Lavery, and Brown]{holmes2016bending}
M.A.J. Holmes, R.~Brown, P.A.L. Wauters, N.P. Lavery, and S.G.R. Brown.
\newblock Bending and twisting friction models in soft-sphere discrete element simulations for static and dynamic problems.
\newblock \emph{Appl. Math. Modelling}, 40\penalty0 (5-6):\penalty0 3655--3670, 2016.
\newblock URL \url{https://doi.org/10.1016/j.apm.2015.10.026}.

\bibitem[Gouache et~al.(2010)Gouache, Brunskill, Scott, Gao, Coste, and Gourinat]{gouache2010regolith}
T.P. Gouache, C.~Brunskill, G.P. Scott, Y.~Gao, P.~Coste, and Y.~Gourinat.
\newblock Regolith simulant preparation methods for hardware testing.
\newblock \emph{Planet. and Space Sci.}, 58\penalty0 (14-15):\penalty0 1977--1984, 2010.
\newblock URL \url{https://doi.org/10.1016/j.pss.2010.09.021}.

\bibitem[Zhang and Makse(2005)]{zhang2005jamming}
H.P. Zhang and H.A. Makse.
\newblock Jamming transition in emulsions and granular materials.
\newblock \emph{Phys. Rev. E}, 72\penalty0 (1):\penalty0 011301, 2005.
\newblock URL \url{https://doi.org/10.1103/PhysRevE.72.011301}.

\bibitem[Wang et~al.(2021)Wang, Lei, Yang, Xu, Xu, Zhao, and Song]{wang2021effects}
J.~Wang, M.~Lei, H.~Yang, K.~Xu, S.~Xu, P.~Zhao, and Y.~Song.
\newblock Effects of coefficient of friction and coefficient of restitution on static packing characteristics of polydisperse spherical pebble bed.
\newblock \emph{Particuology}, 57:\penalty0 1--9, 2021.
\newblock URL \url{https://doi.org/10.1016/j.partic.2020.12.013}.

\bibitem[Knight et~al.(1995)Knight, Fandrich, Lau, Jaeger, and Nagel]{knight1995density}
J.B. Knight, C.G. Fandrich, C.N. Lau, H.M. Jaeger, and S.R. Nagel.
\newblock Density relaxation in a vibrated granular material.
\newblock \emph{Phys. Rev. E}, 51\penalty0 (5):\penalty0 3957, 1995.
\newblock URL \url{https://doi.org/10.1103/PhysRevE.51.3957}.

\bibitem[Mykulyak(2014)]{mykulyak2014features}
S.V. Mykulyak.
\newblock Features of nonlinear wave propagation in a layer of a granular medium.
\newblock \emph{Phys. Mesomech.}, 17\penalty0 (2):\penalty0 157--162, 2014.
\newblock URL \url{https://doi.org/10.1134/S1029959914020088}.

\bibitem[Lee and Santamarina(2005)]{lee2005bender}
J.-S. Lee and J.~C. Santamarina.
\newblock Bender elements: {P}erformance and signal interpretation.
\newblock \emph{J. of Geotech. and Geoenviron. Eng.}, 131\penalty0 (9):\penalty0 1063--1070, 2005.
\newblock URL \url{https://doi.org/10.1061/(ASCE)1090-0241(2005)131:9(1063)}.

\bibitem[Mouraille et~al.(2006)Mouraille, Mulder, and Luding]{mouraille2006sound}
O.~Mouraille, W.A. Mulder, and S.~Luding.
\newblock Sound wave acceleration in granular materials.
\newblock \emph{J. of Stat. Mech.: Theory and Exp.}, 2006\penalty0 (07):\penalty0 P07023, 2006.
\newblock URL \url{https://doi.org/10.1088/1742-5468/2006/07/P07023}.

\bibitem[Li et~al.(2020)Li, Hahn, Yao, Germann, Feng, and Zhang]{li2020grain}
W.~Li, E.N. Hahn, X.~Yao, T.C. Germann, B.~Feng, and X.~Zhang.
\newblock On the grain size dependence of shock responses in nanocrystalline sic ceramics at high strain rates.
\newblock \emph{Acta Materialia}, 200:\penalty0 632--651, 2020.
\newblock URL \url{https://doi.org/10.1016/j.actamat.2020.09.044}.

\bibitem[Ning and Evans(2013)]{ning2013discrete}
Z.~Ning and T.M. Evans.
\newblock Discrete element method study of shear wave propagation in granular soil.
\newblock In \emph{Proceedings of the 18th International Conference on Soil Mechanics and Geotechnical Engineering, France, Paris}, pages 1031--1034, 2013.

\bibitem[Shojaaee et~al.(2012)Shojaaee, Roux, Chevoir, and Wolf]{shojaaee2012shear}
Z.~Shojaaee, J.-N. Roux, F.~Chevoir, and D.E. Wolf.
\newblock Shear flow of dense granular materials near smooth walls. {I}. {S}hear localization and constitutive laws in the boundary region.
\newblock \emph{Phys. Rev. E}, 86\penalty0 (1):\penalty0 011301, 2012.
\newblock URL \url{https://doi.org/10.1103/PhysRevE.86.011301}.

\bibitem[Frizzell(2023{\natexlab{a}})]{frizzell2023code}
E.S. Frizzell.
\newblock Shock induced dilation, validation code, 2023{\natexlab{a}}.
\newblock URL \url{https://doi.org/10.5281/zenodo.7608511}.

\bibitem[Frizzell(2023{\natexlab{b}})]{frizzell2023data}
E.S. Frizzell.
\newblock Shock induced dilation, restart files and single run output code, 2023{\natexlab{b}}.
\newblock URL \url{https://doi.org/10.5281/zenodo.7608668}.

\bibitem[Zhang et~al.(2017)Zhang, Richardson, Barnouin, Maurel, Michel, Schwartz, Ballouz, Benner, Naidu, and Li]{zhang2017creep}
Y.~Zhang, D.C. Richardson, O.S. Barnouin, C.~Maurel, P.~Michel, S.R. Schwartz, R.-L. Ballouz, L.AM. Benner, S.P. Naidu, and J.~Li.
\newblock Creep stability of the proposed {AIDA} mission target 65803 {D}idymos: {I}. {D}iscrete cohesionless granular physics model.
\newblock \emph{Icarus}, 294:\penalty0 98--123, 2017.
\newblock URL \url{https://doi.org/10.1016/j.icarus.2017.04.027}.

\bibitem[Subramaniyan and Sun(2008)]{subramaniyan2008continuum}
A.K. Subramaniyan and C.T. Sun.
\newblock Continuum interpretation of virial stress in molecular simulations.
\newblock \emph{Int. J. of Solids and Struct.}, 45\penalty0 (14-15):\penalty0 4340--4346, 2008.
\newblock URL \url{https://doi.org/10.1016/j.ijsolstr.2008.03.016}.

\bibitem[van~den Wildenberg et~al.(2013)van~den Wildenberg, van Loo, and van Hecke]{van2013shock}
S.~van~den Wildenberg, R.~van Loo, and M.~van Hecke.
\newblock Shock waves in weakly compressed granular media.
\newblock \emph{Phys. Rev. Letters}, 111\penalty0 (21):\penalty0 218003, 2013.
\newblock URL \url{https://doi.org/10.1103/PhysRevLett.111.218003}.

\bibitem[Goddard(1990)]{goddard1990nonlinear}
J.D. Goddard.
\newblock Nonlinear elasticity and pressure-dependent wave speeds in granular media.
\newblock \emph{Proc. of the R. Soc. of Lond. S. A: Math. and Phys. Sci.}, 430\penalty0 (1878):\penalty0 105--131, 1990.
\newblock URL \url{https://doi.org/10.1098/rspa.1990.0083}.

\bibitem[Li and Holt(2002)]{li2002particle}
L.~Li and R.M. Holt.
\newblock Particle scale reservoir mechanics.
\newblock \emph{Oil \& Gas Sci. and Technol.}, 57\penalty0 (5):\penalty0 525--538, 2002.
\newblock URL \url{https://doi.org/10.2516/ogst:2002035}.

\bibitem[Somfai et~al.(2005)Somfai, Roux, Snoeijer, Van~Hecke, and Van~Saarloos]{somfai2005elastic}
E.~Somfai, J.-N. Roux, J.H. Snoeijer, M.~Van~Hecke, and W.~Van~Saarloos.
\newblock Elastic wave propagation in confined granular systems.
\newblock \emph{Phys. Rev. E}, 72\penalty0 (2):\penalty0 021301, 2005.
\newblock URL \url{https://doi.org/10.1103/PhysRevE.72.021301}.

\bibitem[Fa et~al.(2015)Fa, Zhu, Liu, and Plescia]{fa2015regolith}
W.Z. Fa, M.-H. Zhu, T.T. Liu, and J.B. Plescia.
\newblock Regolith stratigraphy at the chang'{E}-3 landing site as seen by lunar penetrating radar.
\newblock \emph{Geophys. Res. Lett.}, 42\penalty0 (23):\penalty0 10--179, 2015.
\newblock URL \url{https://doi.org/10.1002/2015GL066537}.

\bibitem[Mohan et~al.(2002)Mohan, Rao, and Nott]{mohan2002frictional}
L.S. Mohan, K.K. Rao, and P.R. Nott.
\newblock A frictional {C}osserat model for the slow shearing of granular materials.
\newblock \emph{J. of Fluid Mechanics}, 457:\penalty0 377--409, 2002.
\newblock URL \url{https://doi.org/10.1017/S0022112002007796}.

\bibitem[Fleischmann et~al.(2016)Fleischmann, Serban, Negrut, and Jayakumar]{fleischmann2016importance}
J.~Fleischmann, R.~Serban, D.~Negrut, and P.~Jayakumar.
\newblock On the importance of displacement history in soft-body contact models.
\newblock \emph{J. of Comput. and Nonlin. Dyn.}, 11\penalty0 (4), 2016.
\newblock URL \url{https://doi.org/10.1115/1.4031197}.

\bibitem[Mohamed and Gutierrez(2010)]{mohamed2010comprehensive}
A.~Mohamed and M.~Gutierrez.
\newblock Comprehensive study of the effects of rolling resistance on the stress--strain and strain localization behavior of granular materials.
\newblock \emph{Granular Matter}, 12\penalty0 (5):\penalty0 527--541, 2010.
\newblock URL \url{https://doi.org/10.1007/s10035-010-0211-x}.

\bibitem[Jiang et~al.(2015)Jiang, Shen, and Wang]{jiang2015novel}
M.~Jiang, Z.~Shen, and J.~Wang.
\newblock A novel three-dimensional contact model for granulates incorporating rolling and twisting resistances.
\newblock \emph{Comput. and Geotechnics}, 65:\penalty0 147--163, 2015.
\newblock URL \url{https://doi.org/10.1016/j.compgeo.2014.12.011}.

\end{thebibliography}


\end{document}